%% file: main.tex
\pdfoutput=1
\documentclass[11pt]{article}
\input{preamble}
\input{numbers/app_numbers}

\input{numbers/sc_avg_numbers}

\input{numbers/sc_avg_fc_numbers}

\input{numbers/edr_numbers}
\input{numbers/edr_loadings_numbers}

\input{numbers/edr_relabel_numbers}

\input{numbers/edr_placebo_numbers}
\input{numbers/edr_parametric_numbers}

\input{numbers/edr_block_numbers}

\input{numbers/overlap_numbers}

\input{numbers/sims_numbers}
\input{numbers/position_numbers}
\input{numbers/bands_numbers}
\input{numbers/stud_dates_numbers}
\input{numbers/comparators_numbers}

\title{\bf Unit-Level Conformal Inference for Panel Counterfactuals\thanks{Ziyi Liu, PhD candidate, Haas School of Business, University of California, Berkeley. Email: \url{zyliu2023@berkeley.edu}. Yiqing Xu, Associate Professor, Department of Political Science, Stanford University. Email: \url{yiqingxu@stanford.edu}. We thank Beniamino Green and P.~M.~Aronow for detecting an error in the parametric bootstrap procedure for interactive fixed effects estimated by EM (IFE-EM; \citealp{GreenAronow2026}), which partially motivated this paper. The authors used Claude Code (Anthropic) and Codex (OpenAI) as research, coding, and writing assistants, including for independent checks of the mathematics and for preparing this manuscript. All errors remain solely the responsibility of the authors.}}
\author{Ziyi Liu\\(UC Berkeley)\and Yiqing Xu\\(Stanford)}
\date{\today}

\begin{document}
\invisibletableofcontents
\etocdepthtag.toc{mtmain}
\onehalfspacing
\maketitle
\thispagestyle{empty}

\begin{abstract}
\noindent We develop conformal inference for panel counterfactual estimators using the exchangeability of prediction errors across units rather than over time. Existing conformal methods for panels with few treated units rely on temporal properties of treated-unit residuals, which are vulnerable to strong temporal heterogeneity such as nonstationarity. We instead hold out each control unit, hide its relevant outcomes, and predict them using the same rule applied to the treated unit. Target-date estimators, including difference-in-differences and synthetic control, use controls' pre-treatment and target-date outcomes to construct the counterfactual, so both full conformal confidence sets and Jackknife+ intervals apply. Full-path estimators, including factor-model imputation and matrix completion, also use controls' outcomes at other post-treatment dates; for these, we use the Jackknife+. Both procedures yield finite-sample, period-wise confidence sets under this across-unit exchangeability, without requiring stationarity or random assignment. We characterize when the coverage holds and show why a common prediction rule alone does not ensure it. For multiple treated units, a block Jackknife+ covers their average effect, while a relabeling test is exact for sharp nulls over all treated cells. We show that the synthetic control placebo test is a full conformal test of no effect that extends to full-path estimators. We also relate our procedures to a parametric bootstrap and classify common estimators by the inference they allow. We illustrate our procedures in three empirical applications.
\end{abstract}

\doublespacing
\clearpage
\setcounter{page}{1}

\section{Introduction}\label{sec:intro}

Over the past decade, a growing literature has proposed a wide variety of panel estimators for causal inference in
social science. They build the counterfactual for a treated unit in different ways, as a simplex-weighted average
of control units \citep{AbadieDiamondHainmueller2010}, from an interactive fixed-effects model \citep{Bai2009, GobillonMagnac2016, Xu2017}, by low-rank matrix completion \citep{AtheyBayati2021}, or by doubly weighted
synthetic difference-in-differences \citep[SDID;][]{ArkhangelskyAthey2021}. All of them target the treated unit's untreated potential outcome and differ mainly in how they use the control panel to approximate the treated unit's untreated path. For many applications, further gains in point estimation are likely to be
incremental.

Inference for these estimators is more difficult, especially with only one or a few treated units. In the canonical application, a single treated unit is observed
over a short post-treatment window, as with West Germany after reunification
\citep{AbadieDiamondHainmueller2015}. There is no cross-section of treated units to average over and no sampling
frame from which the treated unit was drawn. Standard $\sqrt{N}$
asymptotics thus do not apply.

One class of inference relies on the behavior of the treated unit's residuals over time.
\citet{ChernozhukovWuthrichZhu2021} subtract a hypothesized effect from the treated unit's post-treatment outcomes,
fit the panel estimator, and permute the residuals across dates to test whether the post-treatment residuals stand
out among the pre-treatment ones. This test is exact when the residuals are exchangeable over time and
approximately valid under stationarity and weak dependence. \citet{CattaneoFengTitiunik2021} bound the error in
the estimated weights under conditions that allow nonstationary outcomes, but they need to model the post-treatment
error itself and estimate its distribution from the pre-treatment residuals. The augmented synthetic control method
\citep{BenMichaelFellerRothstein2021} fits its ridge outcome model across units but bases inference on variation over time
in the same way. All three methods rest on properties of the treated unit's residuals over time.

Serial correlation and nonstationarity can undermine these properties. Strong serial correlation can break the weak dependence that approximate validity requires, while nonstationarity, such as unit-root idiosyncratic errors, can make pre-treatment residuals unrepresentative of post-treatment errors.\footnote{Relatedly, \citet{GreenAronow2026} report that,  in the \code{gsynth} package (before v1.3.1), the parametric bootstrap for interactive fixed effects estimated by EM (IFE-EM; \citealp{GobillonMagnac2016}) resampled in-sample residuals rather than out-of-sample prediction errors. In their simulations with serially correlated errors, cross-validation selected too many factors, shrinking residuals and reducing interval coverage. \citet{Xu2026response} explains the mechanism. Extra factors absorb noise shared by held-out dates and nearby fitting dates, lowering validation error and encouraging overfitting. Resampling the shrunken residuals then produces intervals that are too narrow.} Both can violate the temporal exchangeability needed for the test of \citet{ChernozhukovWuthrichZhu2021} to be exact. In our simulations, its coverage falls well below the nominal level when errors become noisier after treatment or follow random walks (Supplementary Materials~\ref{app:sims-comparators}). The conformal literature treats these departures as distribution shifts. Reweighting can address them, at a cost in coverage and effective sample size \citep{AngelopoulosBates2023}.

In this article, we compare the treated unit with the control units rather than with its own past. To give each control unit the treated unit's prediction task, we hold it out, hide its relevant outcomes, and predict them using the same rule. Such a unit is a \emph{held-out unit}, and the units from which a fit learns are its \emph{training units}. A unit's \emph{prediction error} is the difference between its untreated outcome and its out-of-sample prediction. This unit-level conformal inference requires neither random assignment nor a stationary error process. It requires only that the prediction errors be exchangeable across units, meaning that relabeling the units leaves their joint distribution unchanged. Intuitively, given each unit's pre-treatment history, a treated unit with no effect must be no harder to predict than any control. In the study of California's Proposition~99 by \citet{AbadieDiamondHainmueller2010}, for example, we hold out each of the 38 control states in turn, hide its 1995 cigarette sales, and predict them from the remaining training states. Comparing California's prediction error under each hypothesized effect with the 38 controls' errors yields a finite-sample confidence set for the 1995 effect.

The prediction errors of the treated unit and of each control need to be constructed in a symmetric way. Conformal
inference offers two procedures for doing so, full conformal inference
\citep{LeiGSellRinaldoTibshiraniWasserman2018} and the Jackknife+ \citep{BarberCandesRamdasTibshirani2021}.
Consider data points $(\mathbf{X}_i,Y_i)_{i=1,\ldots,J}$ and one further point with attributes $\mathbf{X}_0$ but a
missing outcome $Y_0$. Full conformal inference sets $Y_0$ to each candidate value in turn and computes every point's prediction error with that point held out. The confidence set collects the candidates at which the error on point $0$ is not extreme among the $J+1$ errors. Its guarantee requires these prediction errors, computed with the true value of $Y_0$, to be exchangeable. The Jackknife+ avoids candidate values by holding out each point $i$ \textit{together} with point $0$. It uses the error on point $i$ as a margin around that fold's prediction of $Y_0$ and forms the confidence set from order statistics of the endpoints across folds.

In a panel, $\mathbf{X}_i$ contains unit $i$'s pre-treatment outcomes, and $Y_i$ is its untreated outcome at a given post-treatment date, observed for controls but missing for the treated unit. A \emph{prediction rule} maps a held-out unit's $\mathbf{X}_i$ and its training units' outcomes into a prediction of $Y_i$. We distinguish two classes of panel estimators by which control outcomes they use to predict $Y_i$. \emph{Target-date} estimators, including difference-in-differences (DID) and synthetic control, use only pre-treatment and target-date outcomes to construct the prediction. \emph{Full-path} estimators also use outcomes at \emph{other} post-treatment dates. Examples include low-rank methods, such as generalized synthetic control (GSC; \citealp{Xu2017}), IFE-EM, and matrix completion \citep{AtheyBayati2021}.

Both full conformal inference and our Jackknife+ procedure apply to target-date estimators. For full-path estimators, full conformal inference faces a missing-data problem. When we hold out a control, the treated unit becomes a training unit, but its untreated outcomes at other post-treatment dates are unobserved. Even a confidence set for a single date therefore requires a joint search over all post-treatment dates. Our Jackknife+ procedure avoids this problem by fitting on controls alone and applies to both estimator classes.

Under the required exchangeability conditions, both procedures provide finite-sample, period-wise coverage at the requested level, for the effect at each post-treatment date separately. Under these conditions, the Jackknife+ tends to be conservative. In simulations in which units are exchangeable, both attain their coverage guarantees despite serially correlated errors and nonstationary factors. When the treated unit loads more heavily on a drifting factor than any control, DID and synthetic control undercover, while estimators that model the factor correctly retain higher coverage.

Our approach is related to, but distinct from, design-based inference \citep{LeiSudijono2024, BottmerImbensSpiessWarnick2024, FirpoPossebom2018}, which derives exchangeability from random treatment assignment. We instead start from an outcome model and decompose each unit's untreated outcome into a systematic \emph{signal} and an idiosyncratic error. The signal may be unit and time effects or common factors weighted by unit-specific loadings. Applying a common prediction rule mechanically defines the prediction errors of the treated and control units in the same way, but does not guarantee their exchangeability. Because each prediction error combines the unit's idiosyncratic error with the approximation error in fitting its signal, exchangeability depends on both the error distribution and how well the estimator captures each unit's signal. If the estimator misses differences between the treated unit's signal and the controls', the treated unit may be harder to predict, invalidating coverage. Pre-treatment diagnostics help detect such mismatches.

\begin{table}[!ht]\centering\small
\caption{Three approaches to finite-sample inference for synthetic control}
\begin{tabular}{@{}lR{0.2\textwidth}R{0.27\textwidth}R{0.25\textwidth}@{}}
\toprule
Approach & Exchangeable over & Earned from & Fragile when \\
\midrule
Time & residuals over time & exchangeability over time, or time-series conditions & dependence or drift the conditions exclude \\
Unit, design-based & which unit is treated & the assignment law & non-random adoption \\
Unit, model-based & units' prediction errors & comparable units given the structure & heterogeneous signals the estimator does not remove \\
\bottomrule
\end{tabular}
\floatnote{This article develops the third approach. The in-space placebo test of \citet{AbadieDiamondHainmueller2010} belongs to the second approach when the treated unit is chosen at random and to the third when the units' statistics are exchangeable under no effect.}
\label{tab:axes}
\end{table}

Table~\ref{tab:axes} compares the three approaches to finite-sample inference for panel estimators with few treated units. The in-space placebo test of \citet{AbadieDiamondHainmueller2010} and the first step of the parametric bootstrap in \citet{Xu2017} also compare the treated unit with controls that are refit as if treated. Section~\ref{sec:connections} shows that the placebo test is a full conformal test of no effect. It also shows that the parametric bootstrap pairs the controls' errors with a single prediction for the treated unit, as the plain jackknife does.

The rest of the article is organized as follows. Section~\ref{sec:ordinary-estimator} constructs the two conformal confidence sets. Section~\ref{sec:score-conditions} states when the prediction errors are exchangeable. Section~\ref{sec:connections} relates the procedures to existing methods, and Section~\ref{sec:extensions} extends them to multiple treated units and staggered adoption. Section~\ref{sec:evidence} reports simulations and three applications, and Section~\ref{sec:discussion} offers recommendations. The Supplementary Materials (SM) contain the formal statements and proofs.

\section{Conformal Inference for Panel Estimators}
\label{sec:ordinary-estimator}

In this section, we first define the two estimator classes. We then give
each control the treated unit's prediction task by holding it out and hiding the outcomes that the treated unit lacks.
Finally, we compare the treated unit's prediction error with the controls' errors in two ways, by full conformal
inference and by the Jackknife+.

\subsection{The estimator and the dates it uses}
\label{sec:estimator-support}

We consider one treated unit and one post-treatment date. The panel contains \(N=J+1\) units indexed by
\(\Iset=\{0,1,\ldots,J\}\). Unit \(0\) is treated, and the untreated controls form the set
\(\Cset=\{1,\ldots,J\}\). Let \(\Tpre\) and \(\Tpost\) denote the sets of pre- and post-treatment dates. For
\(t\in\Tpost\), the realized effect on unit \(0\) is
\begin{equation}
  \tau_t=Y_{0t}-Y_{0t}(0),
  \label{eq:target-effect}
\end{equation}
where \(Y_{0t}(0)\) is the missing untreated outcome. We assume no anticipation and no spillovers to the controls,
so all pre-treatment outcomes and all control outcomes are untreated. In the Proposition~99 example, unit \(0\) is
California, \(J=38\), \(\Tpost\) runs from 1989 to 2000, and \(t\) might be 1995.

The estimator predicts \(Y_{0t}(0)\) by a rule
\begin{equation}
  \widehat Y_{0t}(0)
  =
  \mathcal A_t
  \left(
    Y_{0,\Tpre};
    Y_{\Cset,\Tpre\cup\Tpost};
    \widehat\eta_t
  \right),
  \label{eq:ordinary-rule}
\end{equation}
and estimates the effect as \(\widehat\tau_t=Y_{0t}-\widehat Y_{0t}(0)\). The argument $\widehat\eta_t$ collects the tuning choices, such as the number of factors, the regularization penalty, and the convergence criterion. The inference procedures below refit the estimator many times, each time giving a control the treated unit's prediction task, and we hold $\widehat\eta_t$ fixed across these refits. All following probability statements thus condition on it.\footnote{Holding $\widehat\eta_t$ fixed also keeps the distinction between target-date and full-path estimators well defined. A tuning value selected from the controls' outcomes at other post-treatment dates would otherwise carry those outcomes into a target-date prediction.} Section~\ref{sec:score-conditions} discusses how to choose $\widehat\eta_t$ so that holding it fixed does not undermine the coverage guarantees.

\input{figures/figure_temporal_support.tex}

\phantomsection\label{sec:pointwise-estimators}%
Estimators differ in which of the controls' outcomes in $\Tpost$ can affect the prediction at~\(t\). A
\emph{target-date} estimator uses the controls' pre-treatment histories and their outcomes at \(t\), and no
outcome at another post-treatment date:
\begin{equation}
  \widehat Y_{0t}^{\mathrm{td}}(0)
  =
  \mathcal A_t^{\mathrm{td}}
  \left(
    Y_{0,\Tpre};
    Y_{\Cset,\Tpre},Y_{\Cset,t};
    \widehat\eta_t
  \right).
  \label{eq:pointwise-rule}
\end{equation}
Synthetic control \citep{AbadieDiamondHainmueller2010} is a target-date estimator. It fits the treated unit's
pre-treatment path with a weighted average of the control paths and applies those weights to the controls'
outcomes at \(t\). SDID \citep{ArkhangelskyAthey2021} with \(t\) as its only post-treatment target is also a
target-date estimator, and so is saturated event-date two-way fixed effects (TWFE) under balanced block
adoption (SM Lemma~\ref{lem:twfe-pointwise-app}).

\phantomsection\label{sec:pooled-estimators}%
We call an estimator \emph{full-path} if its prediction at \(t\) can use the controls' outcomes at every
post-treatment date:
\begin{equation}
  \widehat Y_{0t}^{\mathrm{fp}}(0)
  =
  \mathcal A_t^{\mathrm{fp}}
  \left(
    Y_{0,\Tpre};
    Y_{\Cset,\Tpre\cup\Tpost};
    \widehat\eta_t
  \right).
  \label{eq:pooled-rule}
\end{equation}
In our full-panel implementations, interactive fixed effects estimated via expectation-maximization (IFE-EM; \citealp{GobillonMagnac2016}), GSC \citep{Xu2017}, and matrix completion \citep{AtheyBayati2021} are full-path estimators, as Figure~\ref{fig:ordinary-temporal-support} shows.

\FloatBarrier
\subsection{Prediction error and held-out units}
\label{sec:role-extension}

Conformal inference compares the treated unit's prediction error at date $t$ with the out-of-sample prediction errors of the controls. In-sample residuals are unsuitable because an estimator may fit the training controls closely yet predict $Y_{0t}(0)$ poorly. Instead, we hold out a set of units $Q\subseteq\Iset$, hide their relevant outcomes, and predict their date-$t$ outcomes from the remaining \emph{training units} $\Iset\setminus Q$. For each \emph{held-out unit} $i\in Q$, we denote the prediction by $\widehat Y_{it}^{[Q]}(0)$ and its prediction error by $Y_{it}(0)-\widehat Y_{it}^{[Q]}(0)$.

Equation~\eqref{eq:ordinary-rule} defines the prediction when only unit~$0$ is held out, so that $Q=\{0\}$. To extend this definition to other held-out sets and treat units symmetrically, we use a \emph{common prediction rule}. Every held-out unit is predicted using the same estimation algorithm, the same dates, and the same fixed tuning choices $\widehat{\eta}_t$. The rule for hiding outcomes is also the same across units. For a fixed $Q$, predictions therefore differ across held-out units only because their pre-treatment histories differ. Figure~\ref{fig:role-symmetric-extensions} illustrates the resulting prediction tasks. The left and right columns show target-date and full-path estimators, respectively; the top and bottom rows show one and two held-out units. The outlined column marks the target date $t$.

\input{figures/figure_role_symmetric_extensions.tex}

Which outcomes must be hidden depends on the estimator. The rule is to match the information available for unit~$0$, whose untreated outcomes are observed only before treatment. Each held-out unit therefore retains its pre-treatment history and hides any post-treatment outcomes that the estimator would otherwise use. For a target-date estimator, only the outcome at $t$ must be hidden. Other post-treatment dates are outside the fit, as panels~(a) and~(c) show. A full-path estimator uses every post-treatment date, so the held-out unit's entire post-treatment path must be hidden, as illustrated by the hatched control outcomes in panels~(b) and~(d). Otherwise, the estimator could use more information about a held-out control than is available for unit~$0$.

The top row of Figure~\ref{fig:role-symmetric-extensions} shows what happens when a control is held out alone, $Q=\{j\}$. Unit~$0$ then remains in the training set, even though its untreated post-treatment outcomes are unobserved. For the target-date estimator in panel~(a), the only missing training outcome is $Y_{0t}(0)$. Full conformal inference completes this training outcome with a candidate value. Panel~(b) shows why the same scalar candidate does not suffice for a full-path estimator. Filling unit~$0$'s target cell leaves its other post-treatment cells, marked with question marks, unresolved. A full-path fit would require candidate values for the entire untreated post-treatment path.

The bottom row shows the alternative, in which control~$j$ is held out together with unit~$0$, so that $Q=\{0,j\}$. Both units are predicted using the remaining controls, whose outcomes are observed at every required date. No candidate values are therefore needed to complete the training data. This construction underlies the Jackknife+ and applies to both target-date estimators, as in panel~(c), and full-path estimators, as in panel~(d). For the full-path case, panel~(d) represents two separate fits, one for each held-out unit, using that unit's pre-treatment history and the same training controls. Sections~\ref{sec:conformal-sets} and~\ref{sec:main-jplus} develop full conformal inference and the Jackknife+, respectively.

\FloatBarrier

\subsection{Full conformal inference for target-date estimators}
\label{sec:conformal-sets}
\label{sec:main-full-conformal}

We use test inversion to obtain the confidence set for $\tau_t$. For each candidate effect $c\in\R$, we test it against the data and keep the candidates that the data do not reject. Specifically, the null is period-wise and can be written as:
\begin{equation}
  H_{0,t}(c):\tau_t=c,
  \qquad\text{equivalently}\qquad
  Y_{0t}(0)=Y_{0t}-c.
  \label{eq:common-null}
\end{equation}

Full conformal inference \citep{Vovk2005,
LeiGSellRinaldoTibshiraniWasserman2018} completes the panel under
$H_{0,t}(c)$:
\begin{equation}
  Y_{it}^{(c)}
  =
  \begin{cases}
    Y_{0t}-c,&i=0,\\
    Y_{it},&i\in\Cset.
  \end{cases}
  \label{eq:candidate-completion-main}
\end{equation}
If $c$ is the true effect, the completed panel contains only untreated
outcomes, and unit~$0$ should be no harder to predict than a control. To obtain every unit's prediction error, we hold out each unit in turn, hide its completed outcome at $t$, and predict it by the prediction rule with one held-out unit.

The obtained absolute
prediction error on unit $i$ is its \emph{score},
\begin{equation}
  S_i(c)
  =
  \left|
    Y_{it}^{(c)}-\widehat Y_{it}^{(-i)}(0;c)
  \right|,
  \label{eq:fc-score-main}
\end{equation}
where $\widehat Y_{it}^{(-i)}(0;c)$ is the prediction when unit $i$
alone is held out, from the panel completed under $c$. We require the
rule to reproduce the estimator when unit~$0$ is held out alone, so
$S_0(c)=|\widehat\tau_t-c|$. A candidate far from the estimate thus
gives unit~$0$ a large score. We rank unit~$0$'s score among all $N$
scores, counting ties in favor of the candidate, and keep the
candidates whose rank $p$-value exceeds the miscoverage level
$\gamma\in[0,1)$:
\begin{equation}
  p_t^{\mathrm{FC}}(c)
  =
  \frac{1}{N}
  \sum_{i\in\Iset}
  \one\{S_i(c)\geq S_0(c)\},
  \qquad
  \widehat C_t^{\mathrm{FC}}(\gamma)
  =
  \{c\in\R:p_t^{\mathrm{FC}}(c)>\gamma\}.
  \label{eq:fc-pvalue-main}
\end{equation}

At the true effect, the test rejects only when unit~$0$'s score is
among the $\lfloor\gamma N\rfloor$ largest. Suppose the score vector
$\bigl(S_0(\tau_t),\ldots,S_J(\tau_t)\bigr)$ is \emph{exchangeable} across
units, that is, its conditional distribution is the same after every
relabeling of the units. Unit~$0$'s score is then no more likely than any
other to be among these largest scores, so
\begin{equation}
  \Pr\!\left\{
    \tau_t\in\widehat C_t^{\mathrm{FC}}(\gamma)
    \mid\Ginfo
  \right\}
  \geq
  1-\frac{\lfloor\gamma N\rfloor}{N}
  \geq1-\gamma,
  \label{eq:fc-coverage-main}
\end{equation}
where $\Ginfo$ denotes the conditioning information, the units' signals at the dates the estimator uses and the
fixed tuning values (SM~\ref{app:fc-proof}).

\subsection{The Jackknife+}
\label{sec:main-jplus}\label{sec:main-pooled-jplus}

Full conformal inference refits every control with a loop over every candidate. The
Jackknife+ of \citet{BarberCandesRamdasTibshirani2021} avoids the loop. It holds out unit~$0$ together with each control in turn, so
unit~$0$ is never a training unit and no candidate is needed for
either estimator class. In fold $j$, we apply the prediction rule with
two held-out units, $Q=\{0,j\}$, and hide both units' outcomes as in
Section~\ref{sec:role-extension}, at $t$ for a target-date estimator
and over the whole post-treatment path for a full-path estimator. The fold records the
two predictions:
\begin{equation}
  \widehat M_{0j}
  =\widehat Y_{0t}^{[\{0,j\}]}(0),
  \qquad
  \widehat M_{j0}
  =\widehat Y_{jt}^{[\{0,j\}]}(0).
  \label{eq:jplus-pair-predictions-main}
\end{equation}

The control's absolute prediction error is thus $R_{j0}=|Y_{jt}-\widehat M_{j0}|$. We use this error as a margin around unit~$0$'s prediction to form its prediction interval in fold~$j$, $[\widehat M_{0j}-R_{j0},\,\widehat M_{0j}+R_{j0}]$, which contains $Y_{0t}(0)$ whenever unit~$0$'s absolute prediction error in fold~$j$ is no larger than $R_{j0}$. Repeating this step across all $J$ controls gives $J$ fold intervals. Write $a_j=\widehat M_{0j}-R_{j0}$ and $b_j=\widehat M_{0j}+R_{j0}$ for the endpoints of fold~$j$, and $a_{(1)}\leq\cdots\leq a_{(J)}$ and $b_{(1)}\leq\cdots\leq b_{(J)}$ for the ordered endpoints. For an integer $1 \leq q \leq \lfloor N/2 \rfloor$, the Jackknife+ prediction interval runs from the $q$-th smallest lower endpoint, $a_{(q)}$, to the $q$-th largest upper endpoint, $b_{(N-q)}$, across these folds. Subtracting it from $Y_{0t}$ gives the effect interval
\begin{equation}
  \widehat C_t^{\tau,\mathrm{J+}}(\gamma)
  =\bigl[\,Y_{0t}-b_{(N-q)},\ Y_{0t}-a_{(q)}\,\bigr].
  \label{eq:jplus-set-main}
\end{equation}

To bound the coverage probability, we consider a hypothetical $N\times N$ table built from the complete untreated panel. Each off-diagonal entry $R_{ik}$ is unit~$i$'s absolute prediction error when units~$i$ and $k$ are held out together.\footnote{When unit~$0$ is not one of the held-out units, the refit uses unit~$0$ as a training unit and requires its unobserved untreated outcomes, so $R_{ik}$ is unobservable and used only in the proof.} In this table, unit~$0$'s prediction error in fold~$j$ is $R_{0j}=|Y_{0t}(0)-\widehat M_{0j}|$, and its paired control's error is $R_{j0}$. By construction, the Jackknife+ interval misses $Y_{0t}(0)$ only if unit~$0$ has the larger error, $R_{0j} > R_{j0}$, in at least $N-q$ of its $J$ folds. The question is therefore how many of the $N$ units in the table can satisfy $R_{ik} > R_{ki}$ against at least $N-q$ other units~$k\neq i$.

For any pair of units~$i$ and $k$, at most one of $R_{ik} > R_{ki}$ and $R_{ki} > R_{ik}$ can hold. When $q=1$, at most one unit can satisfy $R_{ik} > R_{ki}$ against all $N-1$ other units, because two such units could not both have the larger error when paired with each other. More generally, counting the pairwise comparisons shows that at most $2q-1$ units can satisfy $R_{ik} > R_{ki}$ against at least $N-q$ other units (SM~Theorem~\ref{thm:generic-jplus-app}). If the table is exchangeable in
the sense that its conditional distribution stays unchanged when its rows and
columns are relabeled together, unit~$0$ is no more
likely than any other unit to be among these $2q-1$ units. We thus set
$q$ to $q_\gamma$, the largest integer with $(2q-1)_+\leq\gamma N$, so
that
\begin{equation}
  \Pr\!\left\{
    \tau_t\in
    \widehat C_t^{\tau,\mathrm{J+}}(\gamma)
    \mid\Ginfo
  \right\}
  \geq
  1-\frac{(2q_\gamma-1)_+}{N}
  \geq1-\gamma.
  \label{eq:jplus-coverage-main}
\end{equation}
When $q_\gamma=0$, the interval is $\R$. Both
guarantees~\eqref{eq:fc-coverage-main} and~\eqref{eq:jplus-coverage-main} are period-wise. SM~\ref{app:bands} extends them to bands over
the whole effect path, built from each unit's largest error across the
post-treatment dates.

\begin{remark}[Full conformal inference for full-path estimators]\label{sec:pooled-fc-difficulty}
	Full conformal inference can still test the sharp null of \emph{no effect} at every post-treatment date for either estimator class. Under this null, each control refit has the complete panel it needs after plugging in the null for the treated unit. However, building a confidence set for a single date $t$ is harder for a full-path estimator. As shown in Section~\ref{sec:role-extension}, each full-path refit requires a candidate effect at every post-treatment date, not just at date $t$. We would have to search over a $|\Tpost|$-dimensional grid of candidate paths and project the accepted paths onto date $t$ (SM~\ref{app:pooled-fc}). Because this joint search is computationally expensive, we restrict full conformal inference to target-date estimators and use the Jackknife+ for both estimator classes. SM~\ref{app:algorithms} provides pseudocode for both procedures.
\end{remark}

\begin{remark}[Coverage with few units]\label{rem:few-units}
	With few units, the guaranteed coverage can take only a few values. As $\gamma$ varies, the guaranteed coverage of full conformal inference, $1-\lfloor\gamma N\rfloor/N$, moves in steps of $1/N$. That of the Jackknife+, $1-(2q_\gamma-1)_+/N$, moves in steps of $2/N$ after its first step. When the number of control units is small, such steps can be large and the guaranteed coverage at $\gamma=0.10$ may exceed 90 percent. When $\gamma < 1/N$, neither procedure can reject any candidate value, and both return $\R$. Table~\ref{tab:levels} lists the guaranteed coverage levels for several sample sizes.
\end{remark}

\section{Exchangeable Scores: Conditions and Diagnostics}
\label{sec:score-conditions}

Both coverage guarantees require scores to be exchangeable across units. The common prediction rule constructs the scores symmetrically but cannot ensure that they are exchangeable. In this section, we give two sufficient conditions for score exchangeability and use DID to illustrate how coverage can fail without them. We also propose diagnostics comparing unit \(0\)'s pre-treatment prediction errors with the controls' errors.

\subsection{Error and score exchangeability}\label{sec:error-score}

We assume that untreated outcomes decompose into a unit-specific fixed signal \(\mu_{is}\) plus an idiosyncratic error \(e_{is}\), \(Y_{is}(0)=\mu_{is}+e_{is}\). Unit \(i\)'s error path is \(e_i\equiv\{e_{is}\}_{s\in\Tpre\cup\{t\}}\) for a target-date estimator and \(e_i\equiv\{e_{is}\}_{s\in\Tpre\cup\Tpost}\) for a full-path estimator. Let \(\Ginfo\) denote the collection of unit signals, the chosen tuning values \(\widehat\eta_t\), and any other conditioning information. Let \(P_\pi\) be a permutation of the unit coordinates. The basic assumption is that the unit-level error paths \(\mathbf e=(e_0,\ldots,e_J)\) are exchangeable:
\begin{equation} \label{eq:error-exchangeability-main}
\mathbf e\mid\Ginfo\ \stackrel{d}{=}\ P_\pi\mathbf e\mid\Ginfo
\qquad\text{for every unit permutation }\pi.
\end{equation}
This assumption permits serial dependence and nonstationarity in each unit's error path, and it permits cross-unit dependence as long as the joint distribution of the error paths is invariant to reordering of the units.

Assumption~\eqref{eq:error-exchangeability-main} depends on \(\Ginfo\), which includes the tuning values \(\widehat\eta_t\). For example, if we estimate a factor model for the Proposition~99 data with two factors fixed before observing the data, this choice reveals nothing about the realized outcomes. Assumption~\eqref{eq:error-exchangeability-main} then requires only that the unit-level error paths be exchangeable. Now suppose we select two factors by cross-validating on the 38 control states and reuse that choice across subsequent refits. Because the selected value reveals information about the realized error paths of the controls but not of California, conditioning on \(\widehat\eta_t\) breaks the symmetry between California and the controls. Selecting a tuning value from California's pre-treatment fit creates the same asymmetry. SM Proposition~\ref{prop:tuning-app} shows that unit-level error paths remain exchangeable conditional on \(\widehat\eta_t\) only if the selection of \(\widehat\eta_t\) is invariant to permuting those paths across units. Choosing the tuning values in advance guarantees this invariance.

Full conformal inference needs the score vector \(\mathbf S_t(\tau_t)=(S_0(\tau_t),\ldots,S_J(\tau_t))^{\top}\) at
the true effect to keep its conditional distribution when the units are relabeled:
\begin{equation}
  \mathbf S_t(\tau_t)\mid\Ginfo
  \ \stackrel{d}{=}\
  P_\pi\mathbf S_t(\tau_t)\mid\Ginfo,
  \label{eq:score-exchangeability-main}
\end{equation}
and the Jackknife+ needs the hypothetical all-pairs table \(\mathbf R=[R_{ik}]_{i,k\in\Iset}\) to satisfy
\begin{equation}
  \mathbf R\mid\Ginfo
  \ \stackrel{d}{=}\
  P_\pi\mathbf R P_\pi^{\top}\mid\Ginfo
  \qquad\text{for every }\pi.
  \label{eq:matrix-exchangeability-main}
\end{equation}
The rows and columns of \(\mathbf R\) move together because each unit appears twice, as the unit whose error is
recorded and as the other held-out unit in a fold.

\phantomsection\label{sec:bridge-main}%

To see why exchangeable errors alone do not make the scores exchangeable, consider unit \(i\)'s signed prediction
error at the true effect when it is the only held-out unit:
\begin{equation}
  Y_{it}(0)-\widehat Y_{it}^{(-i)}(0;\tau_t)
  =
  e_{it}+d_i,
  \qquad
  d_i
  =
  \mu_{it}-\widehat Y_{it}^{(-i)}(0;\tau_t).
  \label{eq:prediction-discrepancy-main}
\end{equation}
We term \(d_i\), the gap between the unit's signal and its prediction, the \emph{approximation error}. It depends
on the errors through the fit, and the prediction errors from fits with two held-out units decompose in the same way. A
common prediction rule applies one procedure to every unit, but the approximation errors can still differ
systematically across units with different signals.

Score exchangeability therefore needs a further condition, which we term \emph{equal predictability}. With the
signals and tuning fixed, moving the error paths between units and refitting must give scores with the same distribution as the relabeled scores (SM Assumption~\ref{ass:compatibility-app}). Intuitively, the fit must not approximate California's signal systematically better or worse than the control
states' signals. Given exchangeable errors, equal predictability is equivalent to
\eqref{eq:score-exchangeability-main} or \eqref{eq:matrix-exchangeability-main}, as Figure~\ref{fig:bridge}
summarizes (SM~\ref{app:bridge-proof}).

\begin{figure}[t]
\centering
\caption{From exchangeable errors to conformal coverage}
\begin{tikzpicture}[
  font=\small, node distance=4mm and 9mm,
  box/.style={draw=paperaccent!80, rounded corners=2pt, fill=celltrain!30, align=center, inner sep=4pt, minimum height=10mm, execute at begin node=\setstretch{1}},
  res/.style={box, fill=celltrain},
  arr/.style={-{Stealth[length=2.2mm]}, line width=.6pt, draw=paperaccent}]
  \node[box, text width=40mm] (err) {exchangeable\\error paths};
  \node[box, text width=40mm, below=of err] (comp) {prediction errors satisfy\\equal predictability};
  \coordinate (mid) at ($(err.east)!.5!(comp.east)$);
  \node[res, text width=25mm, right=14mm of mid] (sc) {exchangeable\\scores};
  \node[res, text width=25mm, right=of sc] (cov) {conformal\\coverage};
  \draw[arr] (err.east) -- ++(6mm,0) |- (sc.west);
  \draw[arr] (comp.east) -- ++(6mm,0) |- (sc.west);
  \draw[arr] (sc) -- (cov);
\end{tikzpicture}
\floatnote{Given exchangeable error paths, equal predictability is
equivalent to exchangeable scores, which give the coverage guarantees of
Sections~\ref{sec:main-full-conformal}--\ref{sec:main-jplus}.}
\label{fig:bridge}
\end{figure}

\subsection{Sufficient conditions and a counterexample}\label{sec:sufficient-conditions}

Either of two conditions makes the scores exchangeable. The first is \emph{exchangeable units}, meaning that relabeling units leaves the joint distribution of their complete untreated paths unchanged, conditional on shared information such as common factors. Unit-specific signals remain random, so coverage averages over them rather than holding them fixed. Tuning must be fixed in advance or chosen by a rule that treats units alike. A common prediction rule then gives exchangeable scores (SM Proposition~\ref{prop:complete-unit-route-app}). The fitted model need not be correct; it may use one factor when outcomes depend on three.

The second requires the signals to be fixed and the estimator to remove them exactly. Prediction errors must then depend on the error paths alone. Swapping two units' error paths while keeping their signals fixed must also swap their prediction errors. Exchangeable error paths then give exchangeable scores (SM Proposition~\ref{prop:residualization-route-app}). Consistency or approximate unbiasedness alone is insufficient.

Without either condition, a common prediction rule and exchangeable errors need not yield exchangeable scores. For example, with signals \(\mu_{is}=\alpha_i+\lambda_i f_s\), DID removes the unit effect but leaves a prediction error containing
\[
(\lambda_i-\bar\lambda_{-i})(f_t-\bar f_{\Tpre}),
\]
where \(\bar\lambda_{-i}\) is the other units' average loading. This term can be large when a unit's loading differs from the others' and the factor moves away from its pre-treatment mean.

In the Proposition 99 example, suppose only California loads on the factor and \(f_t\ne\bar f_{\Tpre}\). With \(N=J+1\geq3\) units and no errors, California's absolute prediction error is \(N-1\) times each control's. Its full conformal \(p\)-value is \(1/N\), so the test always rejects the true null when \(\gamma\geq1/N\).

\FloatBarrier
\subsection{Pre-treatment diagnostics}
\label{sec:historical-diagnostics}

Before treatment, unit \(0\)'s untreated outcomes are observed, allowing us to compare its prediction errors with those of held-out controls. We use the last \(m\) pre-treatment dates as \emph{diagnostic dates}, with \(m\) fixed in advance. Our simulations and applications use \(m=4\), corresponding to the four years before 1989 for California.

At each diagnostic date \(s\), we predict \(Y_{0s}\) using unit \(0\)'s history before \(s\) and compute
\begin{equation}
  \widehat\tau_s^{\mathrm{pl}}
  =Y_{0s}-\widehat Y_{0s}(0).
  \label{eq:placebo-estimate-main}
\end{equation}
Under no anticipation, the true pseudo-effect is zero. We apply the post-treatment inference procedure, with tuning at \(s\) fixed in advance or chosen by a rule that treats units alike.

For target-date estimators, \(s\) replaces \(t\), and dates before \(s\) replace \(\Tpre\) in Section~\ref{sec:ordinary-estimator}. Full-path fits use both held-out units' outcomes only before \(s\), while training units contribute later outcomes as well. Full-path diagnostics are therefore retrospective, as Figure~\ref{fig:historical-information} shows.

\input{figures/figure_historical_diagnostics.tex}

We report full conformal sets and \(p\)-values for target-date estimators. For both classes, we report Jackknife+ intervals and \(p\)-values that give the same rejection decisions as the intervals (SM~\ref{app:historical-diagnostics}). In event-study plots of the estimates over time, such as Figure~\ref{fig:apps-sc}, each pre-treatment point shows the prediction error in \eqref{eq:placebo-estimate-main}, and its bar shows the hypothetical effects not rejected at that date. These points share a time axis with post-treatment effects but should not be interpreted as event-study leads.

Under the relevant score condition, the sets cover zero at their stated finite-sample rates, and null \(p\)-values fall below any threshold with probability at most that threshold (SM Corollary~\ref{cor:placebo-validity-app}). The \(m\) diagnostics share overlapping histories and may be dependent. Without a multiplicity adjustment, they provide neither a simultaneous band nor a joint pre-trend test. When the score conditions hold, the expected number of rejections at level \(\gamma\) is at most \(\gamma m\), or 0.4 when \(m=4\) and \(\gamma=0.10\). Increasing \(m\) allows more chance rejections and includes earlier dates with shorter fitting histories. We therefore recommend a small \(m\), with enough history before the earliest diagnostic date.

Repeated rejections may indicate problems with the signals, controls, dates used, or prediction rule, but cannot distinguish among them. Intervals containing zero offer little reassurance when they are wide or when too few units are available for informative ranks. Score conditions may hold before treatment and fail afterward, or vice versa. SM~\ref{app:algorithms} gives the algorithm for the pre-treatment diagnostics.

\section{Connections to Existing Methods}\label{sec:connections}

In this section, we relate the two procedures to existing ones. The
in-space placebo test of \citet{AbadieDiamondHainmueller2010} and the first step of the parametric bootstrap in
\citet{Xu2017} both refit the estimator with a control in the treated unit's place. They differ in whether the treated unit is
a training unit when a control is held out. The same difference separates full conformal inference from the
Jackknife+.

\subsection{Synthetic control's placebo test}\label{sec:equiv}

The placebo test reassigns the intervention to each control state and, in the authors' words, shifts the treated
unit to the donor pool \citep[][p.~501]{AbadieDiamondHainmueller2010}. Each control state is thus held out in turn,
and California, with its observed outcomes, is a training unit in each of these fits. Under the null of no effect, the treated unit's outcomes used in each control's fit are untreated. These fits therefore coincide with the one-unit holdouts in Section~\ref{sec:main-full-conformal} at $c=0$, and ranking the treated unit's statistic among the controls' gives a full conformal test.

The usual placebo implementation differs from ours in two respects. First, it ranks the ratio of post- to pre-treatment root mean squared prediction error, testing no effect at all post-treatment dates jointly. Second, it is rarely inverted into a confidence set. Each candidate requires completing the treated outcomes and refitting every control, as in~\eqref{eq:candidate-completion-main}. For full-path estimators such as GSC, the joint test of no effect at all post-treatment dates remains available, but inversion requires a joint search over candidate effects at every post-treatment date.

The placebo test and plot rely on two assumptions that users should state. First, all placebo fits use the same estimator, tuning, and donor-pool rule, keep the treated unit in the donor pool, and exclude no unit for poor pre-treatment fit. Some tests drop placebo states whose pre-treatment fit is much worse than California's. Our validity argument does not cover these tests because the choice of controls depends on the treated unit's fit. It also does not cover removing the treated unit from placebo donor pools. That procedure is the jackknife construction discussed in Section~\ref{sec:bootstrap}. Second, the statistics are exchangeable under the null. Under these two assumptions, the ratio $p$-value is valid in finite samples (SM Theorem~\ref{thm:multi-relabel-app} with one treated unit).

Random assignment can replace the second assumption by making the treated statistic a uniform draw from the $N$ statistics. Validity then holds over assignments, as in Fisher's randomization test. Without random assignment, exchangeability can follow from exchangeable units, the first sufficient condition in Section~\ref{sec:sufficient-conditions} (SM Proposition~\ref{prop:multi-relabel-route-app} with one treated unit). For absolute errors at a single date, exchangeability is the score condition~\eqref{eq:score-exchangeability-main}.

\subsection{The parametric bootstrap in \texorpdfstring{\citet{Xu2017}}{Xu (2017)}}\label{sec:bootstrap}

Unlike the placebo test, the parametric bootstrap of \citet[][Algorithm~2]{Xu2017} leaves the treated unit out of
every control's fit. In each round of its first step, it holds out one control drawn at random as if it were treated. The bootstrap then fits the factors on a resample of
the remaining controls drawn with replacement, and predicts the held-out control from its pre-treatment history.
Without the resampling, these prediction errors would give the control errors \(R_{j0}\) of the Jackknife+ for a
full-path estimator. The parametric bootstrap draws error paths from the prediction errors of these resampled fits and adds them to a single prediction for the treated unit from the fit on all controls. It then refits the estimator on the simulated panel.

Pairing out-of-fold control errors with one prediction for the treated unit is the plain jackknife, which carries
no finite-sample guarantee. The parametric bootstrap adds resampling and refitting but keeps this pairing. The
Jackknife+ keeps the same control errors but recomputes the treated unit's prediction in each fold. It replaces the bootstrap's simulation with order statistics and a rank argument that requires only the all-pairs table condition~\eqref{eq:matrix-exchangeability-main}.

The parametric bootstrap therefore requires different assumptions, which users should state.\footnote{Equal variances, as assumed in \citet{Xu2017}, are insufficient. \citet{Xu2026correction} restates these assumptions.} The interactive fixed effects model must hold for treated and control units alike. The estimator must use at least the true number of factors but few enough to estimate each unit's loadings accurately from pre-treatment data \citep{IwasakiXu2026}. The controls' loadings must also span the factor space, so that every factor that moves the treated unit moves enough controls to be estimated \citep{Bai2009, Xu2017}. Treatment and covariates must be unrelated to errors at every date. Errors must be independent across units, satisfy \citet{Bai2009}'s restrictions on dependence over time, and have the same distribution for treated and control units. Percentile intervals, whose endpoints are quantiles of the bootstrap estimates, also require symmetrically distributed estimation errors. Even then, coverage is approximate, reaching the nominal rate only as both the number of controls and the number of pre-treatment periods grow.

\subsection{A typology of panel estimators}\label{sec:typology}

In Table~\ref{tab:scope}, we classify existing panel estimators by which of our procedures applies to each estimator and what assumptions on the units are needed for the corresponding coverage guarantee.

\afterpage{%
\begin{landscape}
\begin{table}[p]\centering\small
\singlespacing
\caption{Inference supported by common panel estimators}
\setlength{\tabcolsep}{3pt}
\begin{tabular}{@{}R{0.27\linewidth}cR{0.23\linewidth}R{0.37\linewidth}@{}}
\toprule
Estimator & Estimator class & Procedures & What the guarantee requires \\
\midrule
\multicolumn{4}{@{}l}{\normalsize\itshape\bf The DID and TWFE family} \\
Static TWFE ($Y$ on a single treatment dummy), several post-treatment dates & --- & none & the common coefficient uses treated post-treatment outcomes to predict the treated unit; outside the framework (with one post-treatment date it equals DID) \\
Saturated event-date TWFE; DID imputation, one treated unit (controls observed at every date) & target-date & full conformal (exact, affine); Jackknife+ & the estimator removes the signal exactly when the signal is additive \\
Callaway--Sant'Anna (no covariates), never-treated comparison & target-date & full conformal; Jackknife+ & as above; its two-by-two comparisons use the base period and the target date only \\
Callaway--Sant'Anna (no covariates), not-yet-treated comparison & target-date & full conformal; Jackknife+ & as above, if the window ends before comparison units adopt \\
\addlinespace
\multicolumn{4}{@{}l}{\normalsize\itshape\bf The weighting family} \\
Simplex synthetic control & target-date & full conformal (exact, affine; the set is an interval); Jackknife+ & exchangeable units given the conditioning information; the estimator does not remove the signal exactly \\
Augmented synthetic control (ridge) & target-date & full conformal; Jackknife+ & as simplex synthetic control; the ridge outcome model is fitted date by date \\
SDID, target-specific & target-date & full conformal (grid; the set may be disconnected); Jackknife+ & exchangeable units; tuning fixed in advance or chosen by a rule that treats units alike (SM Props.~\ref{prop:complete-unit-route-app} and~\ref{prop:tuning-app}) \\
\addlinespace
\multicolumn{4}{@{}l}{\normalsize\itshape\bf The low-rank family} \\
Static IFE with a treatment dummy & --- & none & as static TWFE \\
GSC & full-path & Jackknife+ & exchangeable units; with fixed heterogeneous loadings, equal predictability \\
IFE-EM; matrix completion & full-path & Jackknife+ (each held-out unit in its own fit) & as GSC \\
\bottomrule
\end{tabular}
\floatnote{\emph{Procedures} lists the confidence sets that apply. Full conformal confidence sets for a full-path estimator
are omitted. The last column gives a sufficient condition for exchangeable prediction
errors.}
\label{tab:scope}
\end{table}
\end{landscape}
}

Table~\ref{tab:scope} shows two patterns. First, whether a procedure applies depends on which outcomes the estimator uses to predict the treated unit, not on whether its fit combines treated and control cells. GSC, IFE-EM, and matrix
completion all use the controls' full post-treatment panel, and all three work with the Jackknife+ once each
held-out unit's post-treatment path is hidden. By contrast, a static regression with a single treatment dummy over several post-treatment dates predicts
the treated unit's untreated outcome from the treated unit's own post-treatment outcomes. No control can be given
the same task, so such an estimator falls outside the framework.

Second, the guarantee always requires exchangeable prediction errors, and the estimators differ in how this
requirement can be met. The table's last column applies the two sufficient conditions of Section~\ref{sec:sufficient-conditions}. Exchangeable units suffice for every estimator, and DID also removes an additive signal exactly. When fixed signals differ across units, synthetic control and factor models with estimated loadings need equal predictability itself.

Equal predictability is more plausible when the treated unit's signal lies within the range of the controls' signals. For a factor model, researchers can inspect whether the treated unit's estimated loadings lie inside the convex hull of the controls' loadings.

\FloatBarrier
\section{Extensions}\label{sec:extensions}

In this section, we extend the procedures to multiple treated units. Section~\ref{sec:multi} considers units treated from
a common date, and Section~\ref{sec:staggered} considers staggered adoption.

\subsection{Multiple treated units}\label{sec:multi}

Suppose \(K\) units begin treatment on the same date and \(J\) units are controls. Either procedure in Section~\ref{sec:ordinary-estimator} applies to one treated unit at one date. For the average effect, or for a sharp null that specifies the effect in every treated cell, we instead treat sets of \(K\) units as blocks. Figure~\ref{fig:multiple-treated-common-date} shows the prediction tasks.

For an individual effect, set the other treated units aside and apply either procedure with \(N=J+1\). Using the other treated units' outcomes falls outside SM Proposition~\ref{prop:multi-cell-app}, even if the fit uses only their pre-treatment histories to estimate factors. Simultaneous sets for \(M\) treated cells follow from a union bound with miscoverage \(\gamma/M\) per cell. These sets are the whole line when \(\gamma/M<1/N\), so they can be informative only if \(M\leq\gamma N\).

\input{figures/figure_multiple_treated_common_date.tex}

Averaging the treated units and applying the single-unit procedure does not guarantee coverage of their average effect. An average of \(K\) prediction errors need not be exchangeable with a single control's error. The \emph{block Jackknife+} instead compares the treated block with control blocks of the same size.

Partition the controls into \(G=\lfloor J/K\rfloor\) blocks of \(K\), at random, independently of the outcomes. Keep leftover controls as training units in every fit. Each fold holds out the treated block and one control block, fits once using the remaining outcomes and the held-out units' histories, and predicts every held-out unit. For full-path estimators, this single fit replaces the separate fits in Section~\ref{sec:role-extension}. The fold's endpoints are the treated block's average prediction plus and minus the absolute value of the control block's average error. Combine these endpoints as before, replacing \(N\) with \(G+1\), including in \(q_\gamma\). Subtract the resulting interval from the treated block's observed average to obtain an interval for the average effect.

No fit uses treated cells, so effects enter only through their average and may differ arbitrarily across units. With exchangeable units, a common prediction rule for the blocks, a random partition, and tuning fixed in advance, coverage is at least \(1-(2q_\gamma-1)_+/(G+1)\) (SM Proposition~\ref{prop:block-jplus-app}). Balancing control blocks on pre-treatment paths can invalidate this bound because the treated block is not formed by the same rule. The interval is the whole line unless \(G+1\geq1/\gamma\), so \(\gamma=0.10\) requires at least nine controls per treated unit.

Full conformal inference requires the treated units to be training units when a control block is held out, with a candidate untreated outcome for each treated unit. For synthetic control and ridge fits, with blocks scored by their average errors, the set for the average effect is the whole line once enough control blocks weight the treated units unequally (SM Remark~\ref{rem:block-fc-app}). Unequal weights are the usual case.

Under a sharp null, subtracting the specified effects completes the untreated panel. A relabeling test assigns the treatment pattern to every set of \(K\) units in turn and refits this panel. It ranks the observed statistic, such as the absolute estimated average effect over treated cells, among the relabeled statistics. The \(p\)-value is exact when these statistics are exchangeable, meaning that the actual treated units are as easy to predict without treatment as any other \(K\) units.

The relabeling test is full conformal inference applied to blocks and reduces to Section~\ref{sec:main-full-conformal} with one treated unit. For full-path estimators, the null supplies all required post-treatment outcomes, so the test also applies to GSC. The number of relabelings can far exceed \(N\), allowing \(p\)-values below \(1/N\). Randomly sampling relabelings gives a valid Monte Carlo version.

Testing candidate effects shared by every treated cell and retaining those not rejected gives a confidence set for a common effect. Its guaranteed coverage moves in finer steps than the block interval's, but it does not guarantee coverage of the average when effects differ across units. The test of no effect remains exact.

\subsection{Staggered adoption}\label{sec:staggered}

We take the \(K\) adopters' treatment dates as given and target the average effect \(r\) periods after adoption among adopters observed at that relative period.

For an individual treated cell, apply either procedure with the adopter as unit \(0\) and controls untreated through the last date used by the fit. Target-date estimators at \(t\) can use never-treated units and adopters treated after \(t\). Full-path estimators using the entire panel can hold out only never-treated units as controls.

For the average effect, partition never-treated units into blocks of \(K\) and match each block's units one-to-one with the adopters. Both partitioning and matching must be at random, independently of the outcomes. Each matched control receives its adopter's date. Each fold hides the adopters' and one control block's outcomes from their actual or assigned dates onward, fits once to the remaining cells, and predicts the hidden cells. The control block's average error at each \(r\) determines the fold's endpoints as in Section~\ref{sec:multi}. The same fits serve all relative periods, as Figure~\ref{fig:staggered-masking} shows.

Adopters' outcomes before their own dates remain in the fit. Estimators that fit all untreated cells also use these not-yet-treated cells as training data in every fit. They include DID imputation \citep{Liu2024, BorusyakJaravelSpiess2024}, IFE-EM, and matrix completion. Synthetic control and GSC with factors estimated from never-treated units use these cells only to estimate each adopter's own weights or loadings.

\input{figures/figure_staggered_masking.tex}

The guaranteed coverage and the condition for an informative interval are those of Section~\ref{sec:multi}. When fewer adopters are observed at later relative periods, partitioning separately for each period, with blocks sized to its observed adopters, gives more blocks but requires separate fits. The guarantee is then also conditional on the data of adopters that do not reach that period.

The relabeling test assigns each relabeled unit an adopter's date. Conditional on these dates, its \(p\)-value is exact under the exchangeability condition in Section~\ref{sec:multi} (SM~\ref{app:staggered}).

\section{Simulations and Applications}\label{sec:evidence}

In this section, we evaluate the two procedures in two settings: simulations, where coverage and width can be
checked against a known truth, and three empirical applications.

\subsection{Simulation evidence}\label{sec:sims}

For each design, we simulate \SimReps{} panels with $N\in\{\SimNa,\SimNb,\SimNc,\SimNd\}$ units, $T_0=\SimTpre$ pre-treatment dates and \SimTpost{} post-treatment dates (designs and full results in SM~\ref{app:sims}). The outcomes combine unit and time effects, \SimNfactors{} factors, and serially correlated errors.
Two of the factors are random walks, one of them with drift. The \SimNdesigns{} designs differ only in
the treated unit's loading on this drifting factor. In the exchangeable design, that loading is drawn from the same distribution as every control's, so the treated unit is exchangeable with
the controls. In the other \SimNshift{}, it equals the controls' maximum
or exceeds it by \SimGapA, \SimGapB{} or \SimGapC.

In the exchangeable design, coverage is at or above its guaranteed level for every estimator, procedure and number of
units, within Monte Carlo error, as Figure~\ref{fig:sims-exchangeable} shows. At $\gamma=\SimGammaMain$, it ranges from
\SimJpExMin{} to \SimJpExMax{} for the Jackknife+ and from \SimFcExMin{} to \SimFcExMax{} for full conformal
inference. The guarantee holds with serially correlated errors and random-walk factors.\footnote{It also holds when every unit's errors become noisier after treatment or follow random walks. In those designs, the test of \citet{ChernozhukovWuthrichZhu2021} covers \CompCwzBC{} against a nominal 0.90 because it ranks the treated unit's post-treatment residual among its own pre-treatment residuals (SM~\ref{app:sims-comparators}).}

Once the treated unit's loading reaches the controls'
maximum, coverage falls as the gap grows, and falls least for estimators that model the factor and use enough factors. At a gap of \SimGapC{} and
$N=\SimNc$, Jackknife+ coverage is \SimGapTwoDid{} for DID and \SimGapTwoSc{} for synthetic control, compared with
\SimGapTwoIfe{} for IFE-EM with \SimRhigh{} factors (SM Figure~\ref{fig:sims-coverage}).

\input{figures/fig_sims_exchangeable_float}

For DID and synthetic control in our simulation, full conformal sets are narrower than Jackknife+ intervals where the Jackknife+ over-covers. At $N=\SimNc$ and
$\SimNd$ with $\gamma=\SimGammaMain$, they are \SimFcNarrowMin--\SimFcNarrowMax\% narrower, whereas elsewhere they are at most
\SimFcWiderMax\% wider.

\subsection{Empirical applications}\label{sec:empirics}

We apply the procedures to three studies. The first two use outcomes-only simplex synthetic control, a target-date estimator that permits both procedures. The third uses GSC, a full-path estimator, with the Jackknife+. Unless noted otherwise, $\gamma=0.10$.

\begin{figure}[t]
\centering
\caption{Period-wise confidence sets in the two synthetic control applications}
\includegraphics[width=\textwidth]{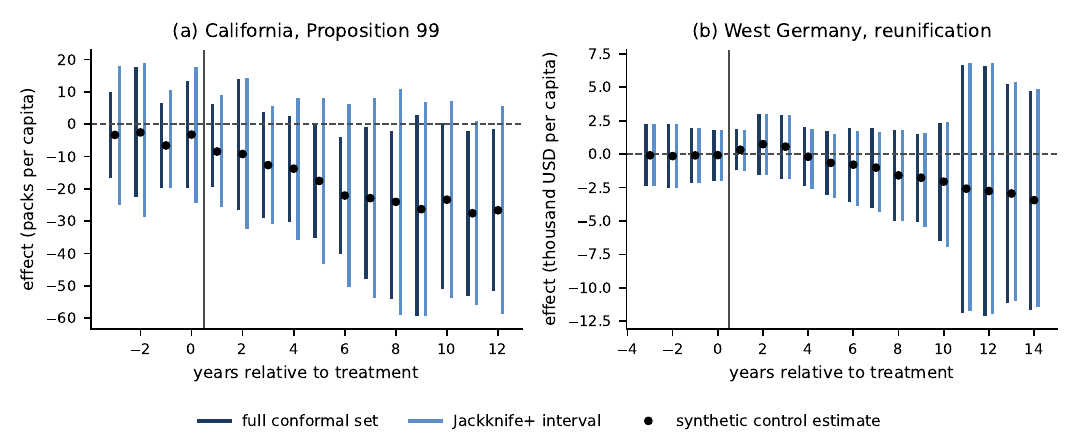}
\floatnote{Dots are synthetic control estimates. Dark bars are full conformal sets and light bars Jackknife+ intervals,
both at $\gamma=0.10$. Dates left of the vertical line are diagnostic dates. Guaranteed coverage is
\AppPropCov{} for California ($N=\AppPropN$) and \AppGerCov{} for West Germany ($N=\AppGerN$).}
\label{fig:apps-sc}
\end{figure}

For the synthetic control studies, we also test the sharp null of no effect in any post-treatment year by ranking absolute average prediction errors from fits that hold out one unit at a time. Jackknife+ intervals for the effect averaged over post-treatment years hold out the treated unit with each control and use the control's absolute average error. The guarantees of both require these absolute average errors to be exchangeable across units. For the Jackknife+, the all-pairs table of these errors must also satisfy the analogue of~\eqref{eq:matrix-exchangeability-main} (SM~\ref{app:sc-tables}).

First, the study of California's Proposition 99 \citep{AbadieDiamondHainmueller2010} has thirty-eight controls, giving $N=\AppPropN$. Both procedures guarantee at least \AppPropCov{} coverage at $\gamma=0.10$, with $q_\gamma=\AppPropQ$ for the Jackknife+. Figure~\ref{fig:apps-sc} shows the period-wise sets. The estimate averages \(\AppPropMean\) packs per capita over \AppPropFirstYear--\AppPropLastYear. Full conformal inference excludes zero in \AppPropFCex{} of \AppPropPost{} post-treatment years, all from the sixth year onward; \ifnum\AppPropJPex=0 no Jackknife+ interval excludes zero\else the Jackknife+ excludes zero in \AppPropJPex{}\fi. The Jackknife+ intervals are wider throughout, mainly because they account for the worst case. When the fold predictions vary little, as here, the interval uses about the $1-\gamma/2$ quantile of the controls' absolute prediction errors rather than the $1-\gamma$ quantile that full conformal inference uses. Neither procedure rejects zero at any of the four diagnostic dates. California's absolute average error ranks \AvgPropRank{} among \AppPropN{} units, giving $p=\AvgPropP$ for the sharp null. The test rejects at 10\%. Inverting it over candidate values of the average effect, as in Section~\ref{sec:main-full-conformal}, gives the full conformal interval $[\AvgPropFcLo,\AvgPropFcHi]$. The Jackknife+ interval is $[\AvgPropJpLo,\AvgPropJpHi]$, with $p=\AvgPropJpP$ (guaranteed coverage \AppPropCov{} for both).

Second, the reunification study \citep{AbadieDiamondHainmueller2015} compares West Germany with sixteen OECD countries, giving $N=\AppGerN$. Both procedures guarantee at least \AppGerCov{} coverage at $\gamma=0.10$. The estimate averages \(\AppGerMean\) thousand dollars per capita over \AppGerFirstYear--\AppGerLastYear. Neither procedure excludes zero in any of the \AppGerPost{} post-treatment years. From the eleventh year onward, the sets are roughly three times as wide as in the first ten years. None of the pre-treatment diagnostics rejects zero. West Germany's absolute average error ranks \AvgGerRank{} among \AppGerN{} units ($p=\AvgGerP$). Its full conformal and Jackknife+ intervals for the average effect are $[\AvgGerFcLo,\AvgGerFcHi]$ and $[\AvgGerJpLo,\AvgGerJpHi]$ (guaranteed coverage \AppGerCov{} for both).

The period-wise result for West Germany depends on the score, which should be chosen in advance. With scores studentized by each unit's pre-treatment root mean squared residual, both procedures exclude zero in \StudGerJpExcl{} of the post-treatment years (SM~\ref{app:bands} and~\ref{app:apps-bands}). This is because West Germany's pre-treatment root mean squared residual is the \StudGerSigRank{} of the \AppGerN{} units'.

\begin{figure}[!thbp]
\centering
\caption{Average turnout effect of election-day registration by elections since adoption}
\includegraphics[width=0.95\textwidth]{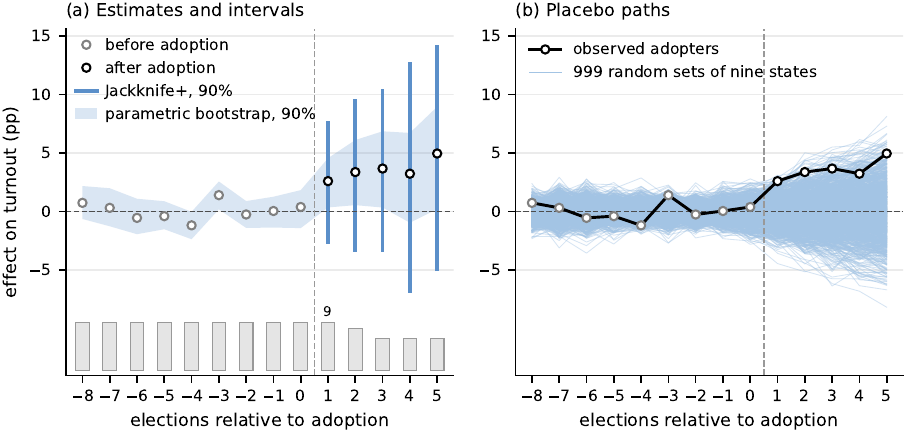}
\floatnote{In panel (a), dots are GSC estimates of the average effect across adopting states, aligned at adoption. Segments are 90 percent intervals that apply the single-unit Jackknife+ to the adopters' average. They carry no
finite-sample guarantee. Table~\ref{tab:edr-block} gives block Jackknife+ intervals, which do. The shaded ribbon is the 90 percent interval from the parametric bootstrap of
\citet[][Algorithm~2]{Xu2017}. Its endpoints are the estimate plus the 5th and 95th percentiles of \EdrParB{} bootstrap draws. Its coverage holds only approximately, in large samples. Bars count the adopting states observed at each election (at most nine). In panel (b), each light blue line shows the path of average effects for one of the \EdrRelB{}
random assignments of the nine adoption paths used in the relabeling test.
The black line shows the observed path.}
\label{fig:apps-edr-att}
\end{figure}

Third, the election-day registration study \citep{Xu2017} includes nine adopting states and thirty-eight never-adopters. We fit GSC with mail-in and motor-voter registration covariates and retain the two factors selected by cross-validation in \citet{Xu2017}. The guarantees require exchangeable scores under this prediction rule, conditional on the selected factor count. Each state-election interval uses $N=\EdrN$ and $q_\gamma=\EdrQ$. Across \EdrCells{} cells, estimates average \EdrMeanEff{} percentage points and \EdrPos{} are positive, but only \EdrExcl{} intervals exclude zero. The median interval is \EdrMedWidth{} points wide. State-level results appear in SM Figure~\ref{fig:apps-edr} and SM Table~\ref{tab:edr-states}.

Figure~\ref{fig:apps-edr-att}(a) shows the aligned estimates with intervals that apply the single-unit Jackknife+ to the adopters' average. As the discussion of multiple treated units explains, such intervals carry no finite-sample guarantee. Table~\ref{tab:edr-block} reports, at each election since adoption, a block Jackknife+ interval for the average effect and a relabeling test of the sharp null of no effect. The block Jackknife+ splits the thirty-eight never-adopting states at random into \numword{\EdrBlockG} blocks of nine, with each state assigned one adopter's adoption path. The two remaining states stay as training units. At $\gamma=0.10$ every interval is the whole line. The 80 percent intervals exclude zero in the first two elections and include it from the third.\footnote{These intervals depend on the partition drawn. Across \EdrBlockPartitions{} random partitions, the first election's interval excludes zero in \EdrBlockExclOne\% of them and the fourth election's in \EdrBlockExclFour\%. SM~\ref{app:edr-multi} reports how much the endpoints vary across partitions.}

The relabeling test is the placebo test of Section~\ref{sec:equiv} with nine treated states. We refit the estimator for \EdrRelB{} random assignments of the nine adoption paths, shown in Figure~\ref{fig:apps-edr-att}(b). In \EdrRelNexceed{} of these assignments, the average effect over all post-adoption state-elections is at least as large in absolute value as the observed one. The test therefore rejects the sharp null of no effect in any state-election ($p=\EdrRelP$).

Figure~\ref{fig:apps-edr-att}(a) also shows the parametric bootstrap of \citet[][Algorithm~2]{Xu2017}. Its 90 percent intervals, $[\EdrParLoOne,\EdrParHiOne]$ points in the first election, exclude zero in \EdrParExclPost{} of the \EdrParNPost{} post-adoption elections shown and in \EdrParExclPre{} of the \EdrParNPre{} pre-adoption elections. Before adoption the effect is zero, so a valid 90 percent interval would exclude zero about one time in ten. The bootstrap intervals are narrow because the bootstrap takes the fitted two-factor model as correct and draws each adopting state's errors independently from the controls' prediction errors. Noise in the average over adopters therefore shrinks as states are added. If errors are positively correlated across states, the bootstrap intervals are anti-conservative.

\begin{table}[!thbp]\centering\small
\caption{Block Jackknife+ intervals for the average turnout effect and relabeling tests of no effect, by elections since adoption}
\begin{tabular}{@{}cccccc@{}}
\toprule
Election since adoption & Adopting states & Estimate & 80\% interval & Block $p$-value & Relabeling $p$-value\\
\midrule
\input{tables/edr_block_table.tex}
\end{tabular}\vspace{-2em}
\floatnote{Estimates are average effects in percentage points, as in Figure~\ref{fig:apps-edr-att}. Intervals are block
Jackknife+ intervals at $\gamma=0.20$, from \numword{\EdrBlockG} blocks of nine never-adopting states drawn in advance. Each
covers the average effect at its election with probability at least 0.80, the highest level at which the treated block and \numword{\EdrBlockG} control blocks give a finite interval. The block $p$-value is the smallest miscoverage level at which the interval excludes zero, so it can only be $1/\EdrBlockNB$,
$3/\EdrBlockNB$ or 1, the last when the interval excludes zero at neither level. The last column is the
relabeling test's $p$-value for the sharp null of no effect in any state-election, from \EdrRelB{} random
assignments of the nine adoption paths (Figure~\ref{fig:apps-edr-att}(b)). The relabeling statistic differs by row. Each row ranks the average effect at that election, except the last row, which ranks the average over all \EdrCells{} post-adoption state-elections, including those after the fifth election.}
\label{tab:edr-block}
\end{table}

We also apply an overlap check, which needs no factor model, rank choice, or post-treatment outcome. After subtracting each unit's pre-treatment mean, we approximate the treated unit's pre-treatment path using nonnegative control weights summing to one. The treated unit's fit error is the root mean squared gap between this path and the weighted average, taken over the pre-treatment dates. We repeat this for each control using the other controls and flag the treated unit when its fit error exceeds the 90th percentile of the controls' fit errors. A flag means that the treated unit's signal may lie outside the controls' range.

Across the three applications, the overlap check flags \OvFlagged{} of \OvTreated{} treated units (SM~\ref{app:overlap-def}; SM Figure~\ref{fig:apps-overlap}). Of the \numword{\EdrLoadNtr} adopting states, \numword{\EdrLoadInHull} have estimated loadings inside the hull of the never-adopting states' loadings. Connecticut is not flagged although its estimated loadings lie just outside this hull (SM Figure~\ref{fig:apps-edr-loadings}), so the two diagnostics can disagree near the boundary.

\section{Recommendations for Applied Researchers}\label{sec:discussion}

First, count the units before choosing the confidence level, since guaranteed coverage moves in steps. With seventeen units, a 95\% set is the whole real line. With twenty units, the Jackknife+'s 90\% and 95\% intervals coincide, both guaranteeing 95\% coverage. For small panels, we recommend reporting the 90\% set alongside its guaranteed coverage (Table~\ref{tab:levels}).

\input{tables/levels_table.tex}

Second, match the procedure to the estimator (Table~\ref{tab:scope}). For target-date estimators, we recommend full conformal inference, which is exact and available in closed form for DID and synthetic control. Its sets are also narrower at $N=\SimNc$ and $\SimNd$ in our simulations, where the Jackknife+ over-covers. With fewer than twenty units, the two procedures guarantee the same coverage at $\gamma=0.10$ and give sets of similar width. The Jackknife+ requires less computation and no candidate values, and we recommend it for full-path estimators.

Third, use a tuning rule that treats all units alike. The simplest option is to fix tuning values in advance and reuse them in every refit, as in our simulations and synthetic control applications. Data-driven tuning also preserves the guarantee if it treats units alike (Section~\ref{sec:sufficient-conditions}). For cross-validation, use only pre-treatment outcomes and hold out the treated unit just as you would each control. Tuning on the treated unit alone or on controls alone can create systematic differences in prediction errors and invalidate coverage, though the change in coverage may be small.

Fourth, choose the score before seeing post-treatment outcomes. Our default, absolute prediction error, suits panels where prediction errors have similar variances across units. When pre-treatment residual sizes vary widely, we recommend a studentized score, which divides each unit's absolute prediction error by its pre-treatment root mean squared residual. In our simulations, coverage with the default score falls below the guaranteed level when the treated unit's error variance exceeds the controls' and rises above it when the variance is lower. The studentized score keeps coverage closer to that level in both cases (SM~\ref{app:sims-position}).

Fifth, check overlap and inspect pre-treatment diagnostics. Run the overlap check of Section~\ref{sec:empirics}, which requires no factor model or rank choice. For factor models, also inspect whether the treated unit's estimated loadings lie inside the convex hull of the controls' loadings. An overlap flag, repeated pre-treatment rejections, or loadings outside this hull warrant investigating whether the treated unit is harder to predict. These findings alone do not establish a failure of exchangeability, and the overlap check has no formal false positive guarantee. Passing the checks is also inconclusive because they can miss units just outside the controls' range and factors that are flat before treatment.

Sixth, report placebo plots alongside confidence sets, as Figure~\ref{fig:apps-edr-att} and Table~\ref{tab:edr-block} do for the turnout study. Under the conditions of Section~\ref{sec:equiv}, the placebo test is a full conformal test of no effect at all post-treatment dates, for target-date and full-path estimators alike. Confidence sets make the test's level explicit and show which effect sizes are not rejected.

With multiple treated units or staggered adoption, the single-unit procedures give a confidence set per treated cell. For average effects, the block Jackknife+ needs at least nine controls per treated unit at $\gamma=0.10$. The relabeling test can use any number of controls and is exact for sharp nulls.

These procedures quantify uncertainty for a chosen estimator. How well it predicts untreated outcomes under nonstationarity remains a question of point estimation. We address this question in the companion paper \citep{LiuXu2026}.

Finally, the R package \code{fect} implements inference for full-path estimators. For GSC, IFE-EM, and matrix completion with an absorbing treatment, \code{vartype = "conformal"} returns Jackknife+ intervals and $p$-values at each post-treatment date, holding tuning values at the original fit. When they are selected by cross-validation in the original fit, the guarantee requires the score condition to hold conditional on the selected values (Section~\ref{sec:error-score}). Options include pre-treatment diagnostics, the block Jackknife+ for average effects (\mbox{\code{conformal.block = TRUE}}), and the relabeling test. Without the block option, average-effect intervals use the single-unit Jackknife+ on the treated units' average, with no finite-sample guarantee.

\clearpage
\vspace{3em}
{\onehalfspacing
\bibliography{refs}
\par}

\clearpage
\ifodd\numexpr\value{page}+1\relax\else\null\thispagestyle{empty}\clearpage\fi
\appendix
\setcounter{page}{1}
\renewcommand{\thepage}{S-\arabic{page}}
\renewcommand{\thesection}{A\arabic{section}}
\setcounter{equation}{0}\renewcommand{\theequation}{A\arabic{equation}}
\setcounter{table}{0}\renewcommand{\thetable}{A\arabic{table}}
\setcounter{figure}{0}\renewcommand{\thefigure}{A\arabic{figure}}
\etocdepthtag.toc{mtsm}
\begin{center}{\Large\bfseries Online Supplementary Materials}\end{center}
\medskip
\bigskip
{\singlespacing
\etocsettocstyle{\noindent{\large\bfseries Contents}\par\medskip}{}
\etocsettagdepth{mtmain}{none}
\etocsettagdepth{mtsm}{subsection}
\makeatletter\renewcommand{\numberline}[1]{\hb@xt@\@tempdima{#1.\hfil}}\renewcommand{\@pnumwidth}{2.2em}\renewcommand{\@tocrmarg}{3em}\makeatother
\tableofcontents}
\clearpage

\onehalfspacing
\section{Setup, results and proofs}
\label{app:formal}

This section gives the formal setup, the definitions and assumptions, and every result in the paper with its proof. A short proof follows its result; a longer one comes at the end of the subsection.

\subsection{Formal setup and the common prediction rule}
\label{app:formal-setup}

Fix \(t\in\Tpost\). The target-date conditioning information is built from
\[
  \boldsymbol\mu^{(t)}
  =\bigl(\mu_{is}:i\in\Iset,
                    s\in\Tpre\cup\{t\}\bigr)
\]
and the tuning values \(\widehat\eta_t\), which are fixed in advance or selected from the data and then held
fixed in every fit. Define
\begin{equation}
  \Ginfo_t=\sigma(\boldsymbol\mu^{(t)},\widehat\eta_t).
  \label{eq:conditioning-information-appendix}
\end{equation}
The shorthand \(\Ginfo\) in the main text refers to
\(\Ginfo_t\). For a full-path procedure, define
\begin{equation}
  \boldsymbol\mu^{\mathrm{all}}
  =\bigl(\mu_{is}:i\in\Iset,
                    s\in\Tpre\cup\Tpost\bigr),
  \qquad
  \Ginfo_t^{\mathrm{all}}
  =\sigma(\boldsymbol\mu^{\mathrm{all}},\widehat\eta_t).
  \label{eq:pooled-conditioning-information-appendix}
\end{equation}

The maintained causal conditions are
\begin{align}
  Y_{0s}&=Y_{0s}(0),
  &&s\in\Tpre,
  \label{eq:no-anticipation-app}\\
  Y_{is}&=Y_{is}(0),
  &&i\in\Cset,
    \ s\in\Tpre\cup\Tpost,
  \label{eq:no-spillovers-app}\noeqref{eq:no-spillovers-app}\\
  Y_{0s}&=Y_{0s}(0)+\tau_s,
  &&s\in\Tpost.
  \label{eq:consistency-app}
\end{align}

For target-date inference, define
\[
  e_i^{(t)}=(e_{is}:s\in\Tpre\cup\{t\}),
  \qquad
  \mathbf e^{(t)}=(e_0^{(t)},\ldots,e_J^{(t)}).
\]
For full-path inference, define
\[
  e_i^{\mathrm{all}}
  =(e_{is}:s\in\Tpre\cup\Tpost),
  \qquad
  \mathbf e^{\mathrm{all}}
  =(e_0^{\mathrm{all}},\ldots,e_J^{\mathrm{all}}).
\]
The conditional error assumptions are, almost surely,
\begin{align}
  \mathbf e^{(t)}\mid\Ginfo_t
  &\stackrel d=
  P_\pi\mathbf e^{(t)}\mid\Ginfo_t,
  \label{eq:conditional-ee-pw-app}\\
  \mathbf e^{\mathrm{all}}\mid\Ginfo_t^{\mathrm{all}}
  &\stackrel d=
  P_\pi\mathbf e^{\mathrm{all}}
  \mid\Ginfo_t^{\mathrm{all}},
  \label{eq:conditional-ee-pool-app}
\end{align}
for every unit permutation \(\pi\). The action of the permutation is
defined by \((P_\pi x)_{\pi(i)}=x_i\). For a matrix, the same relabeling
acts as \(P_\pi R P_\pi^\top\).

For a set \(Q\subseteq\Iset\) of held-out units, define the target-date
and full-path panels available to the fit by
\begin{align}
  D_t^{\mathrm{td}}(Q;Y)
  &=
  \left(
    Y_{\Iset,\Tpre},
    Y_{\Iset\setminus Q,t}
  \right),
  \label{eq:pointwise-partial-panel}\\
  D^{\mathrm{fp}}(Q;Y)
  &=
  \left(
    Y_{\Iset,\Tpre},
    Y_{\Iset\setminus Q,\Tpost}
  \right).
  \label{eq:pooled-partial-panel}
\end{align}
These partial panels contain only the available observations. The hidden cells are omitted from the input rather than replaced by numerical fill values. When the prediction rule uses covariates, every unit's covariates at every date enter every fit, none are hidden, and they are relabeled together with the unit's outcomes. We assume that treatment does not affect them. Where the signals are held fixed, the covariates belong to the conditioning information \(\Ginfo_t\) or \(\Ginfo_t^{\mathrm{all}}\).

For \(\ell\in\{\mathrm{td},\mathrm{fp}\}\), write
\begin{equation}
  D_t^\ell(Q;Y)
  =
  \begin{cases}
    D_t^{\mathrm{td}}(Q;Y),&\ell=\mathrm{td},\\
    D^{\mathrm{fp}}(Q;Y),&\ell=\mathrm{fp}.
  \end{cases}
  \label{eq:support-indexed-partial-panel-app}
\end{equation}

For \(r\in\{1,2\}\),
\(\ell\in\{\mathrm{td},\mathrm{fp}\}\), and \(|Q|=r\), let
\begin{equation}
  \mathcal A_t^{(r,\ell)}
  \left(
    Q;D_t^{\ell}(Q;Y),\widehat\eta_t
  \right)
  \in\R^Q
  \label{eq:formal-extension-app}
\end{equation}
denote the vector of predictions for the held-out units. We call
\(\mathcal A_t^{(r,\ell)}\) the prediction rule with \(r\) held-out units.
The superscript \(\ell\) records whether the rule belongs to a target-date
estimator (\(\mathrm{td}\)) or a full-path estimator (\(\mathrm{fp}\)). In
the formal statements below, we call the rule with one held-out unit the
\emph{singleton} rule and the rule with two held-out units the \emph{pair}
rule. A singleton fit and a pair fit are fits of these two rules.

Each rule includes preprocessing, optimization, convergence checks, and a
finite fallback. It is measurable and finite-valued on every partial panel
used below. All implementations in this paper are deterministic
conditional on \(\widehat\eta_t\), which includes any fixed seed. A
randomized prediction rule would require augmenting the prediction map
with independent randomness and imposing
\eqref{eq:formal-role-symmetry-app} below jointly on the data and that randomness.

The prediction rule is common to all units when relabeling the units
relabels the predictions and changes nothing else. Formally, we call the
rule a \emph{common prediction rule} if, conditional on
\(\widehat\eta_t\),
\begin{equation}
  \left[
  \mathcal A_t^{(r,\ell)}
  \left(
    \pi Q;
    P_\pi D_t^{\ell}(Q;Y),
    \widehat\eta_t
  \right)
  \right]_{\pi(i)}
  =
  \left[
  \mathcal A_t^{(r,\ell)}
  \left(
    Q;D_t^{\ell}(Q;Y),
    \widehat\eta_t
  \right)
  \right]_i
  \label{eq:formal-role-symmetry-app}
\end{equation}
for every admissible panel, every permutation \(\pi\), every relevant
set \(Q\) of held-out units, and every \(i\in Q\). The rule must also be
invariant to the order in which the training units are stored. We impose
\eqref{eq:formal-role-symmetry-app} separately on the singleton and pair
rules. It does not require their predictions to agree.

For target-date full conformal inference, the singleton rule must
reproduce the estimator at the treated unit:
\begin{equation}
  \left[
    \mathcal A_t^{(1,\mathrm{td})}
    \left(
      \{0\};D_t^{\mathrm{td}}(\{0\};Y),
      \widehat\eta_t
    \right)
  \right]_0
  =
  \widehat Y_{0t}^{\mathrm{td}}(0).
  \label{eq:ordinary-extension-agreement-app}
\end{equation}
No analogous equality is required for a pair fit, because the pair fit
also holds out a control that the estimator uses as a training unit.

The estimator in \eqref{eq:ordinary-rule} can be written as
\begin{equation}
  \widehat\theta_t
  =\Theta_t\!\left(Y_{\Cset,\Tpre\cup\Tpost};\widehat\eta_t\right),
  \qquad
  \widehat Y_{0t}(0)
  =m_t\!\left(Y_{0,\Tpre};\widehat\theta_t\right).
  \label{eq:m-theta-representation}
\end{equation}
This notation separates the treated unit's own history from
\(\widehat\theta_t\), what the estimator learns from the controls, without
restricting the model. Because \(\widehat\theta_t\) may contain
the entire control panel and a fitting routine, \(m_t\) may run an EM algorithm or solve an optimization problem on the treated history.

For the full-path singleton rule, and for the pair rule of Section~\ref{sec:role-extension} that fits each held-out unit separately, let
\(r\in\{1,2\}\), \(|Q|=r\), and \(B=\Iset\setminus Q\) be the training
units. The prediction for a held-out unit \(h\in Q\) applies the full-path
estimator to \(h\)'s pre-treatment history and the panel of training
units:
\begin{equation}
  \left[
  \mathcal A_t^{(r,\mathrm{fp})}
  \left(
    Q;D^{\mathrm{fp}}(Q;Y),\widehat\eta_t
  \right)
  \right]_h
  =
  m_t\!\left(
    Y_{h,\Tpre};
    \Theta_t(Y_{B,\Tpre\cup\Tpost};\widehat\eta_t)
  \right).
  \label{eq:canonical-pooled-component-rule}
\end{equation}
When \(r=2\), the two predictions use the same training units but come
from separate fits, and neither fit uses the other held-out unit. When
\(Q=\{0\}\), the rule reproduces the full-path estimator:
\begin{equation}
  \left[
    \mathcal A_t^{(1,\mathrm{fp})}
    \left(
      \{0\};D^{\mathrm{fp}}(\{0\};Y),\widehat\eta_t
    \right)
  \right]_0
  =\widehat Y_{0t}^{\mathrm{fp}}(0).
  \label{eq:ordinary-pooled-extension-agreement-app}
\end{equation}

\subsubsection{Saturated event-date TWFE}

Let \(D_i\) identify the treated units and estimate
\begin{equation}
  Y_{is}
  =\alpha_i+\delta_s
   +\sum_{u\in\Tpost}
      \beta_uD_i\one\{s=u\}+v_{is}.
  \label{eq:saturated-twfe}
\end{equation}
With one treated unit, a balanced panel, equal observation weights, and no
covariates, the coefficient at date \(t\) is
\begin{equation}
  \widehat\beta_t
  =
  \left(Y_{0t}-\overline Y_{\Cset,t}\right)
  -
  \frac{1}{|\Tpre|}
  \sum_{s\in\Tpre}
  \left(Y_{0s}-\overline Y_{\Cset,s}\right),
  \label{eq:twfe-did-equivalence}
\end{equation}
which is the DID comparison of date \(t\) with the pre-treatment average.
Outcomes at the other post-treatment dates enter only their own
event-date coefficients. Under the balanced block-adoption design of
Section~\ref{sec:pointwise-estimators}, the saturated TWFE coefficient
therefore has a target-date form. This equivalence can fail with a pooled
treatment indicator, an unbalanced panel, unequal weights, or additional
covariates.

\begin{lemma}[Target-date form of saturated event-date TWFE]
\label{lem:twfe-pointwise-app}
Suppose the panel is balanced, observations receive equal weight, there is
one treated unit, and the regression in \eqref{eq:saturated-twfe} contains
one treated-by-date indicator for every date in \(\Tpost\). Then
\eqref{eq:twfe-did-equivalence} holds.
\end{lemma}

\begin{proof}[Proof of Lemma~\ref{lem:twfe-pointwise-app}]
For each date \(s\), define the treated-control difference
\[
  g_s=Y_{0s}-\overline Y_{\Cset,s}.
\]
After the unit and date fixed effects are partialled out, the normal
equations for the treatment coefficients are the same as those from
regressing \(g_s\) on an intercept and a separate indicator for every
\(s\in\Tpost\). To see this, let \(\widetilde D_u\) be the indicator of cell \((0,u)\) after two-way
demeaning, and let \(T=|\Tpre|+|\Tpost|\). Then
\(\widetilde D_u^\top\widetilde D_v=(1-1/N)(\one\{u=v\}-1/T)\) and
\(\widetilde D_u^\top Y=(1-1/N)(g_u-\bar g)\), with \(\bar g=T^{-1}\sum_s g_s\), because
\(Y_{0s}-N^{-1}\sum_iY_{is}=(1-1/N)g_s\). After division by \(1-1/N\), the normal equations
\(\sum_{v\in\Tpost}(\one\{u=v\}-1/T)\widehat\beta_v=g_u-\bar g\), \(u\in\Tpost\), are those of the regression
of \(g_s\) on an intercept and the date indicators, with the intercept partialled out.

Write the intercept as \(\kappa\). Because the
post-treatment indicators are saturated, the least-squares objective is
\[
  \sum_{s\in\Tpre}(g_s-\kappa)^2
  +
  \sum_{u\in\Tpost}(g_u-\kappa-\beta_u)^2.
\]
For every \(u\in\Tpost\), the first-order condition for \(\beta_u\) gives
\(\widehat\beta_u=g_u-\widehat\kappa\). Substituting these values removes
the second sum. The first-order condition for \(\kappa\) then gives
\[
  \widehat\kappa
  =\frac1{|\Tpre|}\sum_{s\in\Tpre}g_s.
\]
Therefore
\[
  \widehat\beta_t
  =g_t-\frac1{|\Tpre|}\sum_{s\in\Tpre}g_s,
\]
which is exactly \eqref{eq:twfe-did-equivalence}.
\end{proof}

\subsection{Full conformal coverage}
\label{app:fc-proof}

For a candidate \(c\in\R\), define
\[
  Y_{it}^{(c)}
  =
  \begin{cases}
    Y_{0t}-c,&i=0,\\
    Y_{it},&i\in\Cset.
  \end{cases}
\]
For each \(i\in\Iset\), set
\begin{equation}
  \widehat M_i(c)
  =
  \left[
  \mathcal A_t^{(1,\mathrm{td})}
  \left(
    \{i\};
    Y_{\Iset,\Tpre},
    (Y_{ht}^{(c)})_{h\neq i};
    \widehat\eta_t
  \right)
  \right]_i
  \label{eq:fc-prediction-app}
\end{equation}
and
\begin{equation}
  S_i(c)=|Y_{it}^{(c)}-\widehat M_i(c)|.
  \label{eq:fc-score-app}
\end{equation}
In \eqref{eq:fc-prediction-app}, the second semicolon separates the partial panel
from the fixed tuning values. The rule is the singleton rule of
\eqref{eq:formal-extension-app}.

\begin{assumption}[Conditional full conformal score exchangeability]
\label{ass:fc-score-exchangeability-app}
For every permutation \(\pi\), almost surely,
\begin{equation}
  \mathbf S_t(\tau_t)\mid\Ginfo_t
  \stackrel d=
  P_\pi\mathbf S_t(\tau_t)\mid\Ginfo_t,
  \qquad
  \mathbf S_t(c)=(S_0(c),\ldots,S_J(c))^\top.
  \label{eq:fc-score-exchangeability-app}
\end{equation}
\end{assumption}

\begin{remark}[Conditioning information]
\label{rem:generic-sigma-app}
Theorems~\ref{thm:fc-app}--\ref{thm:pooled-fc-app} and Corollary~\ref{cor:placebo-validity-app} are stated
conditional on \(\Ginfo_t\) or \(\Ginfo_t^{\mathrm{all}}\), or, for the corollary, on the date-\(s\)
\(\sigma\)-fields \(\Ginfo_s^{\mathrm{td}}\) and \(\Ginfo_s^{\mathrm{fp}}\). Their proofs use only that the score vector or table
is exchangeable conditional on the stated \(\sigma\)-field, together with the rank and counting arguments, which
hold for every realization. Each result therefore holds verbatim with that \(\sigma\)-field replaced by any
\(\sigma\)-field \(\mathcal H\) under which the corresponding exchangeability assumption holds. The sufficient
conditions with a fixed signal take \(\mathcal H=\Ginfo_t\). The complete-unit condition of
Proposition~\ref{prop:complete-unit-route-app} takes \(\mathcal H\) to be the common information set, which excludes
a unit-labeled signal.
\end{remark}

\begin{theorem}[Target-date full conformal coverage]
\label{thm:fc-app}
Let \(N=J+1\geq2\) and \(\gamma\in[0,1)\). Suppose
\eqref{eq:no-anticipation-app}--\eqref{eq:consistency-app} hold, the
singleton score rule is measurable and finite-valued, and
Assumption~\ref{ass:fc-score-exchangeability-app} holds. Define
\begin{align}
  p_t^{\mathrm{FC}}(c)
  &=
  \frac1N\sum_{i\in\Iset}
  \one\{S_i(c)\geq S_0(c)\},
  \label{eq:fc-p-app}\\
  \widehat C_t^{\mathrm{FC}}(\gamma)
  &=\{c:p_t^{\mathrm{FC}}(c)>\gamma\}.
  \label{eq:fc-set-app}
\end{align}
Then, almost surely,
\begin{equation}
  \Pr\!\left\{
    \tau_t\in\widehat C_t^{\mathrm{FC}}(\gamma)
    \mid\Ginfo_t
  \right\}
  \geq
  1-\frac{\lfloor\gamma N\rfloor}{N}
  \geq1-\gamma.
  \label{eq:fc-coverage-app}
\end{equation}
\end{theorem}

The full conformal set is an interval when the completed outcome of unit \(0\) enters every control prediction
linearly, that is, when \(\widehat M_j(c)=\alpha_j+w_j(Y_{0t}-c)\) with \(|w_j|\leq1\). To see why, let
\(r_j=Y_{jt}-\alpha_j-w_jY_{0t}\). The difference \(S_j(c)^2-S_0(c)^2=(r_j+w_jc)^2-(\widehat\tau_t-c)^2\) is a
quadratic in \(c\) whose leading coefficient \(w_j^2-1\) is not positive, and it is nonnegative at
\(c=\widehat\tau_t\). Each set \(\{c:S_j(c)\geq S_0(c)\}\) is thus an interval containing \(\widehat\tau_t\). As
\(c\) moves away from \(\widehat\tau_t\) in either direction, the number of these sets that contain \(c\) does not
increase, so \(\widehat C_t^{\mathrm{FC}}(\gamma)\) is an interval.

\begin{proof}[Proof of Theorem~\ref{thm:fc-app}]
At the true candidate, consistency and the absence of spillovers give
\[
  Y_{it}^{(\tau_t)}=Y_{it}(0)
  \qquad\text{for every }i\in\Iset.
\]
Thus \(\mathbf S_t(\tau_t)\) is the untreated score vector covered by
Assumption~\ref{ass:fc-score-exchangeability-app}.

Write \(S_i=S_i(\tau_t)\) and define a reverse-rank $p$-value for every
unit:
\begin{equation}
  p_i
  =\frac1N\sum_{h\in\Iset}\one\{S_h\geq S_i\}.
  \label{eq:reverse-rank-pvalues-app}
\end{equation}
The operational $p$-value at the true candidate is
\(p_t^{\mathrm{FC}}(\tau_t)=p_0\). The map from the score vector to the
vector \((p_0,\ldots,p_J)\) commutes with unit relabeling, so the \(p_i\)'s
are conditionally exchangeable.

Set
\[
  K=\lfloor\gamma N\rfloor,
  \qquad
  A_\gamma=\{i:p_i\leq\gamma\},
  \qquad
  L_\gamma=|A_\gamma|.
\]
We show that \(L_\gamma\leq K\) for every realized score vector. The
claim is immediate if \(A_\gamma\) is empty. Otherwise, choose
\(i^\star\in A_\gamma\) whose score is smallest among the scores in
\(A_\gamma\). Every unit in \(A_\gamma\) then has a score at least as
large as \(S_{i^\star}\), so all \(L_\gamma\) such units are counted in
\(p_{i^\star}\). Therefore
\[
  L_\gamma
  \leq Np_{i^\star}
  \leq\gamma N.
\]
Since \(L_\gamma\) is an integer,
\begin{equation}
  L_\gamma\leq\lfloor\gamma N\rfloor=K.
  \label{eq:rank-count-app}
\end{equation}
The weak comparison in \eqref{eq:reverse-rank-pvalues-app} makes this
argument valid in the presence of ties.

Conditional exchangeability now gives, almost surely,
\begin{align*}
  \Pr\{p_0\leq\gamma\mid\Ginfo_t\}
  &=
  \frac1N
  \operatorname E\!\left[
    \sum_{i\in\Iset}\one\{p_i\leq\gamma\}
    \mathrel{\Big|}\Ginfo_t
  \right]\\
  &=\frac1N\operatorname E[L_\gamma\mid\Ginfo_t]\\
  &\leq\frac{\lfloor\gamma N\rfloor}{N}.
\end{align*}
Finally,
\[
  \tau_t\notin\widehat C_t^{\mathrm{FC}}(\gamma)
  \quad\Longleftrightarrow\quad
  p_t^{\mathrm{FC}}(\tau_t)\leq\gamma
  \quad\Longleftrightarrow\quad
  p_0\leq\gamma.
\]
Taking complements proves the first inequality in
\eqref{eq:fc-coverage-app}; the second follows from
\(\lfloor\gamma N\rfloor\leq\gamma N\).
\end{proof}

\subsection{The bridge from errors to scores}
\label{app:bridge-proof}

The full conformal score vector at the true candidate and the all-pairs table are measurable functions of
the error panel, the fixed signals and the fixed tuning values. When the rule uses covariates, we hold them fixed with the signals and relabel them together. Write
\[
  \mathbf T(\mathbf e^{\ell};\boldsymbol\mu^{\ell},\widehat\eta_t)
\]
for either map. The value of \(\mathbf T\) is the score vector \(\mathbf S_t(\tau_t)\) of the
singleton rule, with \(\ell=\mathrm{td}\), or the table \(\mathbf R^{\ell}\) of the pair rule in
Appendix~\ref{app:jplus-proof}, with \(\ell\in\{\mathrm{td},\mathrm{fp}\}\). Here
\begin{align*}
  (\mathbf e^{\mathrm{td}},\boldsymbol\mu^{\mathrm{td}},\Ginfo_t^{\mathrm{td}})
  &=(\mathbf e^{(t)},\boldsymbol\mu^{(t)},\Ginfo_t),\\
  (\mathbf e^{\mathrm{fp}},\boldsymbol\mu^{\mathrm{fp}},\Ginfo_t^{\mathrm{fp}})
  &=(\mathbf e^{\mathrm{all}},\boldsymbol\mu^{\mathrm{all}},\Ginfo_t^{\mathrm{all}}).
\end{align*}
Relabeling the units acts on a vector as \(P_\pi\mathbf T\) and on a table as \(P_\pi\mathbf TP_\pi^\top\); we
write \(P_\pi\cdot\mathbf T\) for either. Assumptions~\ref{ass:fc-score-exchangeability-app}
and~\ref{ass:jplus-matrix-exchangeability-app} require the output of the map to be exchangeable. In the following condition, we permute the error paths across units, hold the signals fixed, and recompute the scores.

\begin{assumption}[Conditional equal predictability]
\label{ass:compatibility-app}
For every permutation \(\pi\), almost surely,
\begin{equation}
  \mathbf T(P_\pi\mathbf e^{\ell};\boldsymbol\mu^{\ell},\widehat\eta_t)\mid\Ginfo_t^{\ell}
  \ \stackrel d=\
  P_\pi\cdot\mathbf T(\mathbf e^{\ell};\boldsymbol\mu^{\ell},\widehat\eta_t)\mid\Ginfo_t^{\ell}.
  \label{eq:compatibility-app}
\end{equation}
\end{assumption}

The equality holds in distribution and need not hold as a pathwise identity.

\begin{proposition}[Error exchangeability and equal predictability]
\label{prop:bridge-app}
Under conditional error exchangeability, \eqref{eq:conditional-ee-pw-app} for \(\ell=\mathrm{td}\) or
\eqref{eq:conditional-ee-pool-app} for \(\ell=\mathrm{fp}\), Assumption~\ref{ass:compatibility-app} holds for
the score vector if and only if Assumption~\ref{ass:fc-score-exchangeability-app} holds, and for the all-pairs
table if and only if Assumption~\ref{ass:jplus-matrix-exchangeability-app} holds.
\end{proposition}

\begin{proof}[Proof of Proposition~\ref{prop:bridge-app}]
Conditional error exchangeability and measurability imply
\begin{equation}
  \mathbf T(P_\pi\mathbf e^{\ell};\boldsymbol\mu^{\ell},\widehat\eta_t)\mid\Ginfo_t^{\ell}
  \stackrel d=
  \mathbf T(\mathbf e^{\ell};\boldsymbol\mu^{\ell},\widehat\eta_t)\mid\Ginfo_t^{\ell}.
  \label{eq:ee-through-map-app}
\end{equation}
If equal predictability also holds, the left side of \eqref{eq:ee-through-map-app} has the same conditional
distribution as the relabeled score vector or table, and transitivity gives the direct score condition.
Conversely, suppose the direct score condition holds. Equation~\eqref{eq:ee-through-map-app} equates the left
side of \eqref{eq:compatibility-app} with the original scores, and the direct condition equates those with the
relabeled scores, which is \eqref{eq:compatibility-app}.
\end{proof}

To express equal predictability in terms of approximation errors, we write each untreated outcome as its signal plus its error. For the singleton rule, define
\begin{equation}
  d_i(\mathbf e^{(t)})
  =
  \mu_{it}-\widehat M_i(\tau_t).
  \label{eq:singleton-discrepancy-app}
\end{equation}
Then
\begin{equation}
  S_i(\tau_t)
  =
  |e_{it}+d_i(\mathbf e^{(t)})|.
  \label{eq:singleton-score-discrepancy-app}
\end{equation}
For the pair rule, define
\begin{equation}
  d_{ik}^{\ell}(\mathbf e^{\ell})
  =\mu_{it}-\widehat M_{ik}^{\ell}.
  \label{eq:pair-discrepancy-app}
\end{equation}
Then
\begin{equation}
  R_{ik}^{\ell}
  =|e_{it}+d_{ik}^{\ell}(\mathbf e^{\ell})|.
  \label{eq:pair-score-discrepancy-app}
\end{equation}
The following propositions give two sufficient conditions for exchangeable scores.

\begin{proposition}[Exchangeable complete unit objects]
\label{prop:complete-unit-route-app}
For each unit \(i\), let its \emph{complete untreated unit object} \(Z_i\) collect the unit's untreated outcomes over
the dates the scores use and every unit-level input of the declared rule, such as covariates. Write
\(Z=(Z_0,\ldots,Z_J)\). Let \(\mathcal H\) be a \(\sigma\)-field, the \emph{common information set}, and let \(W\) be
\(\mathcal H\)-measurable. Suppose:
\begin{enumerate}
\item \(Z\mid\mathcal H\stackrel d=P_\pi Z\mid\mathcal H\) almost surely, for every unit permutation \(\pi\);
\item the tuning value \(\widehat\eta_t\) is \(\mathcal H\)-measurable, or \(\widehat\eta_t=g(Z,W)\) for a measurable
map \(g\) with \(g(P_\pi z,w)=g(z,w)\) for every \(\pi\) and every \((z,w)\);
\item \setlength{\emergencystretch}{3em}the score vector or table under study equals \(T(Z,W,\widehat\eta_t)\) for a measurable map \(T\) with
\(T(P_\pi z,w,\eta)=P_\pi\cdot T(z,w,\eta)\) for every \(\pi\) and every \((z,w,\eta)\).
\end{enumerate}
Then Assumption~\ref{ass:fc-score-exchangeability-app}, for a score vector, or
Assumption~\ref{ass:jplus-matrix-exchangeability-app}, for a table, holds with \(\mathcal H\) in place of the stated
information, and Theorem~\ref{thm:fc-app} or Theorem~\ref{thm:generic-jplus-app} holds conditional on \(\mathcal H\)
under its other hypotheses (Remark~\ref{rem:generic-sigma-app}). The estimator need not be correctly specified.
\end{proposition}

When \(\widehat\eta_t\) is \(\mathcal H\)-measurable but was selected from the data by a rule that is not of the
second kind in condition~2, condition~1 is an assumption about the units given the selected value.

\begin{proof}[Proof of Proposition~\ref{prop:complete-unit-route-app}]
If \(\widehat\eta_t\) is \(\mathcal H\)-measurable, absorb it into \(W\). Otherwise set
\[
  \widetilde T(z,w)=T(z,w,g(z,w));
\]
conditions~2 and~3 give \(\widetilde T(P_\pi z,w)=P_\pi\cdot\widetilde T(z,w)\).
In either case the scores equal \(\widetilde T(Z,W)\) for a map that is equivariant in its first argument. Conditional
on \(\mathcal H\), \(W\) is fixed and condition~1 gives
\(\widetilde T(Z,W)\stackrel d=\widetilde T(P_\pi Z,W)=P_\pi\cdot\widetilde T(Z,W)\), which is the direct score
condition given \(\mathcal H\).
\end{proof}

\begin{proposition}[Equivariant predictions that remove the signal exactly]
\label{prop:residualization-route-app}
Suppose the error paths are conditionally exchangeable. If, at the true
candidate, the singleton approximation errors in
\eqref{eq:singleton-discrepancy-app} satisfy
\[
  d(P_\pi\mathbf e^{(t)})=P_\pi d(\mathbf e^{(t)}),
\]
then the full conformal scores are conditionally exchangeable. If the
pair approximation errors in \eqref{eq:pair-discrepancy-app} satisfy
\[
  d_{\pi(i),\pi(k)}^{\ell}(P_\pi\mathbf e^{\ell})
  =d_{ik}^{\ell}(\mathbf e^{\ell})
  \qquad(i\neq k),
\]
then the all-pairs table is conditionally exchangeable.
\end{proposition}

\begin{proof}[Proof of Proposition~\ref{prop:residualization-route-app}]
For the singleton rule, \eqref{eq:singleton-score-discrepancy-app} gives
\(S_i(\tau_t)=|e_{it}+d_i(\mathbf e^{(t)})|\),
so the complete score vector is a permutation-equivariant function of the
exchangeable error panel. The complete-unit argument in the proof of
Proposition~\ref{prop:complete-unit-route-app}, with the error panel as
the input, gives the result. In the pair case,
\eqref{eq:pair-score-discrepancy-app} gives
\(R_{ik}^{\ell}=|e_{it}+d_{ik}^{\ell}(\mathbf e^{\ell})|\),
and the displayed equivariance gives the same argument with simultaneous
row-and-column relabeling.
\end{proof}

Error exchangeability and a common prediction rule do not suffice on their own.
Section~\ref{sec:sufficient-conditions} gives a counterexample in which DID does not remove a factor.

When \(\widehat\eta_t\) is selected in the estimator's fit for the treated
unit, we maintain
\eqref{eq:conditional-ee-pw-app} or \eqref{eq:conditional-ee-pool-app},
together with the relevant score condition, conditional on the selected
values. Reusing those values in every conformal fold does not establish
these assumptions. The following result states exactly when conditioning on a
data-selected tuning value preserves the exchangeability of the errors.

\begin{proposition}[Data-selected tuning]
\label{prop:tuning-app}
Fix the signal \(\boldsymbol\mu\) and suppose the error panel \(\mathbf e\) is
exchangeable given \(\boldsymbol\mu\). Let the tuning value be
\(\widehat\eta=g(\boldsymbol\mu+\mathbf e)\) for a deterministic map \(g\), and suppose
\(\mathbf e\) contains every input the selector uses. Then
\(\mathbf e\mid(\boldsymbol\mu,\widehat\eta)\stackrel d=P_\pi\mathbf e\mid(\boldsymbol\mu,\widehat\eta)\)
for every permutation \(\pi\) if and only if
\(g(\boldsymbol\mu+P_\pi\mathbf e)=g(\boldsymbol\mu+\mathbf e)\) almost surely for every \(\pi\).
\end{proposition}

\begin{proof}[Proof of Proposition~\ref{prop:tuning-app}]
Suppose invariance holds. Then
\((P_\pi\mathbf e,\widehat\eta)=(P_\pi\mathbf e,g(\boldsymbol\mu+P_\pi\mathbf e))\), and error
exchangeability makes this pair equal in law to \((\mathbf e,g(\boldsymbol\mu+\mathbf e))=(\mathbf e,\widehat\eta)\).
Conditioning on \(\widehat\eta\) gives the conditional statement. Conversely, the conditional statement gives
\((\mathbf e,g(\boldsymbol\mu+\mathbf e))\stackrel d=(P_\pi\mathbf e,g(\boldsymbol\mu+\mathbf e))\), and error
exchangeability gives \((\mathbf e,g(\boldsymbol\mu+\mathbf e))\stackrel d=(P_\pi\mathbf e,g(\boldsymbol\mu+P_\pi\mathbf e))\).
The two right-hand pairs have the same first coordinate, \(X=P_\pi\mathbf e\), and their second coordinates are
the functions \(g(\boldsymbol\mu+P_\pi^{-1}X)\) and \(g(\boldsymbol\mu+X)\) of it. Because the joint laws are equal and share the first coordinate, the two functions agree almost surely.
\end{proof}

Three cases satisfy the condition. First, the tuning value is fixed in advance. Second, the selector treats all
units alike and every unit has the same signal (\(\mu_{is}=\delta_s\), so that
\(\boldsymbol\mu+P_\pi\mathbf e=P_\pi(\boldsymbol\mu+\mathbf e)\)). Third, a symmetric selector sees the data only
through residuals from which the signal has been removed. For example, symmetric cross-validation can choose a pre-treatment window for DID under an additive signal.

Selecting the tuning value from unit \(0\)'s own fit can violate the condition. With heterogeneous signals, symmetry in the raw outcomes does not
by itself guarantee the condition. If the selector uses data outside \(\mathbf e\), the same argument applies to the
conditional law of \(\widehat\eta\) given \(\mathbf e\).

\subsection{Jackknife+ coverage}
\label{app:jplus-proof}

One theorem covers both a target-date and a full-path estimator. Let
\(\ell\in\{\mathrm{td},\mathrm{fp}\}\). For every unordered pair of
distinct units \(\{i,k\}\), we treat \(i\) and \(k\) as the held-out units, hide their relevant
outcomes, and apply the declared pair rule to the complete untreated panel. Define
\begin{equation}
  (\widehat M_{ik}^{\ell},\widehat M_{ki}^{\ell})
  =
  \begin{cases}
  \mathcal A_t^{(2,\mathrm{td})}
  \left(
    \{i,k\};
    Y_{\Iset,\Tpre}(0),
    Y_{\Iset\setminus\{i,k\},t}(0);
    \widehat\eta_t
  \right),&\ell=\mathrm{td},\\[6pt]
  \mathcal A_t^{(2,\mathrm{fp})}
  \left(
    \{i,k\};
    Y_{\Iset,\Tpre}(0),
    Y_{\Iset\setminus\{i,k\},\Tpost}(0);
    \widehat\eta_t
  \right),&\ell=\mathrm{fp}.
  \end{cases}
  \label{eq:all-pairs-fit-app}
\end{equation}
The first coordinate predicts unit \(i\) and the second predicts unit
\(k\). Let
\begin{equation}
  R_{ik}^{\ell}
  =|Y_{it}(0)-\widehat M_{ik}^{\ell}|,
  \qquad
  R_{ii}^{\ell}=0,
  \label{eq:all-pairs-residual-app}
\end{equation}
and collect the entries in
\(\mathbf R^{\ell}=[R_{ik}^{\ell}]_{i,k\in\Iset}\).

The pair \(\{0,j\}\) used to compute the interval has only observed inputs in
\eqref{eq:all-pairs-fit-app}. Its training units contribute
\(Y_{\Cset\setminus\{j\},t}\) for a target-date estimator and
\(Y_{\Cset\setminus\{j\},\Tpost}\) for a full-path estimator. The computed fits therefore coincide with the \(\{0,j\}\) entries of the hypothetical construction.

For each observed pair \(\{0,j\}\), form
\begin{equation}
  a_j^{\ell}=\widehat M_{0j}^{\ell}-R_{j0}^{\ell},
  \qquad
  b_j^{\ell}=\widehat M_{0j}^{\ell}+R_{j0}^{\ell},
  \label{eq:generic-jplus-endpoints-app}
\end{equation}
where \(R_{j0}^{\ell}=|Y_{jt}-\widehat M_{j0}^{\ell}|\). Let
\(a_{(1)}^{\ell}\leq\cdots\leq a_{(J)}^{\ell}\) and
\(b_{(1)}^{\ell}\leq\cdots\leq b_{(J)}^{\ell}\) be the endpoint order
statistics. Define
\begin{equation}
  q_\gamma
  =
  \max\left\{
    q\in\left\{0,\ldots,\left\lfloor\frac N2\right\rfloor\right\}:
    (2q-1)_+\leq\gamma N
  \right\},
  \qquad
  u_\gamma=N-q_\gamma.
  \label{eq:generic-jplus-calibration-app}
\end{equation}
The prediction interval is
\begin{equation}
  \widehat C_t^{Y,\ell,\mathrm{J+}}(\gamma)
  =
  \begin{cases}
    \R,&q_\gamma=0,\\[2pt]
    [a_{(q_\gamma)}^{\ell},b_{(u_\gamma)}^{\ell}],&q_\gamma\geq1,
  \end{cases}
  \label{eq:generic-jplus-prediction-app}
\end{equation}
and the effect interval is
\begin{equation}
  \widehat C_t^{\tau,\ell,\mathrm{J+}}(\gamma)
  =Y_{0t}-\widehat C_t^{Y,\ell,\mathrm{J+}}(\gamma).
  \label{eq:generic-jplus-effect-app}
\end{equation}
We term the integer \(q_\gamma\) the calibration.

\begin{assumption}[Conditional all-pairs score exchangeability]
\label{ass:jplus-matrix-exchangeability-app}
For every permutation \(\pi\), almost surely,
\begin{equation}
  \mathbf R^{\ell}\mid\Ginfo_t^{\ell}
  \stackrel d=
  P_\pi\mathbf R^{\ell}P_\pi^\top\mid\Ginfo_t^{\ell}.
  \label{eq:jplus-matrix-exchangeability-app}
\end{equation}
\end{assumption}

\begin{theorem}[Target-date and full-path Jackknife+ coverage]
\label{thm:generic-jplus-app}
Let \(N=J+1\geq2\), \(\gamma\in[0,1)\), and
\(\ell\in\{\mathrm{td},\mathrm{fp}\}\). Suppose
\eqref{eq:no-anticipation-app}--\eqref{eq:consistency-app} hold, the
complete all-pairs map is measurable and finite-valued, and
Assumption~\ref{ass:jplus-matrix-exchangeability-app} holds. Then, almost
surely,
\begin{align}
  \Pr\!\left\{
    Y_{0t}(0)\in
    \widehat C_t^{Y,\ell,\mathrm{J+}}(\gamma)
    \mid\Ginfo_t^{\ell}
  \right\}
  &\geq
  1-\frac{(2q_\gamma-1)_+}{N}
  \geq1-\gamma,
  \label{eq:generic-jplus-prediction-coverage-app}\\
  \Pr\!\left\{
    \tau_t\in
    \widehat C_t^{\tau,\ell,\mathrm{J+}}(\gamma)
    \mid\Ginfo_t^{\ell}
  \right\}
  &\geq
  1-\frac{(2q_\gamma-1)_+}{N}
  \geq1-\gamma.
  \label{eq:generic-jplus-effect-coverage-app}
\end{align}
\end{theorem}

The counting step in the proof of Theorem~\ref{thm:generic-jplus-app} follows \citet[Section~6]{BarberCandesRamdasTibshirani2021}. The choice of outcomes that each fold hides and the calibration \(q_\gamma\) are specific to this setting.

\begin{remark}[Leave-two-out tests]
\label{rem:leave-two-out-app}
\citet{LeiSudijono2024} replace the folds that hide the pair \(\{0,j\}\) with folds that hide a triple \(\{0,i,j\}\) and count the
pairs of controls that unit \(0\) beats within its triple. With one control in place of two, the count is the statistic of the test behind the Jackknife+ bound, and the Jackknife+ interval contains the set of candidates that the test keeps.
The counting argument in the proof of Theorem~\ref{thm:generic-jplus-app} extends to triples, because each triple has at most one strict winner. If the table of triple scores is exchangeable in the same sense
as Assumption~\ref{ass:jplus-matrix-exchangeability-app}, the set of candidates at which unit \(0\) beats fewer
than \(w\) of the \(J(J-1)/2\) pairs covers \(\tau_t\) with probability at least \(1-s^*(w)/N\). Here
\(N\geq3\), \(w\) is an integer with \(1\leq w\leq J(J-1)/2\), and \(s^*(w)\) is the largest integer
\(s\in\{1,\ldots,N\}\) with \(w\leq(s-1)(s-2)/6+(N-s)(N-2)/2\). The construction applies to target-date and full-path
estimators alike, with the outcomes hidden as each requires, at the cost of \(J(J-1)/2\) fits.

Although the test's $p$-value takes finer values than \(k/N\), its guarantee does not.
Under exchangeability, a non-randomized test that treats controls symmetrically rejects a true null with probability
equal to the expected number of rejected placements among the \(N\) placements of unit \(0\)'s data, divided by \(N\).
The worst case over exchangeable laws is attained by the uniform law over relabelings of a single dataset and is a multiple of \(1/N\)
\citep[compare][Proposition~B.1]{LeiSudijono2024}. The exact guarantee is therefore \(1-k/N\) for
an integer \(k\); full conformal inference attains each level when scores have no ties.

A guarantee above \(1-1/N\) requires that no exchangeable law give a positive probability of rejecting a true null.
At \(N=17\), for example, a non-randomized procedure of this kind
guarantees 95\% coverage only if no exchangeable law makes it reject the true value. A randomized test can
attain any level.
\end{remark}

Full conformal inference tests \(H_{0,t}(c)\) using the rank $p$-value in \eqref{eq:fc-p-app}. The Jackknife+
constructs a prediction interval for \(Y_{0t}(0)\) and rejects \(c\) when \(Y_{0t}-c\) falls outside it. To report
a Jackknife+ $p$-value, count lower- and upper-tail violations by
\begin{align}
  L_t^{\ell}(c)
  &=\sum_{j\in\Cset}\one\{Y_{0t}-c<a_j^{\ell}\},
  \qquad
  U_t^{\ell}(c)
  =\sum_{j\in\Cset}\one\{Y_{0t}-c>b_j^{\ell}\},
  \label{eq:jplus-tail-counts-app}\\
  m_t^{\ell}(c)
  &=\max\{L_t^{\ell}(c),U_t^{\ell}(c)\},
  \qquad
  p_t^{\ell,\mathrm{J+}}(c)
  =\min\left\{1,\frac{2\{N-m_t^{\ell}(c)\}-1}{N}\right\}.
  \label{eq:jplus-pvalue-app}
\end{align}

\begin{proposition}[Jackknife+ interval duality]
\label{prop:jplus-duality-app}
For every \(c\in\R\), \(\gamma\in[0,1)\) and \(\ell\in\{\mathrm{td},\mathrm{fp}\}\),
\begin{equation}
  c\notin\widehat C_t^{\tau,\ell,\mathrm{J+}}(\gamma)
  \quad\Longleftrightarrow\quad
  p_t^{\ell,\mathrm{J+}}(c)\leq\gamma.
  \label{eq:jplus-duality-app}
\end{equation}
\end{proposition}

Because the comparisons in \eqref{eq:jplus-tail-counts-app} are strict, candidates at the interval's endpoints
are retained; when \(q_\gamma=0\), every candidate is retained. We call this $p$-value the interval-dual $p$-value because it
reproduces the interval's rejection rule exactly, using only the pair fits. It differs from the rank $p$-value of full conformal inference. The pre-treatment diagnostics in Appendix~\ref{app:historical-diagnostics} report this $p$-value at
each diagnostic date.

\begin{proposition}[Failed folds]
\label{prop:failed-folds-app}
Let a finite-valued pair rule, the reference rule, satisfy Assumption~\ref{ass:jplus-matrix-exchangeability-app}
and agree with the implemented rule on every operational fold whose fit succeeds. Let \(F\subseteq\Cset\) be the
folds whose fit fails, and set \(a_j=-\infty\) and \(b_j=+\infty\) for \(j\in F\), keeping \(N\) and \(q_\gamma\)
unchanged. Then the implemented interval contains the reference interval on every dataset, so it has coverage at
least \(1-(2q_\gamma-1)_+/N\); its interval-dual \(p\)-value is at least the reference \(p\)-value; and if
\(|F|\geq q_\gamma\geq1\), the interval is \(\R\).
\end{proposition}

\begin{proof}[Proof of Proposition~\ref{prop:failed-folds-app}]
Lowering an entry of a list can only lower each of its order statistics, and raising an entry can only raise them.
Replacing reference endpoints by \(-\infty\) and \(+\infty\) on \(F\) therefore lowers \(a_{(q)}\) and raises
\(b_{(N-q)}\), so the implemented interval contains the reference interval.
Theorem~\ref{thm:generic-jplus-app} applied to the reference rule gives the coverage bound. In
\eqref{eq:jplus-tail-counts-app}, \(Y_{0t}-c<-\infty\) and \(Y_{0t}-c>+\infty\) never hold, so failed folds add no
strict tail violation. Thus \(m\) cannot increase, and the \(p\)-value, which is nonincreasing in \(m\), cannot fall.
If \(|F|\geq q\), at least \(q\) lower endpoints equal \(-\infty\) and at least \(q\) upper endpoints equal
\(+\infty\), so \(a_{(q)}=-\infty\) and \(b_{(N-q)}=+\infty\).
\end{proof}

Full conformal inference treats failed fits the same way. A control whose fit fails at candidate \(c\) gets score
\(S_j(c)=+\infty\), and a failed unit-\(0\) fit gives \(p_t^{\mathrm{FC}}(c)=1\). Take as reference any
finite-valued singleton score rule that meets the conditions of Theorem~\ref{thm:fc-app} and agrees with the
implemented rule wherever a fit succeeds. Each failed fit can only raise \(p_t^{\mathrm{FC}}(c)\) above the reference
value, so the implemented set contains the reference set and keeps the coverage of Theorem~\ref{thm:fc-app}.

Appendix~\ref{app:alg-jplus-fp} states how Theorem~\ref{thm:generic-jplus-app} applies when units that are
unobserved at the target date, or ineligible, are left out of the pool.

\begin{proof}[Proof of Theorem~\ref{thm:generic-jplus-app}]
If \(q_\gamma=0\), both intervals equal \(\R\), so both coverage
probabilities are one. Suppose henceforth that
\[
  q=q_\gamma\geq1,
  \qquad
  r=N-q.
\]

For \(i\neq k\), define the directed comparison
\begin{equation}
  A_{ik}=\one\{R_{ik}^{\ell}>R_{ki}^{\ell}\},
  \qquad A_{ii}=0.
  \label{eq:tournament-comparison-app}
\end{equation}
A tie gives \(A_{ik}=A_{ki}=0\), so
\begin{equation}
  A_{ik}+A_{ki}\leq1.
  \label{eq:one-directed-win-app}
\end{equation}
Since \(\mathbf A=[A_{ik}]\) changes only by row-and-column relabeling when
we relabel \(\mathbf R^{\ell}\), it is conditionally exchangeable under
Assumption~\ref{ass:jplus-matrix-exchangeability-app}.

Let
\begin{equation}
  D_i=\sum_{k\neq i}A_{ik},
  \qquad
  \mathcal S=\{i:D_i\geq r\}.
  \label{eq:strange-set-app}
\end{equation}
We first bound the number of units in \(\mathcal S\). Write
\(s=|\mathcal S|\). If \(s\leq q\), then
\(s\leq2q-1\) because \(q\geq1\). Suppose \(s\geq q+1\). Each
unit in \(\mathcal S\) wins at least \(r=N-q\) comparisons. At most
\(N-s\) of these wins can be against units outside \(\mathcal S\), so
each unit in \(\mathcal S\) wins at least
\[
  (N-q)-(N-s)=s-q
\]
comparisons against other units in \(\mathcal S\). Summing these internal
wins gives
\begin{equation}
  s(s-q)
  \leq
  \sum_{i\in\mathcal S}
  \sum_{\substack{k\in\mathcal S\\k\neq i}}A_{ik}.
  \label{eq:internal-wins-lower-app}
\end{equation}
By \eqref{eq:one-directed-win-app}, every unordered pair inside
\(\mathcal S\) contributes at most one directed win. Hence
\begin{equation}
  \sum_{i\in\mathcal S}
  \sum_{\substack{k\in\mathcal S\\k\neq i}}A_{ik}
  \leq\binom{s}{2}.
  \label{eq:internal-wins-upper-app}
\end{equation}
Combining the two bounds and dividing by \(s>0\) yields
\[
  s-q\leq\frac{s-1}{2},
\]
and therefore
\begin{equation}
  |\mathcal S|\leq2q-1.
  \label{eq:strange-set-size-app}
\end{equation}

Conditional row-and-column exchangeability makes the indicators
\(\one\{i\in\mathcal S\}\) exchangeable. Thus, almost surely,
\begin{align}
  \Pr\{0\in\mathcal S\mid\Ginfo_t^{\ell}\}
  &=
  \frac1N
  \operatorname E\!\left[
    \sum_{i\in\Iset}\one\{i\in\mathcal S\}
    \mathrel{\Big|}\Ginfo_t^{\ell}
  \right]\\
  &=\frac1N\operatorname E[|\mathcal S|\mid\Ginfo_t^{\ell}]\\
  &\leq\frac{2q-1}{N}.
  \label{eq:strange-probability-app}
\end{align}

Suppose first that the untreated outcome exceeds the upper endpoint:
\[
  Y_{0t}(0)>b_{(r)}^{\ell}.
\]
At least \(r\) controls \(j\) satisfy
\[
  \widehat M_{0j}^{\ell}+R_{j0}^{\ell}<Y_{0t}(0).
\]
For each such \(j\),
\[
  R_{0j}^{\ell}
  =|Y_{0t}(0)-\widehat M_{0j}^{\ell}|
  >R_{j0}^{\ell}.
\]
Hence \(A_{0j}=1\) for at least \(r\) controls, so
\(0\in\mathcal S\).

Suppose instead that the untreated outcome is below the lower endpoint:
\[
  Y_{0t}(0)<a_{(q)}^{\ell}.
\]
At least
\[
  J-q+1=N-q=r
\]
controls \(j\) satisfy
\[
  \widehat M_{0j}^{\ell}-R_{j0}^{\ell}>Y_{0t}(0).
\]
For each of these controls, again
\(R_{0j}^{\ell}>R_{j0}^{\ell}\), so \(0\in\mathcal S\).

The interval is nonempty. For every \(j\),
\(a_j^{\ell}\leq b_j^{\ell}\), which implies
\(a_{(m)}^{\ell}\leq b_{(m)}^{\ell}\) for every order \(m\). Since \(q\leq r\),
\begin{equation}
  a_{(q)}^{\ell}\leq a_{(r)}^{\ell}\leq b_{(r)}^{\ell}.
  \label{eq:jplus-nonempty-app}
\end{equation}
The two tail arguments therefore establish
\[
  \{Y_{0t}(0)\notin
    \widehat C_t^{Y,\ell,\mathrm{J+}}(\gamma)\}
  \subseteq
  \{0\in\mathcal S\}.
\]
Combining this inclusion with \eqref{eq:strange-probability-app} proves
the first inequality in
\eqref{eq:generic-jplus-prediction-coverage-app}.

Consistency gives
\begin{align*}
  \tau_t\in
  \widehat C_t^{\tau,\ell,\mathrm{J+}}(\gamma)
  &\Longleftrightarrow
  Y_{0t}-\tau_t\in
  \widehat C_t^{Y,\ell,\mathrm{J+}}(\gamma)\\
  &\Longleftrightarrow
  Y_{0t}(0)\in
  \widehat C_t^{Y,\ell,\mathrm{J+}}(\gamma),
\end{align*}
so prediction and effect coverage are identical. Finally, the definition
of \(q_\gamma\) gives
\((2q_\gamma-1)_+\leq\gamma N\), proving the second inequality in each
coverage statement.
\end{proof}

\begin{proof}[Proof of Proposition~\ref{prop:jplus-duality-app}]
Suppose first that \(q_\gamma=0\). The interval \eqref{eq:generic-jplus-effect-app} equals \(\R\). Also
\(m_t^\ell(c)\leq J=N-1\), so
\(p_t^{\ell,\mathrm{J+}}(c)\geq1/N>\gamma\). Both events in \eqref{eq:jplus-duality-app} are therefore false.

For the remaining case, set \(q=q_\gamma\geq1\) and \(y(c)=Y_{0t}-c\). Because each endpoint collection contains
\(J=N-1\) values,
\begin{align}
  y(c)<a_{(q)}^{\ell}
  &\quad\Longleftrightarrow\quad
  L_t^\ell(c)\geq N-q,
  \label{eq:jplus-lower-order-app}\\
  y(c)>b_{(N-q)}^{\ell}
  &\quad\Longleftrightarrow\quad
  U_t^\ell(c)\geq N-q.
  \label{eq:jplus-upper-order-app}
\end{align}
Equations~\eqref{eq:jplus-lower-order-app} and \eqref{eq:jplus-upper-order-app} therefore imply
\[
  c\notin\widehat C_t^{\tau,\ell,\mathrm{J+}}(\gamma)
  \quad\Longleftrightarrow\quad
  m_t^\ell(c)\geq N-q.
\]
Because \(q\leq\lfloor N/2\rfloor\), the event on the right is equivalent to
\(p_t^{\ell,\mathrm{J+}}(c)\leq(2q-1)/N\). The $p$-value takes values only at these cutoffs and at one. By
definition, \(q_\gamma\) is the largest \(q\) for which \((2q-1)/N\leq\gamma\). Thus the last event is
equivalent to \(p_t^{\ell,\mathrm{J+}}(c)\leq\gamma\), proving the claim.
\end{proof}

\subsection{Joint full conformal inference for a full-path estimator}
\label{app:pooled-fc}

For a full-path estimator, the fit that holds out a control needs unit \(0\)'s untreated outcomes at every
post-treatment date, as Remark~\ref{sec:pooled-fc-difficulty} notes. The exact construction therefore posits a candidate
for the whole effect path, tests the joint hypothesis \(\boldsymbol\tau=\boldsymbol c\), and projects the accepted
paths onto coordinate \(t\). Restricting the fit to \(\Tpre\cup\{t\}\) would avoid this search only by turning the
estimator into a target-date one.

Let \(\boldsymbol c=(c_s:s\in\Tpost)\) be a candidate effect path and define the completed post-treatment panel by
\begin{equation}
  Y_{is}^{(\boldsymbol c)}
  =
  \begin{cases}
    Y_{0s}-c_s,&i=0,\\
    Y_{is},&i\in\Cset,
  \end{cases}
  \qquad s\in\Tpost,
  \label{eq:path-candidate-completion-app}
\end{equation}
so that at \(\boldsymbol c=\boldsymbol\tau\) every completed value equals its untreated potential outcome. For
each \(i\in\Iset\), hide unit \(i\)'s complete post-treatment path, keep its pre-treatment history, and let
\(\widehat M_i^{\mathrm{fp}}(\boldsymbol c)\) be the full-path singleton prediction from \(Y_{\Iset,\Tpre}\) and
\(Y_{\Iset\setminus\{i\},\Tpost}^{(\boldsymbol c)}\) with tuning \(\widehat\eta_t\). The score at the target date is
\(S_i^{\mathrm{fp}}(\boldsymbol c)=|Y_{it}^{(\boldsymbol c)}-\widehat M_i^{\mathrm{fp}}(\boldsymbol c)|\). By
\eqref{eq:ordinary-pooled-extension-agreement-app}, the prediction for unit \(0\) equals the estimator's
full-path prediction at every candidate path, so
\(S_0^{\mathrm{fp}}(\boldsymbol c)=|\widehat\tau_t^{\mathrm{fp}}-c_t|\) with
\(\widehat\tau_t^{\mathrm{fp}}=Y_{0t}-\widehat Y_{0t}^{\mathrm{fp}}(0)\). Let
\(p_t^{\mathrm{fp,FC}}(\boldsymbol c)\) be the rank \(p\)-value \eqref{eq:fc-p-app} with
\(S_i^{\mathrm{fp}}(\boldsymbol c)\) in place of \(S_i(c)\), and define
\begin{align}
  \widehat{\mathcal C}_t^{\mathrm{path}}(\gamma)
  &=\left\{\boldsymbol c\in\R^{|\Tpost|}:p_t^{\mathrm{fp,FC}}(\boldsymbol c)>\gamma\right\},
  \label{eq:pooled-fc-path-region-app}\\
  \widehat C_t^{\mathrm{fp,FC}}(\gamma)
  &=\left\{c_t:\text{there exists }\boldsymbol c_{-t}\text{ with }
    \boldsymbol c\in\widehat{\mathcal C}_t^{\mathrm{path}}(\gamma)\right\}.
  \label{eq:pooled-fc-projection-app}
\end{align}

\begin{theorem}[Joint full conformal coverage for a full-path estimator]
\label{thm:pooled-fc-app}
Let \(N=J+1\geq2\) and \(\gamma\in[0,1)\). Suppose \eqref{eq:no-anticipation-app}--\eqref{eq:consistency-app} hold,
the full-path singleton score rule is measurable and finite-valued, and the
membership events in \eqref{eq:pooled-fc-path-region-app} and
\eqref{eq:pooled-fc-projection-app} are measurable. Suppose also that, at
the true path, the score vector is conditionally exchangeable:
\begin{equation}
  \mathbf S_t^{\mathrm{fp}}(\boldsymbol\tau)
  \mid\Ginfo_t^{\mathrm{all}}
  \stackrel d=
  P_\pi\mathbf S_t^{\mathrm{fp}}(\boldsymbol\tau)
  \mid\Ginfo_t^{\mathrm{all}}
  \label{eq:pooled-fc-score-exchangeability-app}
\end{equation}
for every permutation \(\pi\). Then, almost surely,
\begin{align}
  \Pr\!\left\{
    \boldsymbol\tau\in
    \widehat{\mathcal C}_t^{\mathrm{path}}(\gamma)
    \mid\Ginfo_t^{\mathrm{all}}
  \right\}
  &\geq
  1-\frac{\lfloor\gamma N\rfloor}{N}\geq1-\gamma,
  \label{eq:pooled-fc-path-coverage-app}\\
  \Pr\!\left\{
    \tau_t\in
    \widehat C_t^{\mathrm{fp,FC}}(\gamma)
    \mid\Ginfo_t^{\mathrm{all}}
  \right\}
  &\geq
  1-\frac{\lfloor\gamma N\rfloor}{N}\geq1-\gamma.
  \label{eq:pooled-fc-projected-coverage-app}
\end{align}
\end{theorem}

The coverage statement applies to the exact path region and its exact
projection.

\begin{proof}[Proof of Theorem~\ref{thm:pooled-fc-app}]
At \(\boldsymbol c=\boldsymbol\tau\), the candidate-completed inputs in
\eqref{eq:path-candidate-completion-app} equal the complete untreated
panel. The score vector is therefore the vector in
\eqref{eq:pooled-fc-score-exchangeability-app}, and the true path is
excluded from \(\widehat{\mathcal C}_t^{\mathrm{path}}(\gamma)\) exactly
when \(p_t^{\mathrm{fp,FC}}(\boldsymbol\tau)\leq\gamma\). The rank
argument in the proof of Theorem~\ref{thm:fc-app}, with
\(S_i^{\mathrm{fp}}(\boldsymbol\tau)\) in place of \(S_i(\tau_t)\) and
\(\Ginfo_t^{\mathrm{all}}\) in place of \(\Ginfo_t\), bounds the
conditional probability of this event by \(\lfloor\gamma N\rfloor/N\),
which proves \eqref{eq:pooled-fc-path-coverage-app}. Because
\(\boldsymbol\tau\in\widehat{\mathcal C}_t^{\mathrm{path}}(\gamma)\)
implies \(\tau_t\in\widehat C_t^{\mathrm{fp,FC}}(\gamma)\) by
\eqref{eq:pooled-fc-projection-app},
\eqref{eq:pooled-fc-projected-coverage-app} follows.
\end{proof}

\subsection{Simultaneous bands over the post-treatment dates}
\label{app:bands}

The guarantees above are period-wise. Over the \(H=|\Tpost|\) post-treatment dates, the period-wise sets
\((\widehat C_s)_{s\in\Tpost}\) can all cover their targets with probability well below \(1-\gamma\). A
\emph{simultaneous band} is a family \((\widehat B_s)_{s\in\Tpost}\) with
\(\Pr\{\tau_s\in\widehat B_s\text{ for all }s\in\Tpost\}\geq1-\beta\). Each band below is the projection onto the dates of a confidence region \(\widehat{\mathcal C}\subseteq\R^H\) for the effect path \(\boldsymbol\tau=(\tau_s)_{s\in\Tpost}\), or a superset of that projection. Because \(\{\boldsymbol\tau\in\widehat{\mathcal C}\}\) implies
\(\{\tau_s\in\widehat B_s\text{ for all }s\}\), the band inherits the region's coverage.

The Bonferroni band runs each period-wise procedure at \(\gamma/H\). By the union bound, its coverage is at least
\(1-\gamma\). This band is the real line at every date unless
\(\gamma/H\geq1/N\), that is, \(N\geq H/\gamma\). At \(\gamma=0.10\), this requires \(N\geq120\) for
Proposition~99 (\(H=12\)) and \(N\geq140\) for German reunification (\(H=14\)), so no application here meets it.

The full conformal and Jackknife+ bands below score each unit's whole post-treatment path. A unit's path score is
its largest weighted absolute prediction error over \(\Tpost\). The weight is computed from that unit's own fit by a rule
that is the same for every unit and is chosen before the results are seen. We use either \(w_i=1\) (equal weights)
or \(w_i=1/\widehat\sigma_i\) (studentized weights), where \(\widehat\sigma_i\) is the root mean squared in-sample
residual of held-out unit \(i\) over \(\Tpre\) in the same fit.

Because weights are computed within each fit, unit \(0\) and control \(j\) each get their own weight in Jackknife+ fold \(j\). Under a target-date estimator, which has one fit per date, we take \(\widehat\sigma_i\) from a pre-treatment fit
that is the same at every date; for the full conformal band, it must also not depend on \(\boldsymbol c\). We also assume
\(\widehat\sigma_i>0\) for every unit, which fails only when a unit's pre-treatment fit is exact.

For a target-date estimator and a candidate path \(\boldsymbol c\in\R^H\), we complete \(Y_{0s}^{(\boldsymbol c)}=Y_{0s}-c_s\) at
every \(s\in\Tpost\), let \(\widehat M_{is}(c_s)\) be the date-\(s\) singleton prediction
\eqref{eq:fc-prediction-app}, which uses only \(c_s\), and set
\[
  S_i(\boldsymbol c)=\max_{s\in\Tpost}w_i\bigl|Y_{is}^{(\boldsymbol c)}-\widehat M_{is}(c_s)\bigr|,
\]
where we define \(p^{\mathrm{FC}}(\boldsymbol c)\) and \(\widehat{\mathcal C}^{\mathrm{FC}}\) by \eqref{eq:fc-p-app} and
\eqref{eq:fc-set-app} with \(S_i(\boldsymbol c)\) in place of \(S_i(c)\). Let \(\widehat B^{\mathrm{FC}}_s\) be
the projection of \(\widehat{\mathcal C}^{\mathrm{FC}}\) onto date \(s\).

For the Jackknife+ with \(\ell\in\{\mathrm{td},\mathrm{fp}\}\), fold \(j\) hides units \(0\) and \(j\)
at every post-treatment date. We use the date-\(s\) pair fit \eqref{eq:all-pairs-fit-app} for each \(s\) under a
target-date estimator, and a single pair fit that hides both paths under a full-path estimator. Write
\(\widehat M_{0j,s}\) and \(\widehat M_{j0,s}\) for the date-\(s\) predictions and \(w_{0}^{(j)}\), \(w_j\) for
the weights from fold \(j\). Set
\[
  R_{j0}=\max_{s\in\Tpost}w_j\bigl|Y_{js}-\widehat M_{j0,s}\bigr|,
  \qquad
  a_{js}=\widehat M_{0j,s}-R_{j0}/w_0^{(j)},
  \qquad
  b_{js}=\widehat M_{0j,s}+R_{j0}/w_0^{(j)},
\]
and, with \(q=q_\gamma\) and \(u=N-q\) from \eqref{eq:generic-jplus-calibration-app},
\(\widehat B^{\mathrm{J+}}_s=Y_{0s}-[a_{(q),s},b_{(u),s}]\) if \(q\geq1\) and \(\widehat B^{\mathrm{J+}}_s=\R\) if
\(q=0\). The all-pairs table \(\mathbf R^{\ell,\mathrm{path}}\) is built by applying the same path score to every pair of units on the complete untreated panel.

\begin{theorem}[Simultaneous bands]
\label{thm:bands-app}
Let \(N=J+1\geq2\) and \(\gamma\in[0,1)\), and suppose \eqref{eq:no-anticipation-app}--\eqref{eq:consistency-app}
hold, the path scores are measurable and finite, and the events below are measurable.
\begin{enumerate}
\item For a target-date estimator, if the vector \((S_0(\boldsymbol\tau),\ldots,S_J(\boldsymbol\tau))\) is
exchangeable conditional on a \(\sigma\)-field \(\mathcal H\), then
\(\Pr\{\tau_s\in\widehat B^{\mathrm{FC}}_s\text{ for all }s\in\Tpost\mid\mathcal H\}\geq
\Pr\{\boldsymbol\tau\in\widehat{\mathcal C}^{\mathrm{FC}}\mid\mathcal H\}\geq1-\lfloor\gamma N\rfloor/N\).
\item If the table \(\mathbf R^{\ell,\mathrm{path}}\) satisfies the exchangeability
\eqref{eq:jplus-matrix-exchangeability-app} conditional on a \(\sigma\)-field \(\mathcal H\), then
\(\Pr\{\tau_s\in\widehat B^{\mathrm{J+}}_s\text{ for all }s\in\Tpost\mid\mathcal H\}\geq1-(2q_\gamma-1)_+/N\).
\end{enumerate}
\end{theorem}

\begin{proof}[Proof of Theorem~\ref{thm:bands-app}]
Part 1 is the rank argument in the proof of Theorem~\ref{thm:fc-app} applied to the scalar scores
\(S_i(\boldsymbol\tau)\), followed by the projection argument of Appendix~\ref{app:bands}. For part 2, the proof of Theorem~\ref{thm:generic-jplus-app} uses only the score comparisons
\(A_{ik}=\one\{R_{ik}>R_{ki}\}\), which are defined for any scalar score. Thus the bound on the strange set
\(\mathcal S\) and \(\Pr\{0\in\mathcal S\mid\mathcal H\}\leq(2q-1)_+/N\) hold without change. Let
\(R_{0j}(\boldsymbol c)=\max_s w_0^{(j)}|Y_{0s}-c_s-\widehat M_{0j,s}|\). If
\(\#\{j:R_{0j}(\boldsymbol\tau)\leq R_{j0}\}<q\), unit \(0\) beats at least \(J-q+1=N-q\) controls, so
\(0\in\mathcal S\). Hence the region \(\{\boldsymbol c:\#\{j:R_{0j}(\boldsymbol c)\leq R_{j0}\}\geq q\}\) covers
\(\boldsymbol\tau\) with probability at least \(1-(2q-1)_+/N\). Since \(R_{0j}(\boldsymbol c)\leq R_{j0}\) holds
exactly when \(c_s\in[Y_{0s}-b_{js},Y_{0s}-a_{js}]\) for every \(s\), at each point of the region and each date
\(s\) the coordinate \(c_s\) lies in at least \(q\) of these \(J\) intervals. Every such value lies between the
\(q\)th smallest left endpoint, \(Y_{0s}-b_{(u),s}\), and the \(q\)th largest right endpoint,
\(Y_{0s}-a_{(q),s}\), that is, in \(\widehat B^{\mathrm{J+}}_s\).
\end{proof}

The exchangeability conditions for the two bands are those of Assumptions~\ref{ass:fc-score-exchangeability-app}
and~\ref{ass:jplus-matrix-exchangeability-app}, applied to the path score. The sufficient condition of
Proposition~\ref{prop:complete-unit-route-app}, exchangeable complete units, implies them for a common prediction
rule with tuning fixed in advance, because the same rule computes each unit's weight from its own fit. With
studentized weights, the condition concerns each unit's pre-treatment fit together with its post-treatment errors.

The relabeling test of Theorem~\ref{thm:multi-relabel-app} can also use the largest weighted absolute estimated effect over the treated cells as its statistic. Inverting this test over candidate effect arrays gives a band over the cells of
multiple treated units, through a search of the same kind as in Appendix~\ref{app:pooled-fc}. With one treated
unit and a target-date estimator, this band is the full conformal band above.

The Jackknife+ band uses the period-wise folds and requires no further fits. Appendix~\ref{app:band-computation} computes the full conformal band for DID and simplex synthetic control.
Appendix~\ref{app:sims-bands} checks the bands in finite samples, and
Appendix~\ref{app:apps-bands} reports them for the three applications.

\subsection{Pre-treatment diagnostics: construction and validity}
\label{app:historical-diagnostics}

Index the last pre-treatment date by \(0\). For a number \(m<|\Tpre|\), fixed before inspecting the diagnostic
output, the diagnostic dates are \(\mathcal S_{\mathrm{pl}}=\{-m+1,\ldots,-1,0\}\). The results below hold for
any such \(m\). For a diagnostic date \(s\in\mathcal S_{\mathrm{pl}}\),
let
\begin{equation}
  \mathcal H_s=\{u\in\Tpre:u<s\},
  \qquad
  \mathcal B_s=\{u\in\Tpre\cup\Tpost:u\geq s\}
  \label{eq:placebo-history-app}
\end{equation}
be the dates before \(s\), a nonempty set, and the dates from \(s\) on. The estimator uses \(Y_{0,\mathcal H_s}\)
to predict the treated unit's outcome at \(s\), and its pre-treatment prediction error is
\(\widehat\tau_s^{\mathrm{pl}}\) in \eqref{eq:placebo-estimate-main}. Under no anticipation, the pseudo-effect
\(\tau_s^{\mathrm{pl}}=Y_{0s}-Y_{0s}(0)\) is zero. For a candidate shift \(c\), the diagnostic null
\(H_{0,s}^{\mathrm{pl}}(c)\) states \(\tau_s^{\mathrm{pl}}=c\), that is, \(Y_{0s}(0)=Y_{0s}-c\); we maintain
\(c=0\).

At each diagnostic date, we use the same estimator family, rule for which controls are eligible, tuning rule, numerical routine, and failure rule as after treatment. Because the available history changes
with \(s\), the tuning value \(\widehat\eta_s\) may differ across diagnostic dates. At each \(s\), it is fixed in
advance or chosen by a rule that meets the condition of Proposition~\ref{prop:tuning-app}. This value remains fixed across all the fits
at that date that hold out a control.

The constructions at a diagnostic date are those of Appendices~\ref{app:fc-proof} and~\ref{app:jplus-proof} with
the target date \(t\), the held-out history \(\Tpre\), the hidden block \(\Tpost\) and the tuning
\(\widehat\eta_t\) replaced by \(s\), \(\mathcal H_s\), \(\mathcal B_s\) and \(\widehat\eta_s\). For a complete
panel \(Z\) and a set \(Q\subseteq\Iset\) of held-out units, the panels available to a fit are
\begin{equation}
  D_s^{\mathrm{pl},\mathrm{td}}(Q;Z)
  =\left(Z_{\Iset,\mathcal H_s},Z_{\Iset\setminus Q,s}\right),
  \qquad
  D_s^{\mathrm{pl},\mathrm{fp}}(Q;Z)
  =\left(Z_{\Iset,\mathcal H_s},Z_{\Iset\setminus Q,\mathcal B_s}\right),
  \label{eq:placebo-pointwise-panel-app}
\end{equation}
written \(D_s^{\mathrm{pl},\ell}(Q;Z)\) for \(\ell\in\{\mathrm{td},\mathrm{fp}\}\). The singleton and pair rules
\(\mathcal A_s^{(r,\ell)}(Q;D_s^{\mathrm{pl},\ell}(Q;Z),\widehat\eta_s)\) apply the same estimation algorithm as
\(\mathcal A_t^{(r,\ell)}\) to these panels, and the declared rules must satisfy
\eqref{eq:formal-role-symmetry-app} after the same substitutions.

For full conformal inference under a target-date rule, a candidate pseudo-effect \(c\) gives the panel
\(Y^{(c)}\), which equals \(Y\) off the column at \(s\), and the singleton scores
\begin{equation}
  Y_{is}^{(c)}=Y_{is}-c\,\one\{i=0\},
  \qquad
  S_{i,s}(c)=\left|Y_{is}^{(c)}-\widehat M_{i,s}(c)\right|,
  \label{eq:placebo-fc-score-app}
\end{equation}
where \(\widehat M_{i,s}(c)\) is the singleton prediction from \(D_s^{\mathrm{pl},\mathrm{td}}(\{i\};Y^{(c)})\). The
$p$-value \(p_s^{\mathrm{pl,FC}}(c)\) and the set \(\widehat C_s^{\mathrm{pl,FC}}(\gamma)\) are
\eqref{eq:fc-p-app} and \eqref{eq:fc-set-app} computed from these scores.

For the Jackknife+ with \(\ell\in\{\mathrm{td},\mathrm{fp}\}\), the all-pairs table is
\begin{equation}
  \mathbf R_s^\ell=[R_{ik,s}^\ell]_{i,k\in\Iset},
  \qquad
  R_{ik,s}^{\ell}=\left|Y_{is}(0)-\widehat M_{ik,s}^{\ell}\right|,
  \qquad
  R_{ii,s}^{\ell}=0,
  \label{eq:placebo-all-pairs-fit-app}
\end{equation}
where \(\widehat M_{ik,s}^{\ell}\) is the prediction for unit \(i\) from
\(\mathcal A_s^{(2,\ell)}(\{i,k\};D_s^{\mathrm{pl},\ell}(\{i,k\};Y(0)),\widehat\eta_s)\). As after treatment in Appendix~\ref{app:jplus-proof}, only the
\(J\) folds \(\{0,j\}\) are computed, and their inputs are observed. The endpoints
\(a_{j,s}^\ell\) and \(b_{j,s}^\ell\), the intervals \(\widehat C_s^{Y,\mathrm{pl},\ell,\mathrm{J+}}(\gamma)\) and
\(\widehat C_s^{\mathrm{pl},\ell,\mathrm{J+}}(\gamma)\), and the $p$-value \(p_s^{\mathrm{pl},\ell,\mathrm{J+}}(c)\) are
\eqref{eq:generic-jplus-endpoints-app}--\eqref{eq:generic-jplus-effect-app} and
\eqref{eq:jplus-tail-counts-app}--\eqref{eq:jplus-pvalue-app} with \(s\) and \(Y_{0s}\) in place of \(t\) and
\(Y_{0t}\). Proposition~\ref{prop:jplus-duality-app} holds after the same substitution.

\paragraph{Period-wise validity.}

Let \(\Ginfo_s^{\mathrm{td}}\) collect the signals of all units on
\(\mathcal H_s\cup\{s\}\), the fixed tuning value
\(\widehat\eta_s\), and any other conditioning information of the target-date diagnostic. Define \(\Ginfo_s^{\mathrm{fp}}\) in the same way for the
full-path rule, with the signals of all units over all dates. In the
full conformal condition below, the singleton scores are evaluated at the
maintained pseudo-effect of zero.

\begin{corollary}[Period-wise pre-treatment diagnostic coverage]
\label{cor:placebo-validity-app}
Let \(N=J+1\geq2\) and \(\gamma\in[0,1)\), and suppose no anticipation and no spillovers hold. If the target-\(s\)
singleton score vector in \eqref{eq:placebo-fc-score-app} is exchangeable
conditional on \(\Ginfo_s^{\mathrm{td}}\), then, almost surely,
\begin{equation}
  \Pr\left\{
    0\in\widehat C_s^{\mathrm{pl,FC}}(\gamma)
    \mid\Ginfo_s^{\mathrm{td}}
  \right\}
  \geq
  1-\frac{\lfloor\gamma N\rfloor}{N}
  \geq1-\gamma.
  \label{eq:placebo-fc-coverage-app}
\end{equation}
If, for \(\ell\in\{\mathrm{td},\mathrm{fp}\}\), the all-pairs table
\(\mathbf R_s^\ell\) is row-and-column exchangeable conditional on
\(\Ginfo_s^\ell\), then, almost surely,
\begin{equation}
  \Pr\left\{
    0\in
    \widehat C_s^{\mathrm{pl},\ell,\mathrm{J+}}(\gamma)
    \mid\Ginfo_s^\ell
  \right\}
  \geq
  1-\frac{(2q_\gamma-1)_+}{N}
  \geq1-\gamma.
  \label{eq:placebo-jplus-coverage-app}
\end{equation}
Under the corresponding score condition, the null $p$-values satisfy, for
every \(x\in[0,1]\),
\begin{align}
  \Pr\left\{
    p_s^{\mathrm{pl,FC}}(0)\leq x
    \mid\Ginfo_s^{\mathrm{td}}
  \right\}
  &\leq x,
  \label{eq:placebo-fc-superuniform-app}\\
  \Pr\left\{
    p_s^{\mathrm{pl},\ell,\mathrm{J+}}(0)\leq x
    \mid\Ginfo_s^\ell
  \right\}
  &\leq x.
  \label{eq:placebo-jplus-superuniform-app}
\end{align}
\end{corollary}

Each result holds at one diagnostic date; Section~\ref{sec:historical-diagnostics} discusses what the
diagnostics do not show.

\begin{proof}[Proof of Corollary~\ref{cor:placebo-validity-app}]
For the target-date constructions, take \(s\) as the target date and
\(\mathcal H_s\), the dates before \(s\), in place of \(\Tpre\). No anticipation and no spillovers give
\[
  Y_{0s}=Y_{0s}(0),
  \qquad
  Y_{is}=Y_{is}(0),\quad i\in\Cset.
\]
At \(c=0\), the completion in \eqref{eq:placebo-fc-score-app} recovers the
complete untreated target slice. The rank argument in
Theorem~\ref{thm:fc-app} then gives
\eqref{eq:placebo-fc-coverage-app}. For \(x<1\), applying the same rank
bound at \(\gamma=x\) gives
\eqref{eq:placebo-fc-superuniform-app}; the claim is immediate at \(x=1\).

For both rules, the table in \eqref{eq:placebo-all-pairs-fit-app} applies the
declared pair map to the complete untreated panel, and the computed
\(\{0,j\}\) folds are entries of it. Conditional row-and-column exchangeability of
\(\mathbf R_s^\ell\) therefore permits the tournament argument in
Theorem~\ref{thm:generic-jplus-app}, which gives \eqref{eq:placebo-jplus-coverage-app}. Prediction and pseudo-effect coverage agree because
\[
  0\in
  Y_{0s}
  -\widehat C_s^{Y,\mathrm{pl},\ell,\mathrm{J+}}(\gamma)
  \quad\Longleftrightarrow\quad
  Y_{0s}=Y_{0s}(0)
  \in\widehat C_s^{Y,\mathrm{pl},\ell,\mathrm{J+}}(\gamma).
\]
For \(x<1\), Proposition~\ref{prop:jplus-duality-app} applied at \(s\), combined
with the Jackknife+ noncoverage bound at \(\gamma=x\), gives
\eqref{eq:placebo-jplus-superuniform-app}; the inequality is immediate at
\(x=1\).
\end{proof}

\subsection{Multiple treated units}
\label{app:multi}

The main text fixes one treated unit, unit \(0\). Now let \(\mathcal K\subseteq\Iset\) be the treated units, with \(K=|\mathcal K|\geq2\), all treated at every date in \(\Tpost\), and let \(\Cset=\Iset\setminus\mathcal K\) be the controls, with \(J=|\Cset|\) and \(n=K+J\). We maintain no anticipation and no spillovers, so a cell \((i,s)\) is treated only if \(i\in\mathcal K\) and \(s\in\Tpost\). Section~\ref{sec:multi} describes the three questions and the constructions that answer them. We state their formal results here. Figure~\ref{fig:multiple-treated-common-date} shows the prediction tasks of the block Jackknife+ and the relabeling test. Appendix~\ref{app:staggered} treats staggered adoption.

\paragraph{One treated cell at a time.}
Fix a treated unit \(k\) and a date \(t\in\Tpost\). The \emph{reduced panel} contains unit \(k\) and the controls \(\Cset\), observed over \(\Tpre\cup\{t\}\) for a target-date estimator and over \(\Tpre\cup\Tpost\) for a full-path one. Unit \(k\) plays the role of unit \(0\).

\begin{proposition}[One treated cell among several]
\label{prop:multi-cell-app}
Fix a treated cell \((k,t)\) and apply the declared procedure to the reduced panel. In that panel, unit \(k\) is the only treated unit and every other outcome is untreated. Consequently, if the score assumption of Theorem~\ref{thm:fc-app} (target-date full conformal inference) or of Theorem~\ref{thm:generic-jplus-app} (the Jackknife+) holds for the reduced panel, conditional on the adoption date and the information stated there, the corresponding set for \(\tau_{kt}\) has the coverage that theorem gives, with \(N=J+1\).
\end{proposition}

\begin{proof}[Proof of Proposition~\ref{prop:multi-cell-app}]
With a common adoption date, the reduced panel contains unit \(k\) and the controls. Every outcome of a control is untreated by no spillovers, and unit \(k\) is untreated on \(\Tpre\) by no anticipation. The reduced panel therefore satisfies the setup of Appendix~\ref{app:formal-setup} with unit \(k\) as unit \(0\) and \(N=J+1\). The cited theorem therefore applies.
\end{proof}

The other treated units stay out of the fit. An implementation that uses their outcomes is a different procedure that the proposition does not cover, for example one that estimates the factors from all units' pre-treatment outcomes. The rank argument then applies only if those units are training units in every fit, whichever unit is held out, and the score assumption must then hold conditional on their data. Sets for several cells combine by a union bound.

\begin{corollary}[Simultaneous statements over several cells]
\label{cor:multi-simultaneous-app}
Let \((k_1,t_1),\ldots,(k_M,t_M)\) be treated cells, and for each \(m\) let \(\widehat C_m\) be a set from Proposition~\ref{prop:multi-cell-app} at miscoverage \(\gamma_m\), computed from a reduced panel of \(N_m\) units, with miscoverage at most \(\beta_m\): \(\beta_m=\lfloor\gamma_mN_m\rfloor/N_m\) for full conformal inference and \(\beta_m=(2q_{\gamma_m}-1)_+/N_m\) for the Jackknife+. Then
\[
  \Pr\!\left\{\tau_{k_mt_m}\in\widehat C_m\text{ for all }m\mid\mathcal A\right\}
  \geq1-\sum_{m=1}^M\beta_m
  \geq1-\sum_{m=1}^M\gamma_m ,
\]
where \(\mathcal A\) denotes the adoption date.
\end{corollary}

\begin{proof}[Proof of Corollary~\ref{cor:multi-simultaneous-app}]
By Proposition~\ref{prop:multi-cell-app}, \(\Pr\{\tau_{k_mt_m}\notin\widehat C_m\mid\mathcal H_m\}\leq\beta_m\)
for each \(m\), where \(\mathcal H_m\) is the information on which cell \(m\)'s theorem conditions. It contains the
adoption dates, so taking expectations given \(\mathcal A\) gives
\(\Pr\{\tau_{k_mt_m}\notin\widehat C_m\mid\mathcal A\}\leq\beta_m\), and \(\beta_m\leq\gamma_m\) by
Theorems~\ref{thm:fc-app} and~\ref{thm:generic-jplus-app}. The union bound given \(\mathcal A\) gives the
result.
\end{proof}

The corollary needs no assumption about how the sets depend on one another.

\paragraph{The average effect at a date.}
Let \(\bar\tau_t=K^{-1}\sum_{k\in\mathcal K}\tau_{kt}\), with the individual effects unrestricted. The Jackknife+ extends to this target by replacing each unit with a block of \(K\) units. We partition \(\Cset\) into \(G=\lfloor J/K\rfloor\) blocks \(B_1,\ldots,B_G\) of size \(K\), at random, independently of the outcomes. Balancing the control blocks on pre-treatment paths can invalidate the bound, because the balanced blocks resemble one another in a way that the treated block does not. Leftover controls are training units in every fit.

Fold \(g\) hides the treated block \(\mathcal K\) and \(B_g\). In one fit, we predict each hidden unit using the remaining outcomes, including the hidden units' histories. A target-date estimator hides the two blocks' target-date outcomes; a full-path estimator hides their post-treatment paths. For a full-path estimator, this single fit differs from the separate fits of Section~\ref{sec:role-extension}, even when \(K=1\). Both are pair rules using data that contain both held-out units' histories, and Theorem~\ref{thm:generic-jplus-app} covers either.

Write \(\widehat Y_{it}^{[\mathcal K\cup B_g]}\) for the predictions and record
\begin{equation}
  M_g=\frac1K\sum_{k\in\mathcal K}\widehat Y_{kt}^{[\mathcal K\cup B_g]},
  \qquad
  R_g=\Bigl|\frac1K\sum_{i\in B_g}\bigl(Y_{it}-\widehat Y_{it}^{[\mathcal K\cup B_g]}\bigr)\Bigr|,
  \label{eq:block-fold-app}
\end{equation}
the treated block's average prediction and the absolute value of the control block's average error. Let \(N_{\mathrm B}=G+1\), compute \(q_\gamma\) from \eqref{eq:generic-jplus-calibration-app} with \(N_{\mathrm B}\) in place of \(N\), and let \(a_g=M_g-R_g\) and \(b_g=M_g+R_g\). The interval for the treated block's average untreated outcome is \([a_{(q_\gamma)},b_{(N_{\mathrm B}-q_\gamma)}]\), or \(\R\) when \(q_\gamma=0\). Subtracting it from the treated units' average outcome \(\bar Y_{\mathcal Kt}\) gives the interval \(\widehat C^{\bar\tau}_t(\gamma)\) for \(\bar\tau_t\). Algorithm~\ref{alg:block-jplus-app} in Appendix~\ref{app:algorithms} gives the steps.

The treated block's error in fold \(g\) is \(\bar Y_{\mathcal Kt}-\bar\tau_t-M_g\), because no fit uses a treated cell. Every fold hides the treated units at \(t\), or over \(\Tpost\) under a full-path estimator, and a target-date estimator uses no other post-treatment date. The effects enter only through their average, so two arrays with the same \(\bar\tau_t\) give the same interval.

Coverage follows from Theorem~\ref{thm:generic-jplus-app} applied to blocks. The proof uses the whole table of \(R_{BB'}\) for blocks \(B\neq B'\) among \(B_0=\mathcal K,B_1,\ldots,B_G\). Here \(R_{BB'}\) is the absolute value of block \(B\)'s average prediction error when \(B\) and \(B'\) are hidden together, at the true untreated outcomes. To compute the interval, we need only the predictions \(M_g\) and the column \(B'=B_0\), which contains the \(R_g\).

\begin{proposition}[Block Jackknife+ coverage]
\label{prop:block-jplus-app}
Suppose \eqref{eq:no-anticipation-app}--\eqref{eq:consistency-app} hold for every treated unit, the block all-pairs map is measurable and finite-valued, and the matrix \([R_{BB'}]\) is exchangeable under permutations of the block labels conditional on a \(\sigma\)-field \(\mathcal H\). Then, almost surely,
\[
  \Pr\bigl\{\bar\tau_t\in\widehat C^{\bar\tau}_t(\gamma)\mid\mathcal H\bigr\}
  \geq1-\frac{(2q_\gamma-1)_+}{N_{\mathrm B}}
  \geq1-\gamma .
\]
The exchangeability holds under three conditions. The complete untreated unit objects of the units in \(\mathcal K\cup\Cset\) are exchangeable conditional on \(\mathcal H\). The partition and the tuning are fixed before the outcomes are seen and contained in \(\mathcal H\). The block all-pairs map is equivariant under the permutations of the units that map blocks onto blocks and fix the leftover controls, as it is when the procedure treats units only through their data and applies the same rule to every block.
\end{proposition}

\begin{proof}[Proof of Proposition~\ref{prop:block-jplus-app}]
The proof of Theorem~\ref{thm:generic-jplus-app} uses only the exchangeability of the all-pairs table conditional on the stated \(\sigma\)-field and the counting bound, which holds for every realization (Remark~\ref{rem:generic-sigma-app}). We apply it with the \(N_{\mathrm B}\) blocks \(B_0,\ldots,B_G\) as the units, the matrix \([R_{BB'}]\) as the table, \(N_{\mathrm B}\) in place of \(N\), and the identity \(\bar Y_{\mathcal Kt}=\bar Y_{\mathcal Kt}(0)+\bar\tau_t\), where \(\bar Y_{\mathcal Kt}(0)\) is the treated block's average untreated outcome, in place of \(Y_{0t}=Y_{0t}(0)+\tau_t\). The computed entries \(R_{B_gB_0}=R_g\) and predictions \(M_g\) correspond to the \(\{0,j\}\) folds in the single-unit case, which gives the bound.

For the second statement, a permutation of the block labels is induced by a permutation of the units that maps each block onto its image in a fixed order and fixes the leftover controls. Conditional on \(\mathcal H\) and the leftover controls' data, exchangeability of the complete unit objects makes the panel law invariant under this permutation. By equivariance, the block map commutes with the permutation. The table is therefore exchangeable under permutations of the block labels conditional on \(\mathcal H\) and that data, hence conditional on \(\mathcal H\).
\end{proof}

The interval is a proper subset of the line only when \(N_{\mathrm B}\geq1/\gamma\), that is, when \(J\geq9K\) at \(\gamma=0.10\).

\begin{remark}[Full conformal inference for the average]
\label{rem:block-fc-app}
Full conformal inference scores a control block as the held-out block and makes the treated units training units in its fit. The fit then needs their untreated outcomes at every cell it uses. It thus needs one candidate per treated unit at date \(t\) for a target-date estimator and one per treated cell for a full-path one. In general the control block's score depends on each candidate separately. A DID fit is the exception, since it uses the completed outcomes only through their average. For the other estimators, restricting the candidates to a common value \(c\) gives a set with a coverage guarantee only when the effects are equal across units.

Searching over the \(K\)-dimensional candidate and projecting the accepted region onto the mean is valid. Suppose blocks are scored by their average error, as above, and the control fits are affine in the candidates, as for synthetic control and ridge. Then the projection is the whole line once at least \(\lfloor\gamma N_{\mathrm B}\rfloor\) control blocks weight the treated units unequally. With fewer such blocks it can be bounded. For fits that are not affine in the candidates, we make no general claim. Differences among the candidates that leave the treated block's score unchanged can inflate a control block's score without limit, unless that block's fit weights every treated unit equally.

When blocks are scored by their largest error, a target-date fit gives a bounded projection once \(\gamma N_{\mathrm B}\geq1\), provided its weights on the treated units are fixed and sum in absolute value to less than one for every control. A full-path fit also needs the score to cover the candidates at the other post-treatment dates. In the cases we have computed, the projection is wider than the interval above. Remark~\ref{sec:pooled-fc-difficulty} describes the same problem for a candidate path.
\end{remark}

\paragraph{All treated cells at once.}
A sharp null specifies every treated effect,
\begin{equation}
  H(\boldsymbol c):\ \tau_{ks}=c_{ks}\quad\text{for all }k\in\mathcal K,\ s\in\Tpost,
  \label{eq:multi-sharp-null-app}
\end{equation}
for a declared array \(\boldsymbol c\). Two cases are \(\boldsymbol c=\mathbf 0\), no effect in any treated cell, and a constant effect, \(c_{ks}=c\). The completed panel \(Y^{(\boldsymbol c)}\) replaces \(Y_{ks}\) by \(Y_{ks}-c_{ks}\) on every treated cell and keeps every other outcome. At \(\boldsymbol c=\boldsymbol\tau\), every cell of \(Y^{(\boldsymbol c)}\) is untreated, so the declared procedure can be run on it with any \(K\) units in the treated role.

A \emph{role assignment} is an injective map \(\rho:\mathcal K\to\Iset\). The units in \(\rho(\mathcal K)\) are treated at every date in \(\Tpost\), and the others form the pool. The observed assignment is the identity, written \(\iota\), and \(\mathcal R\) is the set of all assignments, with \(|\mathcal R|=n!/(n-K)!\). With a common adoption date the treatment pattern depends on \(\rho\) only through the block \(\rho(\mathcal K)\). The test below therefore ranks the treated block among the \(\binom nK\) blocks of \(K\) units, each counted \(K!\) times. A \emph{relabeling statistic} \(T(\rho;y)\in[0,\infty]\) applies the declared procedure, with tuning fixed in advance, to a complete panel \(y\) under the treatment pattern of \(\rho\). An example is the absolute value of the estimator's average effect over the treated cells that \(\rho\) implies. Define
\begin{equation}
  p^{\mathrm{rel}}(\boldsymbol c)
  =\frac{1}{|\mathcal R|}
   \sum_{\rho\in\mathcal R}
   \one\!\left\{T(\rho;Y^{(\boldsymbol c)})\geq T(\iota;Y^{(\boldsymbol c)})\right\}.
  \label{eq:multi-relabel-p-app}
\end{equation}
Ties are counted in favor of the null.

The relabeling test is full conformal inference applied to blocks. The completed panel supplies the treated cells, as the candidate does in Section~\ref{sec:main-full-conformal}, and every block is scored under the same completed panel. The null must therefore give the effect of every treated unit at every date the statistic uses. For each treated unit, some assignment places it in the pool, where its completed outcomes enter the fit, and some places it in the relabeled block, where they enter the score. Let \(\mathcal H\) be a \(\sigma\)-field that contains the adoption date and the fixed tuning values.

\begin{assumption}[Relabeling exchangeability]
\label{ass:multi-relabel-app}
For every permutation \(\pi\) of \(\Iset\), almost surely,
\[
  \bigl(T(\pi\circ\rho;Y^{(\boldsymbol\tau)})\bigr)_{\rho\in\mathcal R}\mid\mathcal H
  \stackrel d=
  \bigl(T(\rho;Y^{(\boldsymbol\tau)})\bigr)_{\rho\in\mathcal R}\mid\mathcal H .
\]
\end{assumption}

The assumption requires that, absent treatment, the actual treated units be as easy to predict as any other \(K\) units.

\begin{theorem}[Relabeling test]
\label{thm:multi-relabel-app}
Suppose no anticipation and no spillovers hold, \(T\) is measurable, and
Assumption~\ref{ass:multi-relabel-app} holds. Then, for every \(\gamma\in[0,1)\), almost surely,
\[
  \Pr\!\left\{p^{\mathrm{rel}}(\boldsymbol\tau)\leq\gamma\mid\mathcal H\right\}
  \leq\frac{\lfloor\gamma|\mathcal R|\rfloor}{|\mathcal R|}
  \leq\gamma .
\]
The same bound, with \(B+1\) in place of \(|\mathcal R|\), holds for the Monte Carlo \(p\)-value
\[
  p^{\mathrm{rel}}_B(\boldsymbol c)
  =\frac{1}{B+1}
   \left(1+\sum_{b=1}^B\one\!\left\{T(\rho_b;Y^{(\boldsymbol c)})\geq T(\iota;Y^{(\boldsymbol c)})\right\}\right),
\]
where \(\rho_1,\ldots,\rho_B\) are drawn independently and uniformly from \(\mathcal R\), independently of the data and of \(\mathcal H\).
\end{theorem}

\begin{proposition}[Exchangeable complete units suffice]
\label{prop:multi-relabel-route-app}
Write \((Q_\pi y)_i=y_{\pi(i)}\) for the panel with unit labels permuted. Suppose the complete untreated unit objects,
each unit's untreated outcomes over \(\Tpre\cup\Tpost\) and every covariate the procedure uses, are exchangeable across
all \(n\) units conditional on \(\mathcal H\), the tuning is fixed in advance, and the statistic is relabeling
equivariant, \(T(\rho;Q_\pi y)=T(\pi\circ\rho;y)\) for all \(\pi\), \(\rho\) and \(y\). Then
Assumption~\ref{ass:multi-relabel-app} holds.
\end{proposition}

\begin{proof}[Proof of Proposition~\ref{prop:multi-relabel-route-app}]
Let \(Z\) be the array of complete untreated unit objects, so that \(Y^{(\boldsymbol\tau)}\) and the covariates are
read from \(Z\). By equivariance, \(T(\pi\circ\rho;Z)=T(\rho;Q_\pi Z)\) for every \(\rho\), and
\(Q_\pi Z\mid\mathcal H\stackrel d=Z\mid\mathcal H\) by exchangeability. Hence the vector
\((T(\pi\circ\rho;Z))_\rho\) has the same conditional law as \((T(\rho;Z))_\rho\).
\end{proof}

Equivariance holds when the procedure uses only units' data and treatment patterns and does not depend on their storage order, as with fixed-rank GSC, DID and synthetic control. Assumption~\ref{ass:multi-relabel-app} is weaker than exchangeability of complete units. Proposition~\ref{prop:multi-relabel-route-app} applies the complete-unit sufficient condition of Proposition~\ref{prop:complete-unit-route-app} to the whole panel. With a DID statistic and exchangeable errors, the assumption also holds for treated units that differ from the others only by a level shift.

The test includes Theorem~\ref{thm:fc-app} as a special case. Allow \(K=1\), with \(\mathcal K=\{0\}\), and with a target-date estimator take \(T(\rho;y)\) to be the absolute singleton prediction error of unit \(\rho(0)\) at \(t\). Then \(T(\rho;Y^{(c)})=S_{\rho(0)}(c)\), \(p^{\mathrm{rel}}\) equals \(p_t^{\mathrm{FC}}\), and Assumption~\ref{ass:multi-relabel-app} is Assumption~\ref{ass:fc-score-exchangeability-app}. Unlike Fisher's randomization test, which averages over a random assignment, Theorem~\ref{thm:multi-relabel-app} conditions on which units are treated. Section~\ref{sec:multi} discusses the number of relabelings and full-path estimators, and Algorithm~\ref{alg:relabeling-app} in Appendix~\ref{app:algorithms} gives the Monte Carlo version.

Inverting the test over a declared family of arrays gives a confidence set \(\{\boldsymbol c:p^{\mathrm{rel}}(\boldsymbol c)>\gamma\}\) whose coverage is at least \(1-\lfloor\gamma|\mathcal R|\rfloor/|\mathcal R|\) when the true array belongs to the family. Over the constant effects it is a set for a common effect, and its coverage moves in the smallest steps of the constructions in this section. When the effects differ across units, no array of the family is true, and the set carries no guarantee for the average. The test of \(\boldsymbol c=\mathbf 0\) remains exact.

\begin{proof}[Proof of Theorem~\ref{thm:multi-relabel-app}]
Write \(V_\rho=T(\rho;Y^{(\boldsymbol\tau)})\) and, for an assignment \(\sigma\),
\(p_\sigma=|\mathcal R|^{-1}\sum_{\rho\in\mathcal R}\one\{V_\rho\geq V_\sigma\}\), so that
\(p^{\mathrm{rel}}(\boldsymbol\tau)=p_\iota\). Fix a permutation \(\pi\) of \(\Iset\). The map
\(\rho\mapsto\pi\circ\rho\) is a bijection of \(\mathcal R\), so applying \(p_\iota\) to the relabeled vector
\((V_{\pi\circ\rho})_\rho\) gives \(|\mathcal R|^{-1}\sum_{\rho}\one\{V_{\pi\circ\rho}\geq V_{\pi\circ\iota}\}=p_{\pi|_{\mathcal K}}\),
where \(\pi|_{\mathcal K}\in\mathcal R\) is the restriction of \(\pi\) to \(\mathcal K\). By
Assumption~\ref{ass:multi-relabel-app}, \(\Pr\{p_\iota\leq\gamma\mid\mathcal H\}=\Pr\{p_{\pi|_{\mathcal K}}\leq\gamma\mid\mathcal H\}\)
for every \(\pi\). Average over all \(n!\) permutations. Each \(\sigma\in\mathcal R\) is the restriction of exactly
\((n-K)!\) of them, so
\[
  \Pr\{p_\iota\leq\gamma\mid\mathcal H\}
  =\operatorname E\!\left[\frac{1}{|\mathcal R|}\sum_{\sigma\in\mathcal R}\one\{p_\sigma\leq\gamma\}\;\Big|\;\mathcal H\right].
\]
For every realization, at most \(\lfloor\gamma|\mathcal R|\rfloor\) assignments satisfy \(p_\sigma\leq\gamma\): if
\(m\) assignments did, the one among them with the smallest value \(V_\sigma\) would have
\(p_\sigma\geq m/|\mathcal R|\), so \(m\leq\gamma|\mathcal R|\). This gives the first bound.

For the Monte Carlo version, let \(\pi\) be a uniformly drawn permutation, independent of the data and of
\(\rho_1,\ldots,\rho_B\). By Assumption~\ref{ass:multi-relabel-app} and the independence of the draws,
\((V_\iota,V_{\rho_1},\ldots,V_{\rho_B})\) has the same conditional law as
\((V_{\pi\circ\iota},V_{\pi\circ\rho_1},\ldots,V_{\pi\circ\rho_B})\). The assignments
\(\pi|_{\mathcal K},\pi\circ\rho_1,\ldots,\pi\circ\rho_B\) are independent and uniform on \(\mathcal R\): given \(\pi\),
each \(\pi\circ\rho_b\) is uniform because \(\rho\mapsto\pi\circ\rho\) is a bijection, and \(\pi|_{\mathcal K}\) is
uniform. They are also independent of the data, so the \(B+1\) statistics are exchangeable given \(\mathcal H\), and
the counting argument above applies with \(B+1\) in place of \(|\mathcal R|\).
\end{proof}

\subsection{Staggered adoption}
\label{app:staggered}

Let adopter \(k\in\mathcal K\) be treated from its own date \(A_k\), so that a cell \((k,s)\) is treated when \(s\geq A_k\), and let \(\Cset\) be the never-treated units, with \(J=|\Cset|\) and \(n=K+J\). The panel covers dates \(1,\ldots,T\). We condition throughout on the adoption dates. We maintain no anticipation and no spillovers for every unit, so a cell \((i,s)\) is treated only if \(i\in\mathcal K\) and \(s\geq A_i\). The three questions of Section~\ref{sec:multi} carry over, and Section~\ref{sec:staggered} describes the constructions.

For the average effect, the target is the average \(r\) periods after adoption,
\[
  \bar\tau_r=\frac1{K_r}\sum_{k\in\mathcal K_r}\tau_{k,A_k+r},
  \qquad
  \mathcal K_r=\{k\in\mathcal K:A_k+r\leq T\},
\]
with \(r=0\) the first treated date, \(K_r=|\mathcal K_r|\), and the individual effects unrestricted. Table~\ref{tab:edr-block} and Figure~\ref{fig:apps-edr-att} count elections since adoption from 1, so their election \(e\) is relative period \(r=e-1\).

\paragraph{One treated cell at a time.}
Fix a treated cell \((k,t)\) and let \(\bar s_{kt}\) be the last date its fit uses, \(t\) for a target-date estimator and \(T\) for a full-path one. The pool is
\begin{equation}
  \mathcal P_{kt}=\Cset\cup\{j\in\mathcal K\setminus\{k\}:A_j>\bar s_{kt}\},
  \label{eq:staggered-pool-app}
\end{equation}
the never-treated units together with the adopters not yet treated by \(\bar s_{kt}\). The reduced panel contains unit \(k\) and the pool, observed over the dates the fit uses, with unit \(k\) in the role of unit \(0\) and \(\Tpre=\{s:s<A_k\}\).

\begin{proposition}[One treated cell under staggered adoption]
\label{prop:staggered-cell-app}
Proposition~\ref{prop:multi-cell-app} holds for this reduced panel, with the adoption dates in place of the adoption date and \(N=|\mathcal P_{kt}|+1\).
\end{proposition}

\begin{proof}[Proof of Proposition~\ref{prop:staggered-cell-app}]
A unit \(j\in\mathcal P_{kt}\) is either never treated or first treated after \(\bar s_{kt}\), the last date the
procedure uses, so by no anticipation and no spillovers every outcome of \(j\) in the reduced panel is untreated. Unit \(k\) is
untreated on \(\{s:s<A_k\}\) for the same reason. The pool depends only on the adoption dates, which are fixed by conditioning. The proof of
Proposition~\ref{prop:multi-cell-app} then applies with \(\Tpre=\{s:s<A_k\}\) and \(\mathcal P_{kt}\) in place
of \(\Cset\).
\end{proof}

Corollary~\ref{cor:multi-simultaneous-app} combines such cells, with \(N_m=|\mathcal P_{k_mt_m}|+1\) and \(\mathcal A\) the adoption dates.

\paragraph{The average effect at a relative period.}
Partition \(\Cset\) into \(G=\lfloor J/K\rfloor\) control blocks \(B_1,\ldots,B_G\) of size \(K\) and match the
units of each block one-to-one with the adopters, both at random, independently of the outcomes, as in
Appendix~\ref{app:multi}. The \(K\) adopters form the
treated block, and leftover never-treated units are training units in every fit. The unit of \(B_g\) matched
with adopter \(k\), written \(\rho_g(k)\), takes the date \(A_k\) and faces the same prediction task.
We apply the construction of Appendix~\ref{app:multi} to these blocks.

The folds are stated for a full-path estimator. Fold \(g\) hides both blocks from their actual or assigned dates on. We fit the estimator once using the two blocks' cells before
those dates and the complete panels of never-treated units outside \(B_g\), then predict the hidden cells.
Figure~\ref{fig:staggered-masking} shows one fold at two relative periods. Write \(\widehat Y_{is}^{[g]}\)
for fold \(g\)'s prediction of unit \(i\)'s untreated outcome at date \(s\) and record
\begin{equation}
  M_{g,r}=\frac1{K_r}\sum_{k\in\mathcal K_r}\widehat Y_{k,A_k+r}^{[g]},
  \qquad
  R_{g,r}=\Bigl|\frac1{K_r}\sum_{k\in\mathcal K_r}\bigl(Y_{\rho_g(k),A_k+r}-\widehat Y_{\rho_g(k),A_k+r}^{[g]}\bigr)\Bigr|,
  \label{eq:staggered-fold-app}
\end{equation}
the treated block's average prediction and the absolute value of the control block's average error, both over the
cells at relative period \(r\). The interval for \(\bar\tau_r\) is built from the endpoints \(M_{g,r}\mp R_{g,r}\)
with \(N_{\mathrm B}=G+1\), as the interval of Appendix~\ref{app:multi} is built from the endpoints of
\eqref{eq:block-fold-app}. The same \(G\) fits serve every relative period. They serve a target-date estimator
as well, since it uses only the two blocks' cells before their dates and the other units' outcomes at the scored
date, none of which any fold hides. Algorithm~\ref{alg:staggered-block-jplus-app} in
Appendix~\ref{app:algorithms} gives the steps for a full-path estimator.

As in Appendix~\ref{app:multi}, no treated cell enters any fit, so the effects enter only through
\(\bar\tau_r\).

\begin{proposition}[Block Jackknife+ under staggered adoption]
\label{prop:staggered-block-app}
List each block's units in the order of the adopters they are matched with, so that the dates attach to positions. Let \(\mathcal H\) be a \(\sigma\)-field that contains the adoption dates, the partition and the matching. Under the conditions of Proposition~\ref{prop:block-jplus-app} for the blocks so listed, conditional on \(\mathcal H\), the interval for \(\bar\tau_r\) covers with probability at least \(1-(2q_\gamma-1)_+/N_{\mathrm B}\), for each \(r\) with \(K_r\geq1\). The exchangeability holds when the complete untreated unit objects of the adopters and of all never-treated units are exchangeable conditional on \(\mathcal H\), the tuning is fixed before the outcomes are seen and contained in \(\mathcal H\), and the fold rule is equivariant in the sense of Proposition~\ref{prop:block-jplus-app}.
\end{proposition}

\begin{proof}[Proof of Proposition~\ref{prop:staggered-block-app}]
Fix \(r\). The fold rule hides every block from the dates attached to its positions onward and fits on the rest, so it applies the same rule to every block. Then \(M_{g,r}\) and \(R_{g,r}\) are the quantities of Proposition~\ref{prop:block-jplus-app} for the cells at relative period \(r\), and that proposition gives the bound. For the second statement, the argument in the proof of Proposition~\ref{prop:block-jplus-app} applies, with the adopters and all never-treated units in place of \(\mathcal K\cup\Cset\) and each block mapped onto its image in the order of the matching, so the dates move with the positions.
\end{proof}

At every relative period, the interval is a proper subset of the line only when \(N_{\mathrm B}\geq1/\gamma\), as
in Appendix~\ref{app:multi}, and the intervals are period-wise. Table~\ref{tab:edr-block} and Appendix~\ref{app:edr-multi} report them
for the turnout application.

\begin{remark}[One partition per relative period]
\label{rem:staggered-per-period-app}
The blocks can instead have size \(K_r\), with a partition for each relative period. The adopters that do not reach \(r\) are then training units in every fold, with their untreated cells, and the guarantee is conditional on their data as well. Later relative periods, which fewer adopters reach, get more blocks, and hence finer steps in the coverage levels that the interval can guarantee. The cost is separate fits for each period and blocks that are not nested across periods.
\end{remark}

\paragraph{All treated cells at once.}
Under a role assignment \(\rho\), unit \(\rho(k)\) is treated from \(A_k\), and the units outside
\(\rho(\mathcal K)\) serve as controls. The sharp null \eqref{eq:multi-sharp-null-app} runs over the treated cells
\(\{(k,s):s\geq A_k\}\), and the completed panel and the relabeling statistic are as before.
Assumption~\ref{ass:multi-relabel-app}, Theorem~\ref{thm:multi-relabel-app} and
Proposition~\ref{prop:multi-relabel-route-app} apply as stated, with \(\mathcal H\) containing the adoption dates
and the unit objects taken over every date. Two assignments that treat the same units can now differ by adoption date.
The \(n!/(n-K)!\) assignments therefore give as many distinct treatment patterns, except when adopters share a date.
In practice, we use the Monte Carlo version. Section~\ref{sec:empirics} reports the test for the turnout application, and Appendix~\ref{app:edr-multi} counts its assignments.

Inverting the test over the common effect paths, \(c_{ks}=c_{s-A_k}\), gives a set for the common path. Its coverage and its limitations are those of the set for a common effect in Appendix~\ref{app:multi}.

\section{Algorithms}
\label{app:algorithms}

In this section, we give pseudocode for three procedures, in the order in which the main text introduces them:
full conformal inference for target-date estimators, the Jackknife+ for target-date and full-path estimators, and
the pre-treatment diagnostics. Appendix~\ref{app:alg-multi} gives the block Jackknife+ and the relabeling test of
Appendices~\ref{app:multi} and~\ref{app:staggered}. Appendix~\ref{app:overlap-def} describes the overlap check, and
Appendix~\ref{app:band-computation} computes the full conformal band of Appendix~\ref{app:bands}.

\subsection{Full conformal inference, target-date estimators}

\begin{algorithm}[H]
\caption{Target-date full conformal $p$-value at a candidate effect}
\label{alg:pointwise-fc-app}
\begin{algorithmic}[1]
\Require Candidate \(c\); target-date prediction rule with one held-out unit; observed pre-treatment
         panel and target-date outcomes
\Ensure \(p_t^{\mathrm{FC}}(c)\)
\State Set \(Y_{0t}^{(c)}\gets Y_{0t}-c\) and
       \(Y_{it}^{(c)}\gets Y_{it}\) for \(i\in\Cset\)
\ForAll{\(i\in\Iset\)}
  \State Hide \(Y_{it}^{(c)}\) and compute
         \(\widehat M_i(c)\) with the fixed tuning values
  \State Set \(S_i(c)\gets|Y_{it}^{(c)}-\widehat M_i(c)|\)
\EndFor
\State \Return
\(N^{-1}\sum_i\one\{S_i(c)\geq S_0(c)\}\)
\end{algorithmic}
\end{algorithm}

The confidence set retains every candidate for which Algorithm~\ref{alg:pointwise-fc-app} returns a value above
\(\gamma\). A direct inversion over \(K\) candidates requires up to \(NK\) fits with one held-out unit. The
prediction for the treated unit does not depend on the candidate, so one fit serves all \(K\) candidates. The
prediction for a held-out control, by contrast, generally depends on the treated unit's completed outcome, which
changes with the candidate. Reusing these predictions across candidates requires additional structure in the estimator. For DID and synthetic control, a held-out control's prediction is affine in the candidate, so we compute the sets for these two estimators exactly. For other estimators, such as SDID, a grid of candidates gives an approximation that does not carry the guarantee of Theorem~\ref{thm:fc-app}. No result here uses a grid.

\subsection{Jackknife+, target-date and full-path estimators}
\label{app:alg-jplus-fp}

\begin{algorithm}[H]
\caption{Jackknife+ interval, target-date or full-path estimator}
\label{alg:pointwise-jplus-app}\label{alg:pooled-jplus-app}
\begin{algorithmic}[1]
\Require Target-date or full-path prediction rule with two held-out units; observed pre-treatment panel; control
         outcomes at \(t\) (target-date) or over \(\Tpost\) (full-path); \(Y_{0t}\); total miscoverage \(\gamma\)
\Ensure Prediction and effect intervals
\State Compute \(q\gets q_\gamma\) from
       \eqref{eq:generic-jplus-calibration-app}
\If{\(q=0\)}
  \State \Return \(\R\) for both intervals
\EndIf
\State Set \(u\gets N-q\)
\ForAll{\(j\in\Cset\)}
  \If{the rule is a target-date rule}
    \State Hide \(Y_{0t}\) and \(Y_{jt}\); fit the prediction rule with two held-out units using
           \(Y_{\Iset,\Tpre}\) and \(Y_{\Cset\setminus\{j\},t}\); record \(\widehat M_{0j}\) and \(\widehat M_{j0}\)
  \Else
    \State Fit the declared full-path prediction rule to held-out unit \(0\) using
           \(Y_{0,\Tpre}\) and \(Y_{\Cset\setminus\{j\},\Tpre\cup\Tpost}\); record \(\widehat M_{0j}\)
    \State Fit the declared full-path prediction rule to held-out unit \(j\) using
           \(Y_{j,\Tpre}\) and \(Y_{\Cset\setminus\{j\},\Tpre\cup\Tpost}\); record \(\widehat M_{j0}\)
  \EndIf
  \State Set \(R_{j0}\gets|Y_{jt}-\widehat M_{j0}|\)
  \State Set \(a_j\gets\widehat M_{0j}-R_{j0}\) and
         \(b_j\gets\widehat M_{0j}+R_{j0}\)
\EndFor
\State Sort the lower and upper endpoints separately
\State Set \(C^Y\gets[a_{(q)},b_{(u)}]\)
\State \Return \(C^Y\) and
       \([Y_{0t}-b_{(u)},Y_{0t}-a_{(q)}]\)
\end{algorithmic}
\end{algorithm}

The Jackknife+ can handle units unobserved at the target date or ineligible in two ways. The first option excludes them before fitting. The second excludes their scores from the interval while keeping them as training units in every fit, for example to estimate factors.
Under either option, \(N\) and \(q_\gamma\) are computed from the retained pool.
Theorem~\ref{thm:generic-jplus-app} requires the all-pairs table of this pool to satisfy
Assumption~\ref{ass:jplus-matrix-exchangeability-app} conditional on the retained pool and the stated information.
The second option also requires conditioning on the training units' data shared by every fit.
\code{fect} uses the second option for a control that is missing at some pre-treatment date of the treated unit's window
or at the target date. In the three applications, every control is observed at every date and no ineligible unit
enters a fit, so the two options coincide.

\subsection{Pre-treatment diagnostics}

\begin{algorithm}[H]
\caption{Rolling pre-treatment diagnostics}
\label{alg:historical-diagnostics-app}
\begin{algorithmic}[1]
\Require Ordered panel; diagnostic dates \(\mathcal S_{\mathrm{pl}}\);
         target-date or full-path estimator and declared prediction rules;
         total miscoverage \(\gamma\)
\Ensure Period-wise pseudo-effect estimates, confidence sets, and
        $p$-values
\ForAll{\(s\in\mathcal S_{\mathrm{pl}}\), in chronological order}
  \State Form \(\mathcal H_s\) and \(\mathcal B_s\) from \eqref{eq:placebo-history-app}, and fix
         \(\widehat\eta_s\) as in Appendix~\ref{app:historical-diagnostics}
  \State In the Jackknife+ steps below, compute the \(J\) fits with two held-out units even when
         \(q_\gamma=0\), so that the interval-dual $p$-value is available
  \If{the estimator is a target-date estimator}
    \State Predict \(Y_{0s}(0)\) using \(Y_{0,\mathcal H_s}\) and
           \(Y_{\Cset,\mathcal H_s\cup\{s\}}\); record
           \(\widehat\tau_s^{\mathrm{pl}}\)
    \State Apply Algorithm~\ref{alg:pointwise-fc-app} with
           \(t\gets s\) and \(\Tpre\gets\mathcal H_s\); record
           \(p_s^{\mathrm{pl,FC}}(0)\) and
           \(\widehat C_s^{\mathrm{pl,FC}}(\gamma)\)
    \State Apply Algorithm~\ref{alg:pointwise-jplus-app} to the same
           slice; record the Jackknife+ interval and compute its
           interval-dual $p$-value from
           \eqref{eq:jplus-pvalue-app}
  \Else
    \State Predict \(Y_{0s}(0)\) using \(Y_{0,\mathcal H_s}\) and
           \(Y_{\Cset,\Tpre\cup\Tpost}\); record
           \(\widehat\tau_s^{\mathrm{pl}}\)
    \State Apply Algorithm~\ref{alg:pooled-jplus-app} with held-out units' histories
           \(\mathcal H_s\) and held-out units' hidden blocks \(\mathcal B_s\);
           retain complete panels for the training controls
    \State Record the target-\(s\) interval and its interval-dual $p$-value
  \EndIf
\EndFor
\State \Return all period-wise outputs without a multiplicity adjustment
\end{algorithmic}
\end{algorithm}

Implementations should compute \(q_\gamma\) and the interval-dual $p$-value with one documented numerical
convention for comparisons, so that the interval and the $p$-value agree exactly at the cutoffs.

\subsection{Multiple treated units and staggered adoption}
\label{app:alg-multi}

Appendices~\ref{app:multi} and~\ref{app:staggered} add two constructions. Algorithm~\ref{alg:block-jplus-app} is the block Jackknife+ for the average effect over the treated units with a common adoption date, and Algorithm~\ref{alg:staggered-block-jplus-app} gives its version by relative period under staggered adoption. Algorithm~\ref{alg:relabeling-app} is the relabeling test in its Monte Carlo form, which covers both cases. The algorithms inherit the conditions of Propositions~\ref{prop:block-jplus-app} and~\ref{prop:staggered-block-app} and Theorem~\ref{thm:multi-relabel-app}.

\code{fect} forms the control blocks from the never-treated controls observed at every date and keeps the other controls as training units, so \(J\) in \(G=\lfloor J/K\rfloor\) counts those complete controls. For a fold whose fit fails, we use the endpoints \(-\infty\) and \(+\infty\), as in Proposition~\ref{prop:failed-folds-app}.

\begin{algorithm}[H]
\caption{Block Jackknife+ interval for the average effect, common adoption date}
\label{alg:block-jplus-app}
\begin{algorithmic}[1]
\Require Treated block \(\mathcal K\) of \(K\) units; partition of \(\Cset\) into \(G=\lfloor J/K\rfloor\) blocks
         \(B_1,\ldots,B_G\) of size \(K\), the leftover controls kept as training units in every fit; declared target-date or
         full-path estimator; target date \(t\); total miscoverage \(\gamma\)
\Ensure Prediction interval for the treated block's average untreated outcome and effect interval for the
        average effect \(\bar\tau_t\) over \(\mathcal K\) at \(t\)
\State Set \(N_{\mathrm B}\gets G+1\) and compute \(q\gets q_\gamma\) from
       \eqref{eq:generic-jplus-calibration-app} with \(N_{\mathrm B}\) in place of \(N\)
\If{\(q=0\)}
  \State \Return \(\R\) for both intervals
\EndIf
\State Set \(u\gets N_{\mathrm B}-q\)
\ForAll{\(g\in\{1,\ldots,G\}\)}
  \State Hide the outcomes of the units in \(\mathcal K\cup B_g\) at \(t\) under a target-date estimator, or
         over \(\Tpost\) under a full-path estimator; fit the declared estimator once, on \(Y_{\Iset,\Tpre}\) and
         \(Y_{\Cset\setminus B_g,t}\) under a target-date estimator or on \(Y_{\mathcal K\cup B_g,\Tpre}\) and
         \(Y_{\Cset\setminus B_g,\Tpre\cup\Tpost}\) under a full-path one, and predict the hidden units
  \State Record \(\widehat Y_{it}^{[\mathcal K\cup B_g]}\) for \(i\in\mathcal K\cup B_g\)
  \State Set \(M_g\gets K^{-1}\sum_{k\in\mathcal K}\widehat Y_{kt}^{[\mathcal K\cup B_g]}\) and
         \(R_g\gets\bigl|K^{-1}\sum_{i\in B_g}(Y_{it}-\widehat Y_{it}^{[\mathcal K\cup B_g]})\bigr|\)
  \State Set \(a_g\gets M_g-R_g\) and \(b_g\gets M_g+R_g\)
\EndFor
\State Sort the lower and upper endpoints separately
\State Set \(C^Y\gets[a_{(q)},b_{(u)}]\)
\State \Return \(C^Y\) and \([\bar Y_{\mathcal Kt}-b_{(u)},\bar Y_{\mathcal Kt}-a_{(q)}]\), where
       \(\bar Y_{\mathcal Kt}=K^{-1}\sum_{k\in\mathcal K}Y_{kt}\)
\end{algorithmic}
\end{algorithm}

\begin{algorithm}[H]
\caption{Block Jackknife+ intervals by relative period under staggered adoption}
\label{alg:staggered-block-jplus-app}
\begin{algorithmic}[1]
\Require Adopters \(\mathcal K\) with dates \(A_k\); never-treated units \(\Cset\); partition of \(\Cset\) into
         \(G=\lfloor J/K\rfloor\) blocks of size \(K\) and, for each block \(B_g\), a matching
         \(\rho_g:\mathcal K\to B_g\); declared full-path estimator; relative periods \(r\geq0\), with
         \(\mathcal K_r\) the adopters whose date \(A_k+r\) lies in the panel and \(K_r=|\mathcal K_r|\geq1\);
         total miscoverage \(\gamma\)
\Ensure Effect interval for the average effect \(\bar\tau_r\) over \(\mathcal K_r\) at every \(r\)
\State Set \(N_{\mathrm B}\gets G+1\) and compute \(q\gets q_\gamma\) from
       \eqref{eq:generic-jplus-calibration-app} with \(N_{\mathrm B}\) in place of \(N\)
\If{\(q=0\)}
  \State \Return \(\R\) for every \(r\)
\EndIf
\State Set \(u\gets N_{\mathrm B}-q\)
\ForAll{\(g\in\{1,\ldots,G\}\)}
  \State Hide \(Y_{ks}\) and \(Y_{\rho_g(k)s}\) for every \(k\in\mathcal K\) and \(s\geq A_k\); fit the
         estimator once on the remaining cells; record the predictions \(\widehat Y_{is}^{[g]}\) of the hidden
         cells
  \ForAll{\(r\)}
    \State Set \(M_{g,r}\gets K_r^{-1}\sum_{k\in\mathcal K_r}\widehat Y_{k,A_k+r}^{[g]}\) and
           \(R_{g,r}\gets\bigl|K_r^{-1}\sum_{k\in\mathcal K_r}
             (Y_{\rho_g(k),A_k+r}-\widehat Y_{\rho_g(k),A_k+r}^{[g]})\bigr|\)
    \State Set \(a_{g,r}\gets M_{g,r}-R_{g,r}\) and \(b_{g,r}\gets M_{g,r}+R_{g,r}\)
  \EndFor
\EndFor
\ForAll{\(r\)}
  \State Sort the lower and upper endpoints over \(g\) separately and set
         \(\bar Y_r\gets K_r^{-1}\sum_{k\in\mathcal K_r}Y_{k,A_k+r}\)
  \State Record \([\bar Y_r-b_{(u),r},\bar Y_r-a_{(q),r}]\)
\EndFor
\State \Return the intervals for every \(r\)
\end{algorithmic}
\end{algorithm}

The variant of Remark~\ref{rem:staggered-per-period-app} runs the algorithm once for each \(r\), with
\(\mathcal K_r\) as the treated block and \(G_r=\lfloor J/K_r\rfloor\) control blocks, and keeps the other adopters
as training units with their untreated cells.

\begin{algorithm}[H]
\caption{Relabeling test, Monte Carlo $p$-value at a declared array}
\label{alg:relabeling-app}
\begin{algorithmic}[1]
\Require Declared array \(\boldsymbol c\); treated units \(\mathcal K\) with adoption dates \(A_k\), one common
         date in Appendix~\ref{app:multi}; relabeling statistic \(T(\rho;y)\) of Appendix~\ref{app:multi}; number
         of draws \(B\)
\Ensure \(p^{\mathrm{rel}}_B(\boldsymbol c)\)
\State Form the completed panel \(Y^{(\boldsymbol c)}\)
\State Compute \(T_\iota\gets T(\iota;Y^{(\boldsymbol c)})\) under the observed treatment pattern
\ForAll{\(b\in\{1,\ldots,B\}\)}
  \State Draw \(\rho_b\) uniformly from \(\mathcal R\), independently of the data, of the other draws and of the conditioning information in Theorem~\ref{thm:multi-relabel-app}; give unit \(\rho_b(k)\) the date \(A_k\) for every \(k\) and the
         other units the role of controls
  \State Compute \(T_b\gets T(\rho_b;Y^{(\boldsymbol c)})\) under this pattern
\EndFor
\State \Return \((B+1)^{-1}\bigl(1+\sum_{b=1}^B\one\{T_b\geq T_\iota\}\bigr)\)
\end{algorithmic}
\end{algorithm}

The test of a sharp null rejects when the returned value is at most \(\gamma\), and inverting it as in
Appendix~\ref{app:multi} gives a confidence set. When the statistic needs one fit per assignment, as with a full-path estimator, the algorithm uses \(B+1\) fits.

\subsection{The overlap check}
\label{app:overlap-def}

The overlap check assesses whether a treated unit's pre-treatment path lies within the
range of the controls' paths. It requires neither a rank nor a factor model and uses no post-treatment outcome. Like the procedures of
Section~\ref{sec:ordinary-estimator}, it ranks the treated unit among the controls.

For a treated unit \(i\) with pre-treatment dates \(\mathcal P_i\), we use the controls observed at all
these dates. We subtract each path's mean over \(\mathcal P_i\) to remove the level differences absorbed by unit fixed
effects, and denote the demeaned paths by \(\tilde y_i\) and \(\tilde y_j\). The check therefore cannot detect differences in levels. Outcomes-only synthetic control does not absorb these differences. The treated unit's fit error is
\[
  e_i=\min_{w\in\Delta}\Big\{|\mathcal P_i|^{-1}\sum_{t\in\mathcal P_i}
  \big(\tilde y_{it}-\textstyle\sum_jw_j\tilde y_{jt}\big)^2\Big\}^{1/2},
  \qquad \Delta=\Big\{w:w_j\geq0,\ \textstyle\sum_jw_j=1\Big\}.
\]
If the treated unit's loadings lay in the convex hull of the controls' loadings and the outcomes had no
idiosyncratic errors, \(e_i\) would be zero. We fit each control in the same way using the other controls
and flag the treated unit when \(e_i\) exceeds the 90th percentile of the controls' errors. The check has no formal size guarantee.

If there are more than 200 controls, we fit a random sample of 200. The check reports a unit as “not assessed,” rather than passing, if it has fewer than four pre-treatment dates or five controls, or if controls predict one another almost perfectly.\footnote{Their median error is below \(10^{-8}\) times the outcome's standard deviation.} We check the observed outcome, optionally subtracting covariates' contribution using coefficients from the main fit. This adjustment makes the check depend on rank because the coefficients do. The check screens for units whose predictions rely heavily on extrapolation. Like any pre-treatment check, it cannot detect factors that are constant before treatment. It supports, but does not establish, the exchangeability needed for our guarantees.

\subsection{Computing the full conformal band}
\label{app:band-computation}

We compute the full conformal band of Appendix~\ref{app:bands} for DID and simplex synthetic
control, with the weights \(w_i\), the path scores \(S_i(\boldsymbol c)\) and \(H=|\Tpost|\) defined there. For
these estimators, unit \(0\)'s completed outcome enters control \(i\)'s date-\(s\) prediction with a coefficient
\(\kappa_i\) that depends on neither \(s\) nor \(\boldsymbol c\). The coefficient is \(\kappa_i=1/J\) for DID and, for synthetic control, unit \(0\)'s weight in control \(i\)'s synthetic control. The equal and studentized weights of
Appendix~\ref{app:bands} use only the pre-treatment fit, so \(w_i\) does not depend on
\(\boldsymbol c\).

Let \(v_{is}\) be control \(i\)'s date-\(s\) error at \(\boldsymbol c=\mathbf 0\). Its error at \(\boldsymbol c\)
is then \(v_{is}+\kappa_ic_s\), and \(S_0(\boldsymbol c)=\max_s w_0|\widehat\tau_s-c_s|\). With
\(k=\lfloor\gamma N\rfloor\), the region is
\(\{\boldsymbol c:\#\{i\in\Cset:S_i(\boldsymbol c)\geq S_0(\boldsymbol c)\}\geq k\}\) (all of \(\R^H\) if
\(k=0\)). If every \(\kappa_i=0\), it is the box \(|c_s-\widehat\tau_s|\leq S_{(k)}/w_0\), where \(S_{(k)}\) is
the \(k\)th largest control score at \(\mathbf 0\).

In general, fix \(t\) and \(c_t\), and let
\(L=w_0|\widehat\tau_t-c_t|\). For \(s\neq t\), let \(g_{is}\) be the largest value of
\(w_i|v_{is}+\kappa_ix|\) over values of \(x\) satisfying \(w_i|v_{is}+\kappa_ix|\geq w_0|\widehat\tau_s-x|\) (an interval
when \(w_i|\kappa_i|<w_0\), and \(g_{is}=\infty\) otherwise). Then the projection onto date \(t\) contains the inner set of
\(c_t\) at which at least \(k\) controls satisfy \(w_i|v_{it}+\kappa_ic_t|\geq L\) or
\(\max_{s\neq t}w_i|v_{is}+\kappa_i\widehat\tau_s|\geq L\). The projection is contained in the outer set at which at least \(k\)
controls satisfy \(w_i|v_{it}+\kappa_ic_t|\geq L\) or \(\max_{s\neq t}g_{is}\geq L\). For the inner set, we fix the
other coordinates at \(\widehat\tau_{-t}\), where \(S_0=L\). The outer set follows because a control with
\(S_i\geq S_0\) has some date \(s\) at which \(w_i|v_{is}+\kappa_ic_s|\geq S_0\geq\max(L,w_0|\widehat\tau_s-c_s|)\).

Suppose first that \(w_i|\kappa_i|<w_0\) for every control. This holds with equal weights for DID when \(N>2\), and for synthetic control unless a control's synthetic control puts all its weight on unit \(0\). Each control's set of
qualifying \(c_t\) is then an interval containing \(\widehat\tau_t\), so each bounding set consists of points covered by at
least \(k\) intervals. A control with \(w_i|\kappa_i|\geq w_0\) instead qualifies at date \(t\) on a ray, two rays
or the whole line. When \(H\geq2\), such a control also meets the condition of the
outer set at every \(c_t\), since \(g_{is}=\infty\) at every other date. The same count applies to these rays and lines. We report the outer set,
which is a valid band.

\section{Monte Carlo simulation results}
\label{app:sims}

We report the Monte Carlo evidence behind Section~\ref{sec:sims}. In each panel, one unit, unit $0$, is treated after date $T_0$. We ask how often each procedure covers this unit's effect, first when the units are exchangeable and then when the treated unit lies outside the controls' range. Appendix~\ref{app:sims-designs} describes the data-generating process, the estimators and the grid of designs. Appendix~\ref{app:sims-results} reports coverage (Tables~\ref{tab:sims-cov-jplus} and~\ref{tab:sims-cov-fc}, Figure~\ref{fig:sims-coverage}), width (Table~\ref{tab:sims-width}) and the pre-treatment diagnostics (Table~\ref{tab:sims-placebo}). Appendix~\ref{app:sims-position} relates coverage to the treated unit's signal (Tables~\ref{tab:sims-position} and~\ref{tab:sims-position-stud}). Appendix~\ref{app:sims-bands} checks the simultaneous bands of Appendix~\ref{app:bands}. Appendix~\ref{app:sims-comparators} compares the unit-level procedures with a time-based test and the parametric bootstrap (Table~\ref{tab:sims-comparators}).

\subsection{Designs}
\label{app:sims-designs}
Each replication draws a panel of $N$ units over $T_0+T_1$ dates, with $T_0=\SimTpre$ and $T_1=\SimTpost$. Untreated outcomes follow $Y_{it}(0)=\alpha_i+\delta_t+\lambda_i^\top f_t+e_{it}$, and unit $0$ receives a constant effect of \SimEffect{} at every post-treatment date. The unit effects $\alpha_i$ are normal with standard deviation \SimAlphaSd. The time effect $\delta_t$ is a linear trend with slope \SimTrend{} plus a seasonal cycle with amplitude \SimSeasonAmp{} and period \SimSeasonPeriod. The \SimNfactors{} factors are a stationary AR(1) process with coefficient \SimFacPhi{} and innovation standard deviation \SimFacSd, a random walk with innovation standard deviation \SimRwSd, and a random walk with drift \SimDrift{} and innovation standard deviation \SimDriftSd. We refer to the third as the drifting factor.

Loadings are independent and uniform on $\SimLoadRange$. Errors are independent across units. Each unit's error follows a stationary AR(1) process with coefficient \SimErrPhi{} and normal innovations with standard deviation \SimErrSd. Its stationary standard deviation is \SimErrSdStat.

The \SimNdesigns{} designs differ only in the treated unit's loading on the drifting factor, which determines how strongly the unit moves with this factor. Within a replication, they share the factors, the errors and the controls' loadings. In the exchangeable design, the treated unit's loading is drawn like every control's. The complete units are then exchangeable given the factor and time-effect paths, so the bounds of Table~\ref{tab:levels} hold for both procedures, even though the outcomes are serially dependent and two factors are nonstationary. In the other \SimNshift{} designs, the treated unit's loading equals the largest control loading on the drifting factor or exceeds it by \SimGapA, \SimGapB{} or \SimGapC. Its prediction errors need not be exchangeable with the controls', and no guarantee applies.

We apply \SimNest{} estimators. The three target-date estimators are DID (saturated event-date TWFE), simplex synthetic control with a ridge penalty of $\SimScEta$ on the weights, and target-specific SDID. The three full-path estimators are GSC with $r=\SimRlow$ factor (fewer than the data contain), GSC with $r=\SimRhigh$, and IFE-EM with $r=\SimRhigh$. We fix every tuning value in advance and reuse it in every fit. SDID's noise level is \SimSdidNoise, about the standard deviation of first differences implied by the design.

Full conformal inference applies to the target-date estimators. For DID and synthetic control, we compute the set exactly by affine inversion and report its width. For SDID, we test only the true effect, so we report coverage without width. The Jackknife+ applies to all \SimNest. When a fit fails, including a \code{fect} fit that reaches its iteration cap, we retain its fold with an infinite score (Proposition~\ref{prop:failed-folds-app}). In the main run, the share of replications with a failed fit was at most \SimFailMax{} in any cell.

The grid crosses the \SimNdesigns{} designs with $N\in\{\SimNa,\SimNb,\SimNc,\SimNd\}$ at $\gamma=\SimGammaMain$ and with $N=\SimNc$ at $\gamma=\SimGammaLow$, with \SimReps{} replications each. Each cell evaluates the \SimTpost{} post-treatment dates and \SimNplacebo{} diagnostic dates. At $\gamma=\SimGammaLow$ with $N=\SimNa$ or $N=\SimNb$, $\lfloor\gamma N\rfloor=0$ and $q_\gamma=0$. Both procedures then return the whole line, which covers with probability one, so we do not run these cells.

The first \SimPilotReps{} replications of SDID, GSC and IFE-EM come from a pilot with the same seeds and code. At $N=\SimNd$, we use GSC and IFE-EM results from an earlier run with the same design and seeds but the previous code version.

\subsection{Coverage and width}
\label{app:sims-results}
Tables~\ref{tab:sims-cov-jplus} and~\ref{tab:sims-cov-fc} show that, with exchangeable units, coverage in every cell is consistent with its guaranteed bound. In \SimExBelow{} of the \SimExCells{} cells, estimated coverage falls below the bound, in each case by no more than \SimExWorstZ{} standard errors.

The two procedures differ at $N=\SimNc$ and $N=\SimNd$ with $\gamma=\SimGammaMain$. At $N=\SimNc$, both bounds equal \SimBoundThirtyNine, but the Jackknife+ covers \SimJpThirtyNineMin--\SimJpThirtyNineMax{} while full conformal sets cover \SimFcThirtyNineMin--\SimFcThirtyNineMax. The same gap appears at $N=\SimNd$ (Tables~\ref{tab:sims-cov-jplus} and~\ref{tab:sims-cov-fc}). The Jackknife+ accounts for the worst case. When the fold predictions vary little, its interval uses about the \(1-\gamma/2\) quantile of the controls' absolute prediction errors, so its coverage approaches \(1-\gamma/2\).

Figure~\ref{fig:sims-coverage} shows coverage when the treated loading lies outside the controls' range, as summarized in Section~\ref{sec:sims}. With the treated loading at the controls' maximum, Jackknife+ coverage at $N=\SimNc$ already ranges from \SimGapZeroDid{} for DID to \SimGapZeroSdid{} for SDID. At a gap of \SimGapC, DID and synthetic control cover \SimGapTwoDidScMin--\SimGapTwoDidScMax{} across $N$. At $N=\SimNc$, GSC with $r=\SimRlow$ falls to \SimGapTwoGscOne{}, and SDID, which reweights both units and dates, falls to \SimGapTwoSdid. GSC and IFE-EM with $r=\SimRhigh$ stay closest to their bounds, at \SimGapTwoFactorMin--\SimGapTwoFactorMax{} across $N$, and have the highest coverage at $N=\SimNd$. Table~\ref{tab:sims-cov-fc} shows that full conformal coverage follows the same order.

Estimators that model the factors give narrower intervals (Table~\ref{tab:sims-width}). With exchangeable units, $N=\SimNc$ and $\gamma=\SimGammaMain$, the median Jackknife+ width is \SimWidthJpDid{} for DID, \SimWidthJpSc{} for synthetic control, \SimWidthJpSdid{} for SDID, \SimWidthJpGscOne{} and \SimWidthJpGscThree{} for GSC with $r=\SimRlow$ and $r=\SimRhigh$, and \SimWidthJpIfe{} for IFE-EM, against a true effect of \SimEffect. Full conformal sets for DID and synthetic control have median widths \SimWidthFcDid{} and \SimWidthFcSc. Section~\ref{sec:sims} reports how much narrower full conformal sets are at $N=\SimNc$ and $N=\SimNd$, where the Jackknife+ over-covers. In the other cells, each Jackknife+ endpoint comes from the most extreme fold. Full conformal sets are at least as wide; for DID, the two procedures coincide.

Some full conformal sets for synthetic control are unbounded on at least one side. Across the five designs at $\gamma=\SimGammaMain$, such sets occur in \SimOpenTen\% of the post-treatment cells at $N=\SimNa$, in \SimOpenSeventeen\% at $N=\SimNb$, and in none at $N=\SimNc$ or $N=\SimNd$. At $N=\SimNc$ with $\gamma=\SimGammaLow$, they occur in \SimOpenThirtyNineLow\%. In each of the \SimOpenChecked{} unbounded sets we inspected, one control's synthetic control fit puts all its weight on unit~$0$. Such a control's prediction moves with the candidate. On one side, its score therefore grows as fast as unit~$0$'s and keeps the $p$-value above $\gamma$. None of the \SimOpenBoundedChecked{} bounded sets we inspected has such a control. With more controls, an unbounded set would need several such fits at once. The tables include these sets, and Table~\ref{tab:sims-width} reports medians.

Table~\ref{tab:sims-placebo} shows that the pre-treatment diagnostics behave as the theory predicts. With exchangeable units, in \SimPlaceboExAboveAny{} of the \SimPlaceboExCells{} cells the rejection rate exceeds its guaranteed maximum, in each case by no more than \SimFlagSE{} standard errors.

With the treated loading \SimGapC{} above the controls' maximum, the diagnostics for DID and synthetic control reject at \SimPlaceboGapTwoMin--\SimPlaceboGapTwoMax{} of diagnostic dates. For these two estimators, the diagnostics reveal the failure before treatment. They reject less often for SDID and GSC with $r=\SimRlow$, although coverage for both still falls below the bound. For GSC and IFE-EM with $r=\SimRhigh$, whose coverage falls least, they reject at no more than \SimPlaceboGapTwoFactorMax{} of diagnostic dates.

{\renewcommand{\tablerowstretch}{1.15}\input{tables/tab_cov_jplus}}
\input{tables/tab_cov_fc}
\input{figures/fig_sims_coverage_float}
{\renewcommand{\tablerowstretch}{1.0}\input{tables/tab_width}}
\input{tables/tab_placebo}

\subsection{Coverage and the treated unit's signal}
\label{app:sims-position}

With exchangeable units, the guaranteed bounds hold on average over the units' signals. With the signals fixed, coverage can fall below the bound when the treated unit is harder to predict than the controls and exceed it when the controls are harder to predict. The approximation errors of Section~\ref{sec:error-score} can differ across units even when the predictions remove every signal exactly, because each unit's prediction puts its own weights on the other units' errors. Suppose the prediction for unit \(i\) puts weights \(w_{ij}\) on the other units and removes unit \(i\)'s signal exactly. Let the weights sum to one and not depend on the errors, and let the errors be independent across units with variance \(\sigma^2\). Unit \(i\)'s prediction error \(e_{it}-\sum_jw_{ij}e_{jt}\) then has variance \(\sigma^2(1+\sum_jw_{ij}^2)\). A unit whose prediction puts its weight on few other units is therefore harder to predict than a unit whose prediction puts equal weights on many.

Panel A of Table~\ref{tab:sims-position} splits the replications of the exchangeable design by whether the treated unit's signal lies within the range of the controls' signals. For the target-date estimators, full conformal coverage is above the bound when the signal lies within this range and below the bound when it does not. The difference is largest for DID and smallest for SDID. Jackknife+ coverage for the full-path estimators is above the bound in both cases, and for GSC with \(r=\SimRhigh\) and IFE-EM the two cases differ little.

For Panel B, we hold the signals fixed, draw new errors in each replication, and apply synthetic control. In the first three rows, every unit shares its signal with at least one other unit. Coverage rises with the number of controls that share the treated unit's signal. With one such control at \(N=\SimNc\), full conformal coverage is \PosOneFcThirtyNine, against a bound of \SimBoundThirtyNine{} with exchangeable units.

In the last three rows, the controls' signals are random draws, as in the exchangeable design, and the treated unit's signal lies within their range. Coverage exceeds the bound in every row and is highest at the controls' average. Most controls' signals lie outside the range of the other units' signals, which can make these controls harder to predict than the treated unit.

Table~\ref{tab:sims-position-stud} compares two scores, the absolute prediction error and the studentized score of Appendix~\ref{app:bands}, in the designs of Panel B of Table~\ref{tab:sims-position} and in three more. In these three, each of the four signals is shared by about a quarter of the units, and only the standard deviation of the treated unit's errors differs across rows. With twice or half the controls' standard deviation, the error paths are not exchangeable. Full conformal coverage with the absolute prediction error at $N=\SimNc$ is then \PosNoisyFcThirtyNine{} and \PosQuietFcThirtyNine. With the studentized score, it is \PosNoisyFcStudThirtyNine{} and \PosQuietFcStudThirtyNine, the second below the bound of \SimBoundThirtyNine. Where one control shares the treated unit's signal, the studentized score raises full conformal coverage from \PosOneFcThirtyNine{} to \PosOneFcStudThirtyNine. At the controls' average, it lowers coverage below the bound. Jackknife+ coverage with the studentized score is at or above the bound in every row.

\input{tables/tab_position}

\input{tables/tab_position_studentized}

\subsection{Simultaneous bands}
\label{app:sims-bands}

To check the bands of Appendix~\ref{app:bands} in finite samples, we simulated
\(Y_{it}(0)=\alpha_i+\xi_t+\lambda_if_t+\varepsilon_{it}\) with \(\alpha_i,\xi_t,\lambda_i\sim N(0,1)\), a trend
factor \(f_t=2t/T_0\), \(T_0=20\), and a zero effect path. The errors are \(N(0,1)\) and independent across units. We use three designs: errors independent over time, stationary AR(1) errors with coefficient \(0.8\) and variance 1, and a treated loading that exceeds the controls' largest by two. The grid was
\(N\in\{17,39\}\), \(H\in\{5,12\}\) and \(\gamma=0.10\), with 2,000 replications per cell. We computed the bands for DID and simplex synthetic control (full conformal and Jackknife+) and for GSC with \(r=1\) (full-path Jackknife+).

In the two exchangeable designs, the period-wise sets covered the whole path in \BandJointMin{} to \BandJointMax{}\% of
replications. Since \(\gamma/H<1/N\) in every cell, the Bonferroni band is the real line (Appendix~\ref{app:bands}).

In the same two designs, the path bands covered the path in \BandEqMin{} to \BandEqMax{}\% of replications
with equal weights and \BandStudMin{} to \BandStudMax{}\% with studentized weights, against bounds of
\BandBoundSeventeen{}\% (\(N=17\)) and \BandBoundThirtyNine{}\% (\(N=39\)). The largest shortfall below a bound was DID with the equal-weight Jackknife+ at \(N=17\) and \(H=5\) under independent errors, \BandShortCov{}\% against \BandShortBound\%, about two Monte Carlo standard errors. When the treated loading
lies outside the controls' range, no band carries a guarantee.

To compare widths within each cell, we average each band's width over dates and take the median over replications.
We divide this median by the corresponding median for the period-wise full conformal or Jackknife+ procedure with the default score. In the two
exchangeable designs, with equal weights the bands were \BandRatioEqFiveMin{} to \BandRatioEqFiveMax{} times as
wide as the period-wise sets at five dates and \BandRatioEqTwelveMin{} to \BandRatioEqTwelveMax{} times at twelve.
With studentized weights the ratios were \BandRatioStudFiveMin{} to \BandRatioStudFiveMax{} and
\BandRatioStudTwelveMin{} to \BandRatioStudTwelveMax.

\subsection{\texorpdfstring{Unit-level and time-based procedures}{Unit-level and time-based procedures}}
\label{app:sims-comparators}

The introduction contrasts unit-level inference with procedures that compare the treated unit's post-treatment residuals with its own pre-treatment residuals. To separate the choice of inference procedure from the choice of estimator, we hold the estimator fixed and vary only the factors and the errors. The signals are those of the exchangeable design, with two random-walk factors, or the same design with all three factors stationary AR(1) series (coefficient 0.65, innovation standard deviation 0.60). The errors are the main design's stationary AR(1) errors, the same errors doubled for every unit at the five post-treatment dates, or random walks built from the same innovations. The treated unit is drawn like every control, so the units are exchangeable in all six designs. The effect is zero, \(N\in\{17,39\}\), \(\gamma=0.10\), and each cell has 1{,}000 replications.

For synthetic control, we compare full conformal inference and the Jackknife+ with the test of \citet{ChernozhukovWuthrichZhu2021}. At each post-treatment date, the test refits the weights under the null using the pre-treatment dates and that date. It ranks the treated unit's absolute residual at that date among the residuals at all \(T_0+1\) dates. With one post-treatment date, moving-block and i.i.d.\ permutations give the same test. For GSC with one factor, we compare the Jackknife+ with the parametric bootstrap of Section~\ref{sec:bootstrap}, computed by \code{fect} with 200 draws. The point estimates of \code{fect} equal ours to within \(10^{\CompFectLibExp}\).

\input{tables/tab_comparators}

Table~\ref{tab:sims-comparators} reports coverage. No unit-level cell falls more than \CompUnitWorstZ{} standard errors below its bound. With stationary factors and stationary errors, the time-based test covers \CompCwzStatA, close to its nominal level. With drifting factors, it covers \CompCwzDriftA. Synthetic control does not remove the random-walk factors exactly, so its residuals inherit their drift. When the errors grow noisier after treatment or follow random walks, coverage of the time-based test falls to \CompCwzBC{} with drifting or stationary factors. The post-treatment residual is then no longer exchangeable with the pre-treatment ones.

The parametric bootstrap covers \CompBoot{} in every design, although its stated assumptions exclude random-walk errors. It adds error paths drawn from the controls' prediction errors at every date, including post-treatment dates. It therefore also compares the treated unit with the controls. Its coverage is approximate. Its intervals are \CompBootNarrow\% narrower than those of the Jackknife+, whose guaranteed bound is higher.
\FloatBarrier

\section{Additional empirical results}
\label{app:empirics}

This section supplements the applications in Section~\ref{sec:empirics}. We report the synthetic control estimates and confidence sets by year for California and West Germany. For the turnout application, we report the estimated factor loadings, the turnout effects of election-day registration by state, simultaneous sets across state-elections, and average effects. We also report the overlap check in all three applications and the simultaneous bands of Appendix~\ref{app:bands}.

\subsection{The synthetic control applications}
\label{app:sc-tables}

We compute the synthetic control weights by solving the simplex problem with a ridge penalty of \(10^{-5}\) on
the weights. The penalty is fixed in advance and selects a unique solution when the controls outnumber the
pre-treatment years. For every date in Figure~\ref{fig:apps-sc}, Tables~\ref{tab:prop99-years}
and~\ref{tab:germany-years} report the synthetic control estimate, the full conformal set, and the Jackknife+
interval at $\gamma=0.10$. The rows with event time up to $0$ are diagnostic dates, at which the effect is zero under no anticipation.

The full conformal interval for the average effect in Section~\ref{sec:empirics} inverts the rank test reported there. A held-out control's weights use only pre-treatment outcomes, so its average error depends on a candidate effect path only through the path's average. The arguments of Appendices~\ref{app:fc-proof} and~\ref{app:jplus-proof} then apply to these average scores unchanged.

\begin{table}[!ht]\centering\small
\caption{California, Proposition 99: estimates and confidence sets by year}
\begin{tabular}{@{}ccrll@{}}
\toprule
Year & Event time & Estimate & Full conformal set & Jackknife+ interval\\
\midrule
\input{tables/sc_table_prop99.tex}
\end{tabular}
\floatnote{Effects on cigarette sales in packs per capita. $N=\AppPropN$; guaranteed coverage \AppPropCov{} for both
procedures. Full conformal sets are computed exactly by affine inversion.}
\label{tab:prop99-years}
\end{table}

\begin{table}[!ht]\centering\small
\caption{West Germany, reunification: estimates and confidence sets by year}
\begin{tabular}{@{}ccrll@{}}
\toprule
Year & Event time & Estimate & Full conformal set & Jackknife+ interval\\
\midrule
\input{tables/sc_table_germany.tex}
\end{tabular}
\floatnote{Effects on GDP in thousand dollars per capita. $N=\AppGerN$; guaranteed coverage \AppGerCov{} for both
procedures. Full conformal sets are computed exactly by affine inversion.}
\label{tab:germany-years}
\end{table}

\subsection{Turnout by state}
\label{app:edr-states}

\begin{figure}[!ht]
\centering
\caption{Turnout effects of election-day registration by state and election}
\includegraphics[width=0.92\textwidth]{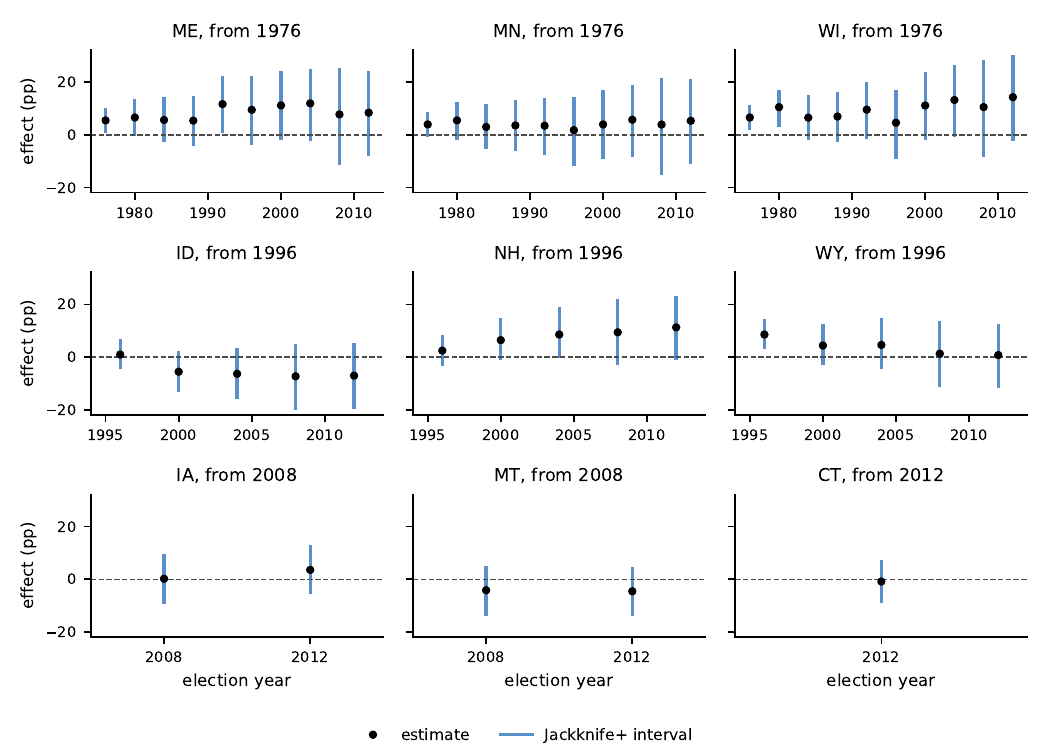}
\floatnote{Each panel is one adopting state, ordered by adoption year. Dots are GSC estimates
of the effect on turnout, in percentage points, in each post-adoption election. Bars are Jackknife+ intervals computed as in Table~\ref{tab:edr-states}.}
\label{fig:apps-edr}
\end{figure}

Table~\ref{tab:edr-states} reports, for each state that adopted election-day registration, the average of its
post-adoption estimates and the estimate and Jackknife+ interval in its first and last post-adoption elections.
Figure~\ref{fig:apps-edr} shows every state-by-election estimate and interval.

\begin{table}[!ht]\centering\small
\caption{Turnout effects of election-day registration by state}
\begin{tabular}{@{}lcccll@{}}
\toprule
State & First EDR election & Elections & Mean effect & First election & Last election\\
\midrule
\input{tables/edr_table.tex}
\end{tabular}
\floatnote{Effects on turnout in percentage points, estimated by GSC with two factors. We fix this number in every fold,
following the cross-validation choice in \citet{Xu2017} for these data. Brackets are Jackknife+ intervals at $\gamma=0.10$ ($N=\EdrN$, $q_\gamma=\EdrQ$), computed by \code{fect}
for one state and one election each. ``Mean effect'' averages a state's post-adoption estimates and has no interval. EDR is election-day registration.}
\label{tab:edr-states}
\end{table}

\subsection{Simultaneous and average effects in the turnout application}
\label{app:edr-multi}

We apply the constructions of Appendices~\ref{app:multi} and~\ref{app:staggered} to the turnout data. Because GSC is
full-path, each adopting state's pool for a single cell comprises the thirty-eight never-treated states, with \(N=39\).
The other adopting states do not enter the fit because \code{fect} estimates factors from never-treated states only. The union bound over the \(M=50\) state-by-election cells gives \(\gamma/M=0.002<1/39\) at
\(\gamma=0.10\), so every cell's set at \(\gamma/M\) is the whole line.

For the average effect by relative period,
the block Jackknife+ of Section~\ref{sec:empirics} has \(q_\gamma=1\) at \(\gamma=0.20\). Its interval is the
range from the smallest lower to the largest upper endpoint of the four folds. With the partition we drew, that range runs from
\EdrBlockLoOne{} to \EdrBlockHiOne{} percentage points at the first post-adoption election, which is \(r=0\), against
\(\EdrFectLoOne\) to \EdrFectHiOne{} for the single-unit Jackknife+ interval of Figure~\ref{fig:apps-edr-att}. Across \EdrBlockPartitions{} random
partitions, the 10th and 90th percentiles of each endpoint lie about two percentage points apart at that election and four at
the fifth.

With one partition per relative period, as in Remark~\ref{rem:staggered-per-period-app}, the number of control blocks rises from four at the first post-adoption election to six at the third. Through the fifth election, the interval at \(\gamma=0.10\) is still the whole line. From the sixth election on, only the three
states that adopted in 1976 remain, and twelve control blocks give a finite interval at \(\gamma=0.10\).

For the relabeling test, \(n=47\) and \(K=9\) give about \(4.9\times10^{14}\) assignments, or \(6.9\times10^{12}\)
patterns, since three adopters share each of two dates and two share a third.

\subsection{Loading overlap in the turnout application}
\label{app:edr-loadings}

\begin{figure}[!ht]
\centering
\caption{Estimated factor loadings, adopting and never-adopting states}
\includegraphics[width=0.8\textwidth]{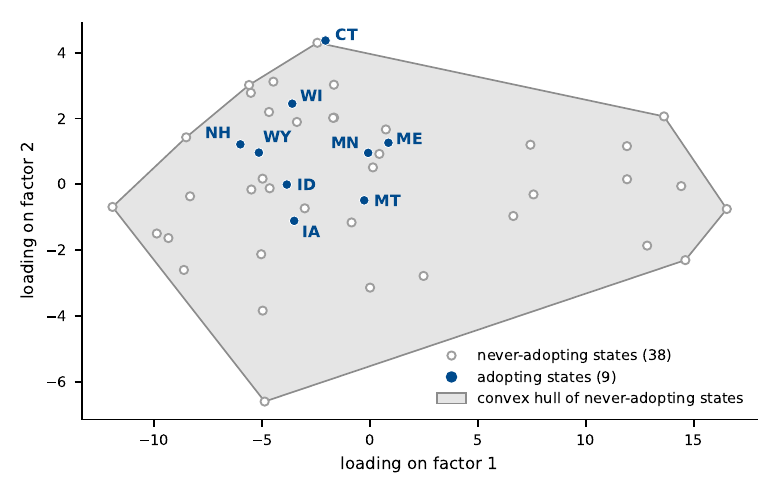}
\floatnote{Each point is one state's estimated loadings on the two factors of the GSC fit.
Hollow gray points are the \EdrLoadNco{} never-adopting states, and the shaded region is their convex hull; filled points
are the \EdrLoadNtr{} adopting states.}
\label{fig:apps-edr-loadings}
\end{figure}

Figure~\ref{fig:apps-edr-loadings} plots each state's estimated loadings on the two factors.
Of the \EdrLoadNtr{} adopting states, \EdrLoadInHull{} lie inside the convex hull of the \EdrLoadNco{} never-adopting states' loadings. \EdrLoadOutHullNames{} lies just outside it.
Two features limit what the figure shows. First, loadings are identified only up to rotation. We therefore report
whether a point lies inside the hull, which rotation does not change, rather than coordinate ranges. Second, an
adopting state's loadings are estimated from its pre-adoption elections only, so they are less precise than the
controls'.

\par

\subsection{The overlap check in the three applications}
\label{app:overlap-check}

Figure~\ref{fig:apps-overlap} shows the overlap check of Appendix~\ref{app:overlap-def} for the \OvTreated{} treated units. The check flags \OvFlagged{} of them. The window is the unit's pre-treatment
period, which is 1970--1988 for California, 1960--1989 for West Germany, and the elections before adoption for each
turnout state. Turnout states that adopted in the same year have the same window and so share their controls'
errors. For the turnout application, we use observed turnout. Subtracting the contributions of the two covariates, using the GSC coefficients,
leaves all flags unchanged.

\begin{figure}[!ht]
\centering
\caption{The overlap check in the three applications}
\includegraphics[width=0.95\textwidth]{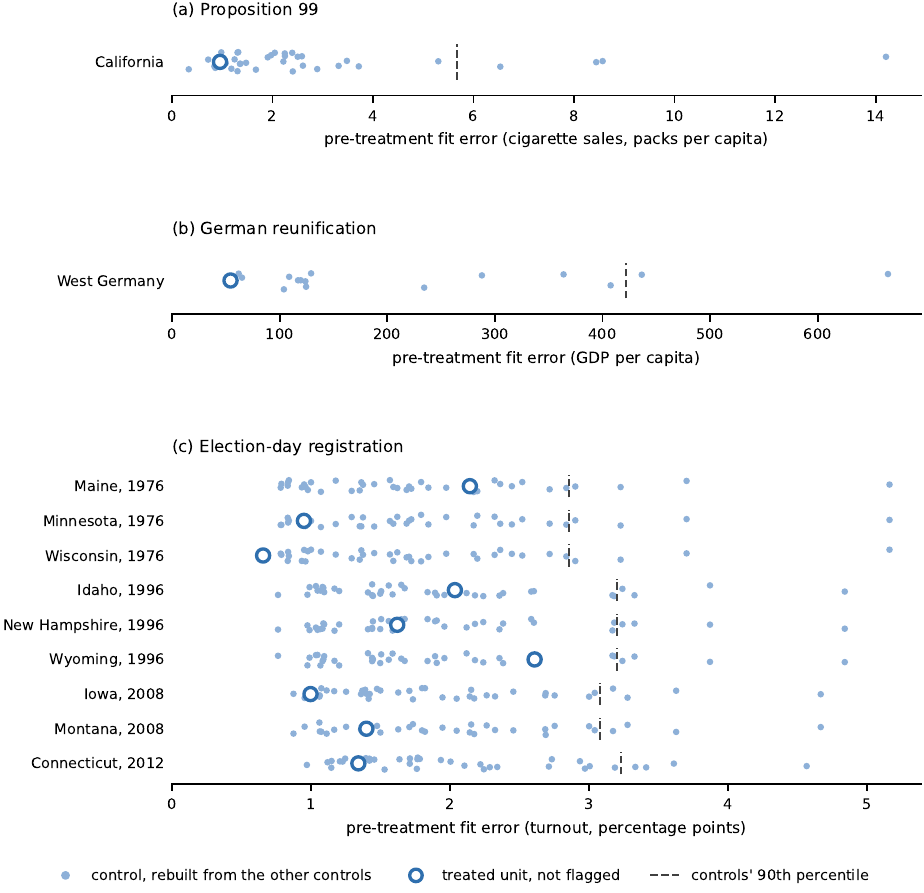}
\floatnote{Each row is a treated unit; turnout states are labeled with their first election under election-day registration. Dots are the controls' pre-treatment fit errors when each control is predicted from the other controls
(Appendix~\ref{app:overlap-def}). Open circles are the treated units' errors. The dashed line marks the
controls' 90th percentile, above which the check flags a unit.}
\label{fig:apps-overlap}
\end{figure}

\subsection{Simultaneous bands in the three applications}
\label{app:apps-bands}

We computed the bands of Appendix~\ref{app:bands} for Proposition~99, German reunification and Minnesota, one of
the nine states in the election-day registration study of Section~\ref{sec:empirics}. With equal weights, no band
excludes zero at any date. With studentized weights, the Jackknife+ band for German reunification excludes zero
in \BandGerJpFirst--\BandGerJpLast, with mean width \BandGerJpWidth. The full conformal outer band of Appendix~\ref{app:band-computation} for German reunification excludes zero in \BandGerFcYears. For Proposition~99, the full conformal outer band excludes zero in \BandPropFcYears{}, and the Jackknife+ band at no date. For Minnesota, no band excludes zero.

Where a band excludes zero and the period-wise sets of Section~\ref{sec:empirics} do not, the difference comes from the studentized scores. For a given score, a band covers every date at once, so it is never narrower than the period-wise sets. With studentized scores and one date at a time, the Jackknife+ intervals for German reunification have mean width \StudGerJpWidth, against \BandGerDateWidth{} with the default score. Both procedures exclude zero in \StudGerJpExcl{} of the \StudGerDates{} years (\StudGerYears). West Germany's pre-treatment root mean squared residual is the \StudGerSigRank{} of the \StudGerN{} units', so studentizing enlarges its score relative to the controls'. For Proposition~99, studentized sets exclude zero in \StudPropFcExcl{} of the \StudPropDates{} years with full conformal inference and in \StudPropJpExcl{} with the Jackknife+, against \StudPropFcExclEq{} and \StudPropJpExclEq{} with the default score. For Minnesota, no studentized Jackknife+ interval excludes zero. Both scores carry the same guarantee, so researchers must choose between them before seeing the outcomes.

\end{document}

%% file: preamble.tex
\usepackage[utf8]{inputenc}
\usepackage[T1]{fontenc}
\usepackage{lmodern}
\usepackage{microtype}
\usepackage[margin=1in]{geometry}
\usepackage{setspace}
\usepackage{amsmath,amssymb,mathtools,amsthm}
\mathtoolsset{showonlyrefs=true}
\usepackage{bm}
\usepackage{booktabs}
\usepackage{array}
\newcolumntype{R}[1]{>{\raggedright\arraybackslash}p{#1}}
\usepackage{graphicx}
\usepackage{float}
\usepackage{pdflscape}
\usepackage{afterpage}
\usepackage{placeins}
\usepackage{tikz}
\usetikzlibrary{patterns,decorations.pathreplacing,positioning,calc,arrows.meta}
\usepackage{enumitem}
\usepackage{algorithm}
\usepackage{algpseudocode}
\usepackage[round]{natbib}
\usepackage{titlesec}
\titleformat{\section}{\normalfont\fontsize{15}{15}\bfseries}{\thesection.}{0.5em}{}
\titlelabel{\thetitle.\ }
\usepackage{xcolor}
\definecolor{CardinalRed}{cmyk}{0,1,0.65,0.34}
\definecolor{LinkNavy}{HTML}{1F3A5F}
\usepackage[flushmargin]{footmisc}
\usepackage{etoolbox}
\makeatletter
\patchcmd{\maketitle}{\parindent 1em\noindent\hb@xt@1.8em{\hss\@textsuperscript{\normalfont\@thefnmark}}}{\parindent 0pt\noindent\@textsuperscript{\normalfont\@thefnmark}}{}{\PackageWarning{preamble}{title footnote indent patch failed}}
\makeatother
\usepackage[colorlinks, linkcolor=LinkNavy, citecolor=LinkNavy, urlcolor=LinkNavy, hyperfootnotes=false, pdftitle={Unit-Level Conformal Inference for Panel Counterfactuals}, pdfauthor={Ziyi Liu and Yiqing Xu}]{hyperref}

\setdisplayskipstretch{1}
\makeatletter
\g@addto@macro\normalsize{%
  \setlength\abovedisplayskip{10.3125pt plus 1.875pt minus 1.875pt}%
  \setlength\belowdisplayskip{10.3125pt plus 1.875pt minus 1.875pt}%
  \setlength\abovedisplayshortskip{5.625pt plus 1.875pt minus 0.9375pt}%
  \setlength\belowdisplayshortskip{5.625pt plus 1.875pt minus 0.9375pt}}
\makeatother

\usepackage{caption}
\usepackage{etoc}
\usepackage{etoolbox}
\makeatletter\patchcmd{\l@section}{1.5em}{2.6em}{}{}\patchcmd{\l@subsection}{2.3em}{3.4em}{}{}\makeatother
\newcommand{\tablerowstretch}{1.3}
\AtBeginEnvironment{table}{}
\newcommand{\floatnote}[1]{\par\vspace{6pt}{\footnotesize\setlength{\parindent}{0pt}\noindent\parbox{\linewidth}{\emph{Note:} #1}\par}}

\newcommand{\numword}[1]{\ifcase\numexpr#1\relax zero\or one\or two\or three\or four\or five\or six\or seven\or eight\or nine\or ten\or eleven\or twelve\else\number\numexpr#1\relax\fi}

\theoremstyle{plain}
\newtheorem{theorem}{Theorem}
\newtheorem{proposition}{Proposition}
\newtheorem{lemma}{Lemma}
\newtheorem{corollary}{Corollary}
\newtheorem{assumption}{Assumption}
\newtheoremstyle{remarkbold}{0.5\topsep}{0.5\topsep}{\normalfont}{}{\bfseries\itshape}{.}{.5em}{}
\theoremstyle{remarkbold}
\newtheorem{remark}{Remark}

\newcommand{\R}{\mathbb R}
\newcommand{\Iset}{\mathcal I}
\newcommand{\Cset}{\mathcal C}

\newcommand{\Tpre}{\mathcal T_0}
\newcommand{\Tpost}{\mathcal T_1}
\newcommand{\one}{\mathbf 1}
\newcommand{\Ginfo}{\mathcal G}

\newcommand{\code}[1]{\texttt{#1}}

\definecolor{paperblue}{HTML}{2F6FAE}
\definecolor{papertan}{HTML}{E2B791}
\definecolor{papergreen}{HTML}{16451A}
\definecolor{paperrose}{HTML}{DB8B95}
\definecolor{papergray}{HTML}{EEEEEC}
\definecolor{paperinactive}{HTML}{D9D9D7}
\definecolor{paperpink}{HTML}{E06C9F}
\definecolor{papersky}{HTML}{9FCBE8}
\definecolor{paperink}{HTML}{333333}
\definecolor{cellpredicted}{HTML}{002B55}
\definecolor{cellquery}{HTML}{4F86C0}
\definecolor{celltrain}{HTML}{A3C4E4}
\definecolor{cellinactive}{HTML}{E6E6E6}
\definecolor{paperaccent}{HTML}{004A8C}
\tikzset{
  panel cell/.style={draw=white,line width=.45pt},
  panel query/.style={panel cell,fill=cellquery},
  panel train/.style={panel cell,fill=celltrain},
  panel predicted/.style={panel cell,draw=paperaccent,line width=.8pt,fill=cellpredicted},
  panel hidden/.style={panel cell,preaction={fill=white},pattern=north east lines,pattern color=paperaccent},
  panel inactive/.style={panel cell,fill=cellinactive},
  panel unavailable/.style={panel cell,preaction={fill=white},pattern=north west lines,pattern color=gray!60!black},
  panel candidate/.style={panel cell,draw=paperaccent,line width=.8pt,fill=celltrain},
  panel divide/.style={draw=paperink,line width=.7pt},
  panel target/.style={draw=paperaccent,line width=.8pt}
}
\newcommand{\PanelCell}[3]{\path[#3] (#1,#2) rectangle ++(.96,-.72);}
\newcommand{\PanelPrediction}[2]{%
  \PanelCell{#1}{#2}{panel predicted}
  \node[font=\scriptsize\bfseries,text=white] at (#1+.48,#2-.36) {\(\widehat Y\)};}
\newcommand{\PanelCandidate}[2]{%
  \PanelCell{#1}{#2}{panel candidate}
  \node[font=\scriptsize\bfseries,text=paperink] at (#1+.48,#2-.36) {\(c\)};}
\newcommand{\PanelLegend}[1]{\tikz[baseline=-.55ex,x=.34cm,y=.30cm]\path[#1] (0,0) rectangle (.96,.72);}
\newcommand{\PanelLegendCandidate}{\tikz[baseline=-.55ex,x=.34cm,y=.30cm]{\path[panel candidate] (0,0) rectangle (.96,.72);\node[font=\tiny\bfseries,text=paperink] at (.48,.36) {\(c\)};}}
\newcommand{\PanelWideHeaders}{%
  \node[font=\scriptsize,text=gray!75!black] at (3.5,.85) {Pre-treatment};
  \node[font=\scriptsize,text=gray!75!black] at (9.5,.85) {Post-treatment};
  \foreach \x/\lab in {0/{-6},1/{-5},2/{-4},3/{-3},4/{-2},5/{-1},6/{0},7/{1},8/{2},9/{3},10/{4},11/{5}}{
    \node[font=\tiny,text=gray!70!black] at (\x+.48,.28) {\(\lab\)};}}

%% file: numbers/app_numbers.tex
\newcommand{\AppGerCov}{0.941}

\newcommand{\AppGerFirstYear}{1990}

\newcommand{\AppGerLastYear}{2003}
\newcommand{\AppGerMean}{-1.30}
\newcommand{\AppGerN}{17}
\newcommand{\AppGerPost}{14}

\newcommand{\AppPropCov}{0.923}

\newcommand{\AppPropFCex}{5}
\newcommand{\AppPropFirstYear}{1989}
\newcommand{\AppPropJPex}{0}
\newcommand{\AppPropLastYear}{2000}
\newcommand{\AppPropMean}{-19.51}
\newcommand{\AppPropN}{39}
\newcommand{\AppPropPost}{12}
\newcommand{\AppPropQ}{2}
\newcommand{\EdrCells}{50}
\newcommand{\EdrExcl}{5}

\newcommand{\EdrMeanEff}{4.90}
\newcommand{\EdrMedWidth}{19.4}
\newcommand{\EdrN}{39}
\newcommand{\EdrPos}{43}
\newcommand{\EdrQ}{2}

%% file: numbers/sc_avg_numbers.tex
\newcommand{\AvgPropRank}{3}
\newcommand{\AvgPropP}{0.077}
\newcommand{\AvgPropJpP}{0.128}
\newcommand{\AvgPropJpLo}{-45.0}
\newcommand{\AvgPropJpHi}{6.0}
\newcommand{\AvgGerRank}{7}
\newcommand{\AvgGerP}{0.412}

\newcommand{\AvgGerJpLo}{-4.7}
\newcommand{\AvgGerJpHi}{2.2}

%% file: numbers/sc_avg_fc_numbers.tex
\newcommand{\AvgPropFcLo}{-35.1}
\newcommand{\AvgPropFcHi}{-4.0}
\newcommand{\AvgGerFcLo}{-4.7}
\newcommand{\AvgGerFcHi}{2.1}

%% file: numbers/edr_loadings_numbers.tex
\newcommand{\EdrLoadNtr}{9}
\newcommand{\EdrLoadNco}{38}
\newcommand{\EdrLoadInHull}{8}

\newcommand{\EdrLoadOutHullNames}{Connecticut}

%% file: numbers/edr_relabel_numbers.tex
\newcommand{\EdrRelB}{999}

\newcommand{\EdrRelP}{0.015}

%% file: numbers/edr_placebo_numbers.tex
\newcommand{\EdrRelNexceed}{14}

%% file: numbers/edr_parametric_numbers.tex
\newcommand{\EdrParB}{1,000}
\newcommand{\EdrParLoOne}{0.34}
\newcommand{\EdrParHiOne}{4.55}
\newcommand{\EdrParExclPost}{four}
\newcommand{\EdrParNPost}{five}
\newcommand{\EdrParExclPre}{two}
\newcommand{\EdrParNPre}{nine}

%% file: numbers/edr_block_numbers.tex
\newcommand{\EdrBlockG}{4}
\newcommand{\EdrBlockNB}{5}
\newcommand{\EdrBlockPartitions}{200}
\newcommand{\EdrBlockLoOne}{0.90}
\newcommand{\EdrBlockHiOne}{4.04}

\newcommand{\EdrBlockExclOne}{83}
\newcommand{\EdrBlockExclFour}{42}
\newcommand{\EdrFectLoOne}{-2.76}
\newcommand{\EdrFectHiOne}{7.76}

%% file: numbers/overlap_numbers.tex
\newcommand{\OvTreated}{11}
\newcommand{\OvFlagged}{none}

%% file: numbers/sims_numbers.tex
\newcommand{\SimTpre}{40}
\newcommand{\SimTpost}{5}
\newcommand{\SimNplacebo}{4}
\newcommand{\SimReps}{1{,}000}
\newcommand{\SimPilotReps}{200}
\newcommand{\SimEffect}{2}
\newcommand{\SimFacPhi}{0.65}
\newcommand{\SimFacSd}{0.60}
\newcommand{\SimRwSd}{0.18}
\newcommand{\SimDriftSd}{0.12}
\newcommand{\SimDrift}{0.035}
\newcommand{\SimAlphaSd}{0.5}
\newcommand{\SimErrPhi}{0.35}
\newcommand{\SimErrSd}{0.35}
\newcommand{\SimTrend}{0.015}
\newcommand{\SimSeasonAmp}{0.40}
\newcommand{\SimSeasonPeriod}{12}
\newcommand{\SimScEta}{10^{-5}}
\newcommand{\SimSdidNoise}{0.60}
\newcommand{\SimRlow}{1}
\newcommand{\SimRhigh}{3}
\newcommand{\SimGapA}{0.5}
\newcommand{\SimGapB}{1}
\newcommand{\SimGapC}{2}
\newcommand{\SimNa}{10}
\newcommand{\SimNb}{17}
\newcommand{\SimNc}{39}
\newcommand{\SimNd}{70}
\newcommand{\SimGammaMain}{0.10}
\newcommand{\SimGammaLow}{0.05}
\newcommand{\SimNfactors}{three}
\newcommand{\SimNest}{six}
\newcommand{\SimNshift}{four}
\newcommand{\SimNdesigns}{five}
\newcommand{\SimLoadRange}{[-1,1]}
\newcommand{\SimFlagSE}{two}
\newcommand{\SimErrSdStat}{0.37}

\newcommand{\SimBoundThirtyNine}{0.923}

\newcommand{\SimJpExMin}{0.889}
\newcommand{\SimJpExMax}{0.953}

\newcommand{\SimFcExMin}{0.889}
\newcommand{\SimFcExMax}{0.945}

\newcommand{\SimExCells}{45}

\newcommand{\SimExWorstZ}{1.8}

\newcommand{\SimExBelow}{11}
\newcommand{\SimJpThirtyNineMin}{0.948}
\newcommand{\SimJpThirtyNineMax}{0.953}
\newcommand{\SimFcThirtyNineMin}{0.919}
\newcommand{\SimFcThirtyNineMax}{0.928}
\newcommand{\SimGapTwoDid}{0.303}
\newcommand{\SimGapZeroDid}{0.883}
\newcommand{\SimGapTwoSc}{0.210}

\newcommand{\SimGapTwoSdid}{0.675}
\newcommand{\SimGapZeroSdid}{0.942}
\newcommand{\SimGapTwoGscOne}{0.485}

\newcommand{\SimGapTwoIfe}{0.876}

\newcommand{\SimGapTwoDidScMin}{0.206}
\newcommand{\SimGapTwoDidScMax}{0.329}
\newcommand{\SimGapTwoFactorMin}{0.686}
\newcommand{\SimGapTwoFactorMax}{0.903}

\newcommand{\SimWidthJpDid}{3.32}
\newcommand{\SimWidthJpSc}{2.23}
\newcommand{\SimWidthJpSdid}{1.94}
\newcommand{\SimWidthJpGscOne}{2.43}
\newcommand{\SimWidthJpGscThree}{1.83}
\newcommand{\SimWidthJpIfe}{1.88}
\newcommand{\SimWidthFcDid}{3.03}
\newcommand{\SimWidthFcSc}{1.98}
\newcommand{\SimFcNarrowMin}{2}
\newcommand{\SimFcNarrowMax}{15}
\newcommand{\SimFcWiderMax}{10}

\newcommand{\SimPlaceboExAboveAny}{10}
\newcommand{\SimPlaceboExCells}{36}
\newcommand{\SimPlaceboGapTwoMin}{0.622}
\newcommand{\SimPlaceboGapTwoMax}{0.791}
\newcommand{\SimPlaceboGapTwoFactorMax}{0.279}

\newcommand{\SimOpenTen}{1.8}
\newcommand{\SimOpenSeventeen}{0.7}

\newcommand{\SimOpenThirtyNineLow}{0.2}
\newcommand{\SimOpenChecked}{120}

\newcommand{\SimOpenBoundedChecked}{120}

\newcommand{\SimFailMax}{0.001}

%% file: numbers/position_numbers.tex
\newcommand{\PosOneFcThirtyNine}{0.892}

\newcommand{\PosOneFcStudThirtyNine}{0.930}

\newcommand{\PosNoisyFcThirtyNine}{0.675}
\newcommand{\PosQuietFcThirtyNine}{0.997}
\newcommand{\PosNoisyFcStudThirtyNine}{0.943}
\newcommand{\PosQuietFcStudThirtyNine}{0.885}

%% file: numbers/bands_numbers.tex
\newcommand{\BandJointMin}{52}
\newcommand{\BandJointMax}{89}
\newcommand{\BandEqMin}{92.9}
\newcommand{\BandEqMax}{97.4}
\newcommand{\BandStudMin}{91.3}
\newcommand{\BandStudMax}{96.3}
\newcommand{\BandBoundSeventeen}{94.1}
\newcommand{\BandBoundThirtyNine}{92.3}
\newcommand{\BandShortCov}{93.0}
\newcommand{\BandShortBound}{94.1}
\newcommand{\BandRatioEqFiveMin}{1.19}
\newcommand{\BandRatioEqFiveMax}{1.37}
\newcommand{\BandRatioEqTwelveMin}{1.33}
\newcommand{\BandRatioEqTwelveMax}{1.56}
\newcommand{\BandRatioStudFiveMin}{1.12}
\newcommand{\BandRatioStudFiveMax}{1.49}
\newcommand{\BandRatioStudTwelveMin}{1.22}
\newcommand{\BandRatioStudTwelveMax}{1.75}
\newcommand{\BandGerJpFirst}{2000}
\newcommand{\BandGerJpLast}{2003}
\newcommand{\BandGerJpWidth}{5.0}
\newcommand{\BandGerDateWidth}{9.0}
\newcommand{\BandGerFcYears}{2003}
\newcommand{\BandPropFcYears}{1997, 1999 and 2000}

%% file: numbers/stud_dates_numbers.tex
\newcommand{\StudGerDates}{fourteen}
\newcommand{\StudGerJpExcl}{nine}

\newcommand{\StudGerYears}{1991, 1992 and 1997--2003}
\newcommand{\StudGerJpWidth}{2.5}

\newcommand{\StudPropDates}{twelve}
\newcommand{\StudPropFcExcl}{seven}
\newcommand{\StudPropFcExclEq}{five}
\newcommand{\StudPropJpExcl}{one}
\newcommand{\StudPropJpExclEq}{zero}

\newcommand{\StudGerSigRank}{second smallest}
\newcommand{\StudGerN}{seventeen}

%% file: numbers/comparators_numbers.tex
\newcommand{\CompUnitWorstZ}{1.3}

\newcommand{\CompCwzDriftA}{0.87}

\newcommand{\CompCwzStatA}{0.88--0.90}

\newcommand{\CompCwzBC}{0.72--0.76}
\newcommand{\CompBoot}{0.89--0.91}
\newcommand{\CompBootNarrow}{15--19}
\newcommand{\CompFectLibExp}{-12}

%% file: figures/figure_temporal_support.tex
\begin{figure}[t]
\caption{Information used to predict the treated unit's untreated outcome}
\centering
\begin{minipage}[t]{0.49\textwidth}
\centering
{\small (a) Target-date estimator}\par\vspace{0.25em}
\begin{tikzpicture}[x=.42cm,y=.62cm]
  \PanelWideHeaders
  \node[anchor=east,font=\scriptsize,align=right] at (-.18,-.35) {Unit 0\\treated};
  \node[anchor=east,font=\scriptsize] at (-.18,-1.35) {Control 1};
  \node[anchor=east,font=\scriptsize] at (-.18,-2.35) {Control 2};
  \node[anchor=east,font=\scriptsize] at (-.18,-3.35) {Other controls};
  \foreach \x in {0,...,6}{
    \PanelCell{\x}{0}{panel query}
    \foreach \r in {1,2,3}{\PanelCell{\x}{-\r}{panel train}}
  }
  \foreach \x in {7,8,10,11}{
    \PanelCell{\x}{0}{panel unavailable}
    \foreach \r in {1,2,3}{\PanelCell{\x}{-\r}{panel inactive}}
  }
  \PanelPrediction{9}{0}
  \foreach \r in {1,2,3}{\PanelCell{9}{-\r}{panel train}}
  \draw[panel divide] (7,.05)--(7,-3.73);
  \draw[panel target] (9,.03) rectangle (10,-3.70);
\end{tikzpicture}

\vspace{0.35em}
{\footnotesize Post-treatment control information is limited to the target date.}
\end{minipage}\hfill
\begin{minipage}[t]{0.49\textwidth}
\centering
{\small (b) Full-path estimator}\par\vspace{0.25em}
\begin{tikzpicture}[x=.42cm,y=.62cm]
  \PanelWideHeaders
  \node[anchor=east,font=\scriptsize,align=right] at (-.18,-.35) {Unit 0\\treated};
  \node[anchor=east,font=\scriptsize] at (-.18,-1.35) {Control 1};
  \node[anchor=east,font=\scriptsize] at (-.18,-2.35) {Control 2};
  \node[anchor=east,font=\scriptsize] at (-.18,-3.35) {Other controls};
  \foreach \x in {0,...,6}{
    \PanelCell{\x}{0}{panel query}
    \foreach \r in {1,2,3}{\PanelCell{\x}{-\r}{panel train}}
  }
  \foreach \x in {7,8,10,11}{
    \PanelCell{\x}{0}{panel unavailable}
    \foreach \r in {1,2,3}{\PanelCell{\x}{-\r}{panel train}}
  }
  \PanelPrediction{9}{0}
  \foreach \r in {1,2,3}{\PanelCell{9}{-\r}{panel train}}
  \draw[panel divide] (7,.05)--(7,-3.73);
  \draw[panel target] (9,.03) rectangle (10,-3.70);
\end{tikzpicture}

\vspace{0.35em}
{\footnotesize All post-treatment control outcomes may affect the prediction at \(t=3\).}
\end{minipage}

\vspace{0.7em}
{\scriptsize
\begin{tabular}{@{}ll@{\qquad}ll@{\qquad}ll@{}}
  \PanelLegend{panel query} & treated history used &
  \PanelLegend{panel train} & control outcome used &
  \PanelLegend{panel predicted} & target predicted \\
  \PanelLegend{panel unavailable} & untreated value unavailable &
  \PanelLegend{panel inactive} & observed but outside the fit &
  &
\end{tabular}
}

\floatnote{The outlined column is the target date \(t=3\). A target-date estimator, panel (a), uses the controls'
post-treatment outcomes only at that date. A full-path estimator, panel (b), can use their whole post-treatment
paths. Unit 0's untreated outcomes at the other post-treatment dates are unavailable.}
\label{fig:ordinary-temporal-support}
\end{figure}

%% file: figures/figure_role_symmetric_extensions.tex
\begin{figure}[t]
\centering
\caption{The prediction task each unit is given under each procedure}
\begin{minipage}[t]{0.485\textwidth}\centering
{\small (a) One held-out unit, target-date}\par\vspace{0.3em}
\resizebox{\linewidth}{!}{%
\begin{tikzpicture}[x=.50cm,y=.60cm]
  \PanelWideHeaders
  \node[anchor=east,font=\scriptsize,align=right] at (-.18,-.35) {Held-out unit \(i\)\\(any unit)};
  \node[anchor=east,font=\scriptsize] at (-.18,-1.35) {Training unit 1};
  \node[anchor=east,font=\scriptsize] at (-.18,-2.35) {Training unit 2};
  \node[anchor=east,font=\scriptsize] at (-.18,-3.35) {Other training units};
  \foreach \x in {0,...,6}{
    \PanelCell{\x}{0}{panel query}
    \foreach \r in {1,2,3}{\PanelCell{\x}{-\r}{panel train}}
  }
  \foreach \x in {7,8,10,11}{
    \foreach \r in {0,1,2,3}{\PanelCell{\x}{-\r}{panel inactive}}
  }
  \PanelPrediction{9}{0}
  \foreach \r in {1,2,3}{\PanelCell{9}{-\r}{panel train}}
  \draw[panel divide] (7,.05)--(7,-3.73);
  \draw[panel target] (9,.03) rectangle (10,-3.70);
\end{tikzpicture}}
\end{minipage}\hfill
\begin{minipage}[t]{0.485\textwidth}\centering
{\small (b) One held-out unit, full-path}\par\vspace{0.3em}
\resizebox{\linewidth}{!}{%
\begin{tikzpicture}[x=.50cm,y=.60cm]
  \PanelWideHeaders
  \node[anchor=east,font=\scriptsize] at (-.18,-.35) {Held-out control \(j\)};
  \node[anchor=east,font=\scriptsize] at (-.18,-1.35) {Unit 0, training};
  \node[anchor=east,font=\scriptsize] at (-.18,-2.35) {Training unit 1};
  \node[anchor=east,font=\scriptsize] at (-.18,-3.35) {Other training units};
  \foreach \x in {0,...,6}{
    \PanelCell{\x}{0}{panel query}
    \foreach \r in {1,2,3}{\PanelCell{\x}{-\r}{panel train}}
  }
  \foreach \x in {7,8,10,11}{
    \PanelCell{\x}{0}{panel hidden}
    \PanelCell{\x}{-1}{panel unavailable}
    \node[font=\scriptsize\bfseries,text=paperaccent] at (\x+.48,-1.36) {?};
    \PanelCell{\x}{-2}{panel train}
    \PanelCell{\x}{-3}{panel train}
  }
  \PanelPrediction{9}{0}
  \PanelCandidate{9}{-1}
  \PanelCell{9}{-2}{panel train}
  \PanelCell{9}{-3}{panel train}
  \draw[panel divide] (7,.05)--(7,-3.73);
  \draw[panel target] (9,.03) rectangle (10,-3.70);
\end{tikzpicture}}
\end{minipage}

\vspace{1em}
\begin{minipage}[t]{0.485\textwidth}\centering
{\small (c) Two held-out units, target-date}\par\vspace{0.3em}
\resizebox{\linewidth}{!}{%
\begin{tikzpicture}[x=.50cm,y=.60cm]
  \PanelWideHeaders
  \node[anchor=east,font=\scriptsize] at (-.18,-.35) {Held-out unit 0};
  \node[anchor=east,font=\scriptsize] at (-.18,-1.35) {Held-out unit \(j\)};
  \node[anchor=east,font=\scriptsize] at (-.18,-2.35) {Training unit 1};
  \node[anchor=east,font=\scriptsize] at (-.18,-3.35) {Other training units};
  \foreach \x in {0,...,6}{
    \PanelCell{\x}{0}{panel query}
    \PanelCell{\x}{-1}{panel query}
    \PanelCell{\x}{-2}{panel train}
    \PanelCell{\x}{-3}{panel train}
  }
  \foreach \x in {7,8,10,11}{
    \PanelCell{\x}{0}{panel unavailable}
    \foreach \r in {1,2,3}{\PanelCell{\x}{-\r}{panel inactive}}
  }
  \PanelPrediction{9}{0}
  \PanelPrediction{9}{-1}
  \PanelCell{9}{-2}{panel train}
  \PanelCell{9}{-3}{panel train}
  \draw[panel divide] (7,.05)--(7,-3.73);
  \draw[panel target] (9,.03) rectangle (10,-3.70);
\end{tikzpicture}}
\end{minipage}\hfill
\begin{minipage}[t]{0.485\textwidth}\centering
{\small (d) Two held-out units, full-path}\par\vspace{0.3em}
\resizebox{\linewidth}{!}{%
\begin{tikzpicture}[x=.50cm,y=.60cm]
  \PanelWideHeaders
  \node[anchor=east,font=\scriptsize] at (-.18,-.35) {Held-out unit 0};
  \node[anchor=east,font=\scriptsize] at (-.18,-1.35) {Held-out unit \(j\)};
  \node[anchor=east,font=\scriptsize] at (-.18,-2.35) {Training unit 1};
  \node[anchor=east,font=\scriptsize] at (-.18,-3.35) {Other training units};
  \foreach \x in {0,...,6}{
    \PanelCell{\x}{0}{panel query}
    \PanelCell{\x}{-1}{panel query}
    \PanelCell{\x}{-2}{panel train}
    \PanelCell{\x}{-3}{panel train}
  }
  \foreach \x in {7,8,10,11}{
    \PanelCell{\x}{0}{panel unavailable}
    \PanelCell{\x}{-1}{panel hidden}
    \PanelCell{\x}{-2}{panel train}
    \PanelCell{\x}{-3}{panel train}
  }
  \PanelPrediction{9}{0}
  \PanelPrediction{9}{-1}
  \PanelCell{9}{-2}{panel train}
  \PanelCell{9}{-3}{panel train}
  \draw[panel divide] (7,.05)--(7,-3.73);
  \draw[panel target] (9,.03) rectangle (10,-3.70);
\end{tikzpicture}}
\end{minipage}

\vspace{0.8em}
{\scriptsize
\begin{tabular}{@{}ll@{\qquad}ll@{\qquad}ll@{}}
  \PanelLegend{panel query} & held-out unit's history, used &
  \PanelLegend{panel train} & training unit's outcome, used &
  \PanelLegend{panel predicted} & target, predicted \\
  \PanelLegend{panel hidden} & held-out unit's outcome, hidden &
  \PanelLegend{panel unavailable} & untreated value unavailable &
  \PanelLegend{panel inactive} & observed but outside the fit \\
  \PanelLegendCandidate & \multicolumn{5}{@{}l@{}}{unit 0's target cell, completed with a candidate value}
\end{tabular}}

\floatnote{Rows are units and columns are dates. The outlined column is the target
date $t=3$. (a) Any unit can be held out. When a control is held out, unit~$0$
is a training unit with its target cell set to a candidate value. (b) A
full-path fit also needs unit~$0$'s other post-treatment cells, marked
with question marks, and so needs a candidate at every post-treatment date
(SM~\ref{app:pooled-fc}). (c) Unit~$0$ and a control are held out
together. (d) Each held-out unit hides its post-treatment path and is
predicted in a separate fit.}
\label{fig:role-symmetric-extensions}
\end{figure}

%% file: figures/figure_historical_diagnostics.tex
\begin{figure}[t]
\caption{Information used by the pre-treatment diagnostics at diagnostic date \(s=-2\)}
\centering
\begin{minipage}[t]{0.49\textwidth}
\centering
{\small (a) Target-date estimator}\par\vspace{0.25em}
\begin{tikzpicture}[x=.42cm,y=.62cm]
  \PanelWideHeaders
  \node[anchor=east,font=\scriptsize] at (-.18,-.35) {Held-out unit 0};
  \node[anchor=east,font=\scriptsize] at (-.18,-1.35) {Held-out unit \(j\)};
  \node[anchor=east,font=\scriptsize] at (-.18,-2.35) {Training unit 1};
  \node[anchor=east,font=\scriptsize] at (-.18,-3.35) {Other training units};
  \foreach \x in {0,...,3}{
    \PanelCell{\x}{0}{panel query}
    \PanelCell{\x}{-1}{panel query}
    \PanelCell{\x}{-2}{panel train}
    \PanelCell{\x}{-3}{panel train}
  }
  \PanelPrediction{4}{0}
  \PanelPrediction{4}{-1}
  \PanelCell{4}{-2}{panel train}
  \PanelCell{4}{-3}{panel train}
  \foreach \x in {5,6}{
    \foreach \r in {0,1,2,3}{\PanelCell{\x}{-\r}{panel inactive}}
  }
  \foreach \x in {7,...,11}{
    \PanelCell{\x}{0}{panel unavailable}
    \foreach \r in {1,2,3}{\PanelCell{\x}{-\r}{panel inactive}}
  }
  \draw[panel divide] (7,.05)--(7,-3.73);
  \draw[panel target] (4,.03) rectangle (5,-3.70);
\end{tikzpicture}

\vspace{0.35em}
{\footnotesize Every fit excludes the dates after \(s\).}
\end{minipage}\hfill
\begin{minipage}[t]{0.49\textwidth}
\centering
{\small (b) Full-path estimator}\par\vspace{0.25em}
\begin{tikzpicture}[x=.42cm,y=.62cm]
  \PanelWideHeaders
  \node[anchor=east,font=\scriptsize] at (-.18,-.35) {Held-out unit 0};
  \node[anchor=east,font=\scriptsize] at (-.18,-1.35) {Held-out unit \(j\)};
  \node[anchor=east,font=\scriptsize] at (-.18,-2.35) {Training unit 1};
  \node[anchor=east,font=\scriptsize] at (-.18,-3.35) {Other training units};
  \foreach \x in {0,...,3}{
    \PanelCell{\x}{0}{panel query}
    \PanelCell{\x}{-1}{panel query}
    \PanelCell{\x}{-2}{panel train}
    \PanelCell{\x}{-3}{panel train}
  }
  \PanelPrediction{4}{0}
  \PanelPrediction{4}{-1}
  \PanelCell{4}{-2}{panel train}
  \PanelCell{4}{-3}{panel train}
  \foreach \x in {5,6}{
    \PanelCell{\x}{0}{panel hidden}
    \PanelCell{\x}{-1}{panel hidden}
    \PanelCell{\x}{-2}{panel train}
    \PanelCell{\x}{-3}{panel train}
  }
  \foreach \x in {7,...,11}{
    \PanelCell{\x}{0}{panel unavailable}
    \PanelCell{\x}{-1}{panel hidden}
    \PanelCell{\x}{-2}{panel train}
    \PanelCell{\x}{-3}{panel train}
  }
  \draw[panel divide] (7,.05)--(7,-3.73);
  \draw[panel target] (4,.03) rectangle (5,-3.70);
\end{tikzpicture}

\vspace{0.35em}
{\footnotesize Held-out units hide their outcomes from \(s\) onward, and training units keep complete panels.}
\end{minipage}

\vspace{0.7em}
{\scriptsize
\begin{tabular}{@{}ll@{\qquad}ll@{\qquad}ll@{}}
  \PanelLegend{panel query} & held-out unit's history, used &
  \PanelLegend{panel train} & training unit's outcome, used &
  \PanelLegend{panel predicted} & diagnostic date, predicted \\
  \PanelLegend{panel hidden} & held-out unit's outcome, hidden &
  \PanelLegend{panel unavailable} & untreated value unavailable &
  \PanelLegend{panel inactive} & observed but outside the fit
\end{tabular}
}

\floatnote{The two held-out units reveal only their histories through \(-3\). Under a target-date estimator, panel (a),
the training units add their outcomes at \(-2\) only. Under a full-path estimator, panel (b), they keep complete
panels, so the diagnostic uses outcomes observed after \(-2\).}

\label{fig:historical-information}
\end{figure}

%% file: figures/figure_multiple_treated_common_date.tex
\providecommand{\MtLabels}{%
  \node[anchor=east,font=\scriptsize] at (-.18,-.35) {Treated \(k_1\)};
  \node[anchor=east,font=\scriptsize] at (-.18,-1.35) {Treated \(k_2\)};
  \node[anchor=east,font=\scriptsize] at (-.18,-2.35) {Treated \(k_3\)};
  \node[anchor=east,font=\scriptsize] at (-.18,-3.35) {Control 1};
  \node[anchor=east,font=\scriptsize] at (-.18,-4.35) {Control 2};
  \node[anchor=east,font=\scriptsize] at (-.18,-5.35) {Control 3};
  \node[anchor=east,font=\scriptsize] at (-.18,-6.35) {Other controls};}
\providecommand{\MtCandidate}[3]{%
  \PanelCell{#1}{#2}{panel candidate}%
  \node[font=\scriptsize\bfseries,text=paperink] at (#1+.48,#2-.36) {\(#3\)};}
\providecommand{\MtTreatedQueryTd}[1]{%
  \foreach \x in {0,...,6}{\PanelCell{\x}{-#1}{panel query}}%
  \foreach \x in {7,8,10,11}{\PanelCell{\x}{-#1}{panel unavailable}}%
  \PanelPrediction{9}{-#1}}
\providecommand{\MtControlQueryTd}[1]{%
  \foreach \x in {0,...,6}{\PanelCell{\x}{-#1}{panel query}}%
  \foreach \x in {7,8,10,11}{\PanelCell{\x}{-#1}{panel inactive}}%
  \PanelPrediction{9}{-#1}}
\providecommand{\MtControlQueryPath}[1]{%
  \foreach \x in {0,...,6}{\PanelCell{\x}{-#1}{panel query}}%
  \foreach \x in {7,8,10,11}{\PanelCell{\x}{-#1}{panel hidden}}%
  \PanelPrediction{9}{-#1}}
\providecommand{\MtTrainTd}[1]{%
  \foreach \x in {0,...,6}{\PanelCell{\x}{-#1}{panel train}}%
  \foreach \x in {7,8,10,11}{\PanelCell{\x}{-#1}{panel inactive}}%
  \PanelCell{9}{-#1}{panel train}}
\providecommand{\MtTrain}[1]{\foreach \x in {0,...,11}{\PanelCell{\x}{-#1}{panel train}}}
\providecommand{\MtCandTd}[2]{%
  \foreach \x in {0,...,6}{\PanelCell{\x}{-#1}{panel train}}%
  \foreach \x in {7,8,10,11}{\PanelCell{\x}{-#1}{panel unavailable}}%
  \MtCandidate{9}{-#1}{#2}}
\providecommand{\MtQueryFp}[1]{%
  \foreach \x in {0,...,6}{\PanelCell{\x}{-#1}{panel query}}%
  \foreach \x in {7,...,11}{\PanelPrediction{\x}{-#1}}}
\providecommand{\MtCandFp}[1]{%
  \foreach \x in {0,...,6}{\PanelCell{\x}{-#1}{panel train}}%
  \foreach \x in {7,...,11}{\PanelCandidate{\x}{-#1}}}
\providecommand{\MtFrame}{%
  \path (12.6,0);
  \draw[panel divide] (7,.05)--(7,-6.73);}
\providecommand{\MtTarget}{\draw[panel target] (9,.03) rectangle (10,-6.70);}
\begin{figure}[!htbp]
\centering
\caption{Prediction tasks with multiple treated units and a common adoption date}
\begin{minipage}[t]{0.485\textwidth}\centering
{\small (a) Block Jackknife+: two blocks held out}\par\vspace{0.3em}
\resizebox{\linewidth}{!}{%
\begin{tikzpicture}[x=.50cm,y=.55cm]
  \PanelWideHeaders
  \MtLabels
  \MtTreatedQueryTd{0}
  \MtTreatedQueryTd{1}
  \MtTreatedQueryTd{2}
  \MtControlQueryTd{3}
  \MtControlQueryTd{4}
  \MtControlQueryTd{5}
  \MtTrainTd{6}
  \MtFrame
  \MtTarget
\end{tikzpicture}}
\end{minipage}\hfill
\begin{minipage}[t]{0.485\textwidth}\centering
{\small (b) Relabeling test under a sharp null}\par\vspace{0.3em}
\resizebox{\linewidth}{!}{%
\begin{tikzpicture}[x=.50cm,y=.55cm]
  \PanelWideHeaders
  \MtLabels
  \MtCandFp{0}
  \MtCandFp{1}
  \MtQueryFp{2}
  \MtQueryFp{3}
  \MtQueryFp{4}
  \MtTrain{5}
  \MtTrain{6}
  \MtFrame
\end{tikzpicture}}
\end{minipage}

\vspace{0.8em}
{\scriptsize
\begin{tabular}{@{}ll@{\qquad}ll@{\qquad}ll@{}}
  \PanelLegend{panel query} & held-out unit's history, used &
  \PanelLegend{panel train} & training unit's outcome, used &
  \PanelLegend{panel predicted} & target, predicted \\
  \PanelLegend{panel unavailable} & untreated value unavailable &
  \PanelLegend{panel inactive} & observed but outside the fit &
  \PanelLegendCandidate & completed under the null
\end{tabular}}

\floatnote{Units \(k_1\) to \(k_3\) are treated from date \(1\). (a) The outlined column is the target date \(t=3\). The treated block and a block of three controls are held out together, and each block is scored by the absolute value of
its average error. One fit predicts all six held-out units. (b) Under a sharp null, the
relabeling test gives the treatment pattern to another set of three units, here \(k_3\) with controls 1 and 2, and
refits.}
\label{fig:multiple-treated-common-date}
\end{figure}

%% file: figures/figure_staggered_masking.tex
\providecommand{\SmHeaders}{%
  \node[font=\scriptsize,text=gray!75!black] at (6,.85) {Calendar date};
  \foreach \x in {0,...,11}{%
    \pgfmathtruncatemacro{\SmDate}{\x+1}%
    \node[font=\tiny,text=gray!70!black] at (\x+.48,.28) {\(\SmDate\)};}}
\providecommand{\SmLabels}{\SmLabelsCore\node[anchor=east,font=\scriptsize] at (-.18,-12.35) {Never treated 9, leftover};}
\providecommand{\SmLabelsNoLeftover}{\SmLabelsCore\node[anchor=east,font=\scriptsize] at (-.18,-12.35) {Never treated 9\phantom{, leftover}};}
\providecommand{\SmLabelsCore}{%
  \node[anchor=east,font=\scriptsize] at (-.18,-.35) {Adopter \(k_1\), \(A=7\)};
  \node[anchor=east,font=\scriptsize] at (-.18,-1.35) {Adopter \(k_2\), \(A=7\)};
  \node[anchor=east,font=\scriptsize] at (-.18,-2.35) {Adopter \(k_3\), \(A=10\)};
  \node[anchor=east,font=\scriptsize] at (-.18,-3.35) {Adopter \(k_4\), \(A=12\)};
  \node[anchor=east,font=\scriptsize] at (-.18,-4.35) {Never treated 1};
  \node[anchor=east,font=\scriptsize] at (-.18,-5.35) {Never treated 2};
  \node[anchor=east,font=\scriptsize] at (-.18,-6.35) {Never treated 3};
  \node[anchor=east,font=\scriptsize] at (-.18,-7.35) {Never treated 4};
  \node[anchor=east,font=\scriptsize] at (-.18,-8.35) {Never treated 5};
  \node[anchor=east,font=\scriptsize] at (-.18,-9.35) {Never treated 6};
  \node[anchor=east,font=\scriptsize] at (-.18,-10.35) {Never treated 7};
  \node[anchor=east,font=\scriptsize] at (-.18,-11.35) {Never treated 8};}
\providecommand{\SmTick}[2]{\draw[panel divide] (#1,{.05-#2}) -- (#1,{-.77-#2});}
\providecommand{\SmDash}[2]{\draw[panel divide,dashed] (#1,{.05-#2}) -- (#1,{-.77-#2});}
\providecommand{\SmLegendTick}{\tikz[baseline=-.55ex,x=.34cm,y=.30cm]\draw[panel divide] (.48,-.1) -- (.48,.82);}
\providecommand{\SmLegendDash}{\tikz[baseline=-.55ex,x=.34cm,y=.30cm]\draw[panel divide,dashed] (.48,-.1) -- (.48,.82);}
\providecommand{\SmTreated}[2]{%
  \SmPre{#1}{panel query}\SmNytCells{#1}{#2}{panel query}%
  \foreach \x in {#2,...,11}{\PanelCell{\x}{-#1}{panel unavailable}}}
\providecommand{\SmTreatedPred}[3]{\SmTreated{#1}{#2}\PanelPrediction{#3}{-#1}}
\providecommand{\SmControl}[2]{%
  \SmPre{#1}{panel query}\SmNytCells{#1}{#2}{panel query}%
  \foreach \x in {#2,...,11}{\PanelCell{\x}{-#1}{panel hidden}}}
\providecommand{\SmControlPred}[3]{\SmControl{#1}{#2}\PanelPrediction{#3}{-#1}}
\providecommand{\SmNyt}[2]{%
  \SmPre{#1}{panel train}\SmNytCells{#1}{#2}{panel train}%
  \foreach \x in {#2,...,11}{\PanelCell{\x}{-#1}{panel unavailable}}}
\providecommand{\SmPre}[2]{\foreach \x in {0,...,5}{\PanelCell{\x}{-#1}{#2}}}
\providecommand{\SmNytCells}[3]{\ifnum#2>6 \pgfmathtruncatemacro{\SmLast}{#2-1}\foreach \x in {6,...,\SmLast}{\PanelCell{\x}{-#1}{#3}}\fi}
\providecommand{\SmTrain}[1]{\foreach \x in {0,...,11}{\PanelCell{\x}{-#1}{panel train}}}
\providecommand{\SmFrame}{\path (12.6,0);}
\begin{figure}[!htbp]
\centering
\caption{Prediction tasks in the block Jackknife+ under staggered adoption}
\begin{minipage}[t]{0.485\textwidth}\centering
{\small (a) One fold, read at \(r=1\)}\par\vspace{0.3em}
\resizebox{\linewidth}{!}{%
\begin{tikzpicture}[x=.50cm,y=.50cm]
  \SmHeaders
  \SmLabels
  \SmTreatedPred{0}{6}{7}\SmTick{6}{0}
  \SmTreatedPred{1}{6}{7}\SmTick{6}{1}
  \SmTreatedPred{2}{9}{10}\SmTick{9}{2}
  \SmTreated{3}{11}\SmTick{11}{3}
  \SmControlPred{4}{6}{7}\SmDash{6}{4}
  \SmControlPred{5}{6}{7}\SmDash{6}{5}
  \SmControlPred{6}{9}{10}\SmDash{9}{6}
  \SmControl{7}{11}\SmDash{11}{7}
  \SmTrain{8}\SmTrain{9}\SmTrain{10}\SmTrain{11}\SmTrain{12}
  \SmFrame
\end{tikzpicture}}
\end{minipage}\hfill
\begin{minipage}[t]{0.485\textwidth}\centering
{\small (b) The same fold, read at \(r=5\)}\par\vspace{0.3em}
\resizebox{\linewidth}{!}{%
\begin{tikzpicture}[x=.50cm,y=.50cm]
  \SmHeaders
  \SmLabels
  \SmTreatedPred{0}{6}{11}\SmTick{6}{0}
  \SmTreatedPred{1}{6}{11}\SmTick{6}{1}
  \SmTreated{2}{9}\SmTick{9}{2}
  \SmTreated{3}{11}\SmTick{11}{3}
  \SmControlPred{4}{6}{11}\SmDash{6}{4}
  \SmControlPred{5}{6}{11}\SmDash{6}{5}
  \SmControl{6}{9}\SmDash{9}{6}
  \SmControl{7}{11}\SmDash{11}{7}
  \SmTrain{8}\SmTrain{9}\SmTrain{10}\SmTrain{11}\SmTrain{12}
  \SmFrame
\end{tikzpicture}}
\end{minipage}

\vspace{0.8em}
{\scriptsize
\begin{tabular}{@{}ll@{\qquad}ll@{\qquad}ll@{}}
  \PanelLegend{panel query} & held-out unit's history, used &
  \PanelLegend{panel train} & training unit's outcome, used &
  \PanelLegend{panel predicted} & cell predicted and scored \\
  \PanelLegend{panel hidden} & held-out unit's outcome, hidden &
  \PanelLegend{panel unavailable} & untreated value unavailable &
  \SmLegendTick\,\SmLegendDash & actual, assigned adoption date
\end{tabular}}

\floatnote{Adopters \(k_1\) and \(k_2\) adopt at date 7, \(k_3\) at 10 and \(k_4\) at 12 (solid bars). Never-treated units
1 to 4 form one control block, each matched to an adopter and given its date (dashed bars). (a) One fit hides both
blocks from their dates on and scores \((k_1,8)\), \((k_2,8)\), \((k_3,11)\) and their matches at \(r=1\). (b) The
same fit scores \((k_1,12)\) and \((k_2,12)\) and their matches at \(r=5\).}
\label{fig:staggered-masking}
\end{figure}

%% file: figures/fig_sims_exchangeable_float.tex
\begin{figure}[!htbp]\centering
\caption{Coverage when the treated unit is exchangeable with the controls}
\includegraphics[width=0.95\textwidth]{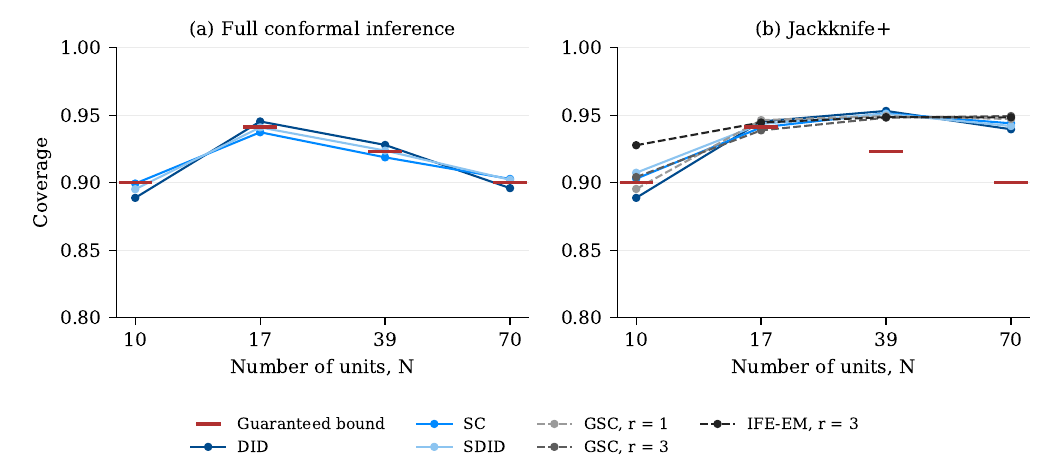}
\label{fig:sims-exchangeable}
\floatnote{Coverage at $\gamma=\SimGammaMain$, averaged over \SimTpost{} post-treatment dates and \SimReps{} replications,
when the treated unit's loadings come from the controls' distribution ($T_0=\SimTpre$). Red ticks mark the
guaranteed bound of Table~\ref{tab:levels}. Full conformal inference applies to the target-date estimators (solid
lines), the Jackknife+ to all six. SC is synthetic control, and $r$ is the number of factors.}
\end{figure}

%% file: tables/edr_block_table.tex
1 & 9 & 2.59 & [0.90, 4.04] & 0.2 & 0.024 \\
2 & 8 & 3.37 & [0.54, 4.91] & 0.2 & 0.035 \\
3 & 6 & 3.67 & [$-$0.30, 6.94] & 0.6 & 0.053 \\
4 & 6 & 3.23 & [$-$3.73, 8.62] & 0.6 & 0.134 \\
5 & 6 & 4.95 & [$-$1.42, 9.67] & 0.6 & 0.047 \\
\addlinespace
All after adoption & 9 & 4.90 & --- & --- & 0.015 \\
\bottomrule

%% file: tables/levels_table.tex
\begin{table}[!ht]\centering\small
\caption{Guaranteed coverage of period-wise sets by the number of units}
\begin{tabular}{@{}rcccc@{}}
\toprule
 & \multicolumn{2}{c}{$\gamma=0.10$} & \multicolumn{2}{c}{$\gamma=0.05$}\\
\cmidrule(lr){2-3}\cmidrule(l){4-5}
$N$ & Full conformal & Jackknife+ & Full conformal & Jackknife+\\
\midrule
10 & 0.900 & 0.900 & --- & --- \\
17 & 0.941 & 0.941 & --- & --- \\
20 & 0.900 & 0.950 & 0.950 & 0.950 \\
30 & 0.900 & 0.900 & 0.967 & 0.967 \\
39 & 0.923 & 0.923 & 0.974 & 0.974 \\
50 & 0.900 & 0.900 & 0.960 & 0.980 \\
\bottomrule
\end{tabular}
\floatnote{Entries are the guaranteed coverage, $1-\lfloor\gamma N\rfloor/N$ for full conformal inference and
$1-(2q_\gamma-1)_+/N$ for the Jackknife+, where $N$ counts the treated unit and the controls. A dash marks the whole
real line.}
\label{tab:levels}
\end{table}

%% file: tables/tab_cov_jplus.tex
\begin{table}[!ht]\centering\footnotesize
\caption{Coverage of Jackknife+ intervals by design, estimator, $N$ and $\gamma$}
\label{tab:sims-cov-jplus}
\setlength{\tabcolsep}{4pt}
\begin{tabular}{@{}lcccccc@{}}
\toprule
 & \multicolumn{4}{c}{$\gamma=0.10$} & \multicolumn{2}{c}{$\gamma=0.05$}\\
\cmidrule(lr){2-5}\cmidrule(l){6-7}
Design and estimator & $N=10$ & $N=17$ & $N=39$ & $N=70$ & $N=39$ & $N=10,17$\\
\midrule
\multicolumn{1}{@{}l}{\emph{Guaranteed bound}} & 0.900 & 0.941 & 0.923 & 0.900 & 0.974 & 1\\
\midrule
\multicolumn{7}{@{}l}{\emph{Exchangeable units}}\\
\quad DID & 0.889 (0.006) & 0.945 (0.004) & 0.953 (0.004) & 0.940 (0.005) & 0.978 (0.003) & $\dagger$\\
\quad SC & 0.903 (0.006) & 0.941 (0.005) & 0.951 (0.004) & 0.944 (0.004) & 0.977 (0.003) & $\dagger$\\
\quad SDID & 0.907 (0.005) & 0.944 (0.004) & 0.951 (0.003) & 0.942 (0.004) & 0.975 (0.002) & $\dagger$\\
\quad GSC, $r=1$ & 0.895 (0.006) & 0.946 (0.004) & 0.949 (0.004) & 0.947 (0.004) & 0.977 (0.002) & $\dagger$\\
\quad GSC, $r=3$ & 0.904 (0.005) & 0.939 (0.004) & 0.948 (0.004) & 0.949 (0.004) & 0.977 (0.002) & $\dagger$\\
\quad IFE-EM, $r=3$ & 0.928 (0.004) & 0.945 (0.004) & 0.949 (0.004) & 0.948 (0.004) & 0.974 (0.003) & $\dagger$\\
\midrule
\multicolumn{7}{@{}l}{\emph{Treated loading at the controls' maximum}}\\
\quad DID & 0.859 (0.007) & 0.883 (0.006) & 0.883 (0.006) & 0.857 (0.007) & 0.938 (0.005) & $\dagger$\\
\quad SC & 0.870 (0.007) & 0.905 (0.006) & 0.901 (0.006) & 0.910 (0.005) & 0.946 (0.004) & $\dagger$\\
\quad SDID & 0.892 (0.005) & 0.932 (0.004) & 0.942 (0.004) & 0.933 (0.004) & 0.970 (0.003) & $\dagger$\\
\quad GSC, $r=1$ & 0.878 (0.006) & 0.913 (0.005) & 0.906 (0.005) & 0.905 (0.006) & 0.949 (0.004) & $\dagger$\\
\quad GSC, $r=3$ & 0.889 (0.005) & 0.934 (0.004) & 0.936 (0.004) & 0.947 (0.004) & 0.969 (0.003) & $\dagger$\\
\quad IFE-EM, $r=3$ & 0.919 (0.005) & 0.947 (0.004) & 0.938 (0.004) & 0.945 (0.004) & 0.973 (0.003) & $\dagger$\\
\midrule
\multicolumn{7}{@{}l}{\emph{Treated loading 0.5 above the maximum}}\\
\quad DID & 0.714 (0.010) & 0.730 (0.010) & 0.701 (0.010) & 0.661 (0.011) & 0.800 (0.008) & $\dagger$\\
\quad SC & 0.706 (0.010) & 0.740 (0.010) & 0.719 (0.010) & 0.715 (0.010) & 0.813 (0.009) & $\dagger$\\
\quad SDID & 0.839 (0.007) & 0.892 (0.006) & 0.902 (0.005) & 0.890 (0.005) & 0.944 (0.004) & $\dagger$\\
\quad GSC, $r=1$ & 0.788 (0.008) & 0.830 (0.008) & 0.797 (0.009) & 0.788 (0.009) & 0.876 (0.007) & $\dagger$\\
\quad GSC, $r=3$ & 0.847 (0.007) & 0.910 (0.005) & 0.927 (0.004) & 0.938 (0.004) & 0.958 (0.003) & $\dagger$\\
\quad IFE-EM, $r=3$ & 0.907 (0.005) & 0.935 (0.004) & 0.931 (0.004) & 0.936 (0.004) & 0.964 (0.003) & $\dagger$\\
\midrule
\multicolumn{7}{@{}l}{\emph{Treated loading 1 above the maximum}}\\
\quad DID & 0.546 (0.011) & 0.560 (0.011) & 0.522 (0.012) & 0.481 (0.012) & 0.615 (0.012) & $\dagger$\\
\quad SC & 0.486 (0.012) & 0.533 (0.012) & 0.490 (0.012) & 0.479 (0.012) & 0.581 (0.012) & $\dagger$\\
\quad SDID & 0.773 (0.008) & 0.823 (0.007) & 0.839 (0.007) & 0.820 (0.007) & 0.901 (0.005) & $\dagger$\\
\quad GSC, $r=1$ & 0.670 (0.010) & 0.720 (0.010) & 0.682 (0.011) & 0.658 (0.011) & 0.764 (0.010) & $\dagger$\\
\quad GSC, $r=3$ & 0.792 (0.008) & 0.876 (0.006) & 0.906 (0.005) & 0.927 (0.004) & 0.947 (0.004) & $\dagger$\\
\quad IFE-EM, $r=3$ & 0.886 (0.006) & 0.921 (0.004) & 0.916 (0.005) & 0.929 (0.004) & 0.954 (0.004) & $\dagger$\\
\midrule
\multicolumn{7}{@{}l}{\emph{Treated loading 2 above the maximum}}\\
\quad DID & 0.326 (0.012) & 0.329 (0.012) & 0.303 (0.012) & 0.282 (0.011) & 0.368 (0.012) & $\dagger$\\
\quad SC & 0.235 (0.011) & 0.250 (0.011) & 0.210 (0.010) & 0.206 (0.010) & 0.275 (0.012) & $\dagger$\\
\quad SDID & 0.625 (0.010) & 0.673 (0.009) & 0.675 (0.009) & 0.656 (0.010) & 0.767 (0.008) & $\dagger$\\
\quad GSC, $r=1$ & 0.502 (0.011) & 0.532 (0.012) & 0.485 (0.012) & 0.467 (0.012) & 0.563 (0.012) & $\dagger$\\
\quad GSC, $r=3$ & 0.686 (0.009) & 0.798 (0.007) & 0.860 (0.006) & 0.897 (0.005) & 0.916 (0.005) & $\dagger$\\
\quad IFE-EM, $r=3$ & 0.851 (0.006) & 0.877 (0.006) & 0.876 (0.006) & 0.903 (0.005) & 0.926 (0.005) & $\dagger$\\
\bottomrule
\end{tabular}
\floatnote{Coverage of the true effect, averaged over the five post-treatment dates and 1{,}000 replications; standard errors in parentheses are clustered by replication. $N$ counts the treated unit and the controls; $T_0=40$ pre-treatment dates. The guaranteed bound is that of Table~\ref{tab:levels}; it applies to the exchangeable design only. No exchangeable cell is more than two standard errors below its bound. $\dagger$: at $\gamma=0.05$ with $N=10$ or $17$, no candidate can be excluded, so the set is the whole line and coverage is 1 (not computed). At $N=70$, GSC and IFE-EM come from an earlier run with the previous version of the code (Appendix~\ref{app:sims-designs}).}
\end{table}

%% file: tables/tab_cov_fc.tex
\begin{table}[!ht]\centering\footnotesize
\caption{Coverage of full conformal sets by design, estimator, $N$ and $\gamma$}
\label{tab:sims-cov-fc}
\setlength{\tabcolsep}{4pt}
\begin{tabular}{@{}lcccccc@{}}
\toprule
 & \multicolumn{4}{c}{$\gamma=0.10$} & \multicolumn{2}{c}{$\gamma=0.05$}\\
\cmidrule(lr){2-5}\cmidrule(l){6-7}
Design and estimator & $N=10$ & $N=17$ & $N=39$ & $N=70$ & $N=39$ & $N=10,17$\\
\midrule
\multicolumn{1}{@{}l}{\emph{Guaranteed bound}} & 0.900 & 0.941 & 0.923 & 0.900 & 0.974 & 1\\
\midrule
\multicolumn{7}{@{}l}{\emph{Exchangeable units}}\\
\quad DID & 0.889 (0.006) & 0.945 (0.004) & 0.928 (0.005) & 0.896 (0.006) & 0.978 (0.003) & $\dagger$\\
\quad SC & 0.899 (0.007) & 0.937 (0.005) & 0.919 (0.005) & 0.903 (0.005) & 0.977 (0.003) & $\dagger$\\
\quad SDID & 0.895 (0.005) & 0.941 (0.004) & 0.924 (0.004) & 0.902 (0.005) & 0.975 (0.002) & $\dagger$\\
\midrule
\multicolumn{7}{@{}l}{\emph{Treated loading at the controls' maximum}}\\
\quad DID & 0.859 (0.007) & 0.883 (0.006) & 0.828 (0.008) & 0.786 (0.009) & 0.938 (0.005) & $\dagger$\\
\quad SC & 0.865 (0.007) & 0.903 (0.006) & 0.862 (0.007) & 0.850 (0.007) & 0.943 (0.004) & $\dagger$\\
\quad SDID & 0.877 (0.006) & 0.928 (0.004) & 0.908 (0.005) & 0.885 (0.005) & 0.968 (0.003) & $\dagger$\\
\midrule
\multicolumn{7}{@{}l}{\emph{Treated loading 0.5 above the maximum}}\\
\quad DID & 0.714 (0.010) & 0.730 (0.010) & 0.637 (0.011) & 0.565 (0.011) & 0.800 (0.008) & $\dagger$\\
\quad SC & 0.695 (0.011) & 0.733 (0.010) & 0.659 (0.011) & 0.641 (0.011) & 0.809 (0.009) & $\dagger$\\
\quad SDID & 0.816 (0.007) & 0.886 (0.006) & 0.857 (0.006) & 0.829 (0.007) & 0.944 (0.004) & $\dagger$\\
\midrule
\multicolumn{7}{@{}l}{\emph{Treated loading 1 above the maximum}}\\
\quad DID & 0.546 (0.011) & 0.560 (0.011) & 0.465 (0.012) & 0.407 (0.012) & 0.615 (0.012) & $\dagger$\\
\quad SC & 0.478 (0.012) & 0.526 (0.013) & 0.431 (0.012) & 0.410 (0.012) & 0.577 (0.012) & $\dagger$\\
\quad SDID & 0.730 (0.008) & 0.805 (0.007) & 0.783 (0.008) & 0.744 (0.008) & 0.897 (0.006) & $\dagger$\\
\midrule
\multicolumn{7}{@{}l}{\emph{Treated loading 2 above the maximum}}\\
\quad DID & 0.326 (0.012) & 0.329 (0.012) & 0.265 (0.011) & 0.237 (0.011) & 0.368 (0.012) & $\dagger$\\
\quad SC & 0.231 (0.011) & 0.244 (0.011) & 0.180 (0.010) & 0.170 (0.009) & 0.271 (0.011) & $\dagger$\\
\quad SDID & 0.543 (0.010) & 0.632 (0.010) & 0.605 (0.010) & 0.578 (0.010) & 0.750 (0.009) & $\dagger$\\
\bottomrule
\end{tabular}
\floatnote{Definitions and marks are as in Table~\ref{tab:sims-cov-jplus}. Coverage is the rate at which the true effect is not rejected.}
\end{table}

%% file: figures/fig_sims_coverage_float.tex
\begin{figure}[!htbp]\centering
\caption{Jackknife+ coverage and the treated unit's loading}
\includegraphics[width=0.95\textwidth]{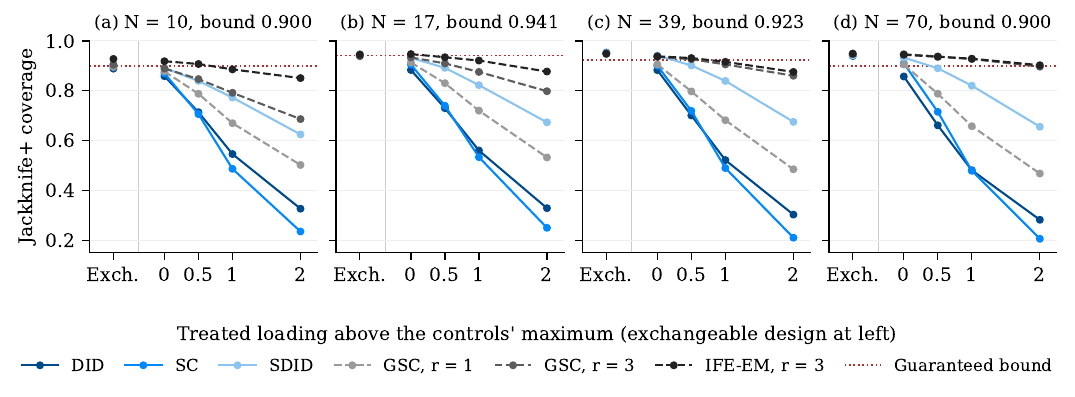}
\label{fig:sims-coverage}
\floatnote{Coverage of the true effect at $\gamma=\SimGammaMain$, averaged over the \SimTpost{} post-treatment dates and
\SimReps{} replications. The horizontal axis is the treated unit's loading on the drifting factor minus the controls'
maximum; the point at the left is the exchangeable design. The dotted line is the guaranteed bound of
Table~\ref{tab:levels}, which applies to the exchangeable design only. Solid lines: target-date estimators;
dashed lines: full-path estimators.}
\end{figure}

%% file: tables/tab_width.tex
\begin{table}[!ht]\centering\footnotesize
\caption{Median width of full conformal sets and Jackknife+ intervals}
\label{tab:sims-width}
\setlength{\tabcolsep}{4pt}
\begin{tabular}{@{}llcccccc@{}}
\toprule
 &  & \multicolumn{4}{c}{$\gamma=0.10$} & \multicolumn{2}{c}{$\gamma=0.05$}\\
\cmidrule(lr){3-6}\cmidrule(l){7-8}
Design and procedure & Estimator & $N=10$ & $N=17$ & $N=39$ & $N=70$ & $N=39$ & $N=10,17$\\
\midrule
\multicolumn{8}{@{}l}{\emph{Exchangeable units}}\\
\quad Full conformal & DID & 3.17 & 3.42 & 3.03 & 2.80 & 3.83 & $\infty$\\
\quad  & SC & 3.21 & 2.89 & 1.98 & 1.69 & 2.75 & $\infty$\\
\quad Jackknife+ & DID & 3.17 & 3.42 & 3.32 & 3.20 & 3.83 & $\infty$\\
\quad  & SC & 3.01 & 2.80 & 2.23 & 1.98 & 2.71 & $\infty$\\
\quad  & SDID & 2.31 & 2.20 & 1.94 & 1.79 & 2.28 & $\infty$\\
\quad  & GSC, $r=1$ & 2.56 & 2.64 & 2.43 & 2.29 & 2.82 & $\infty$\\
\quad  & GSC, $r=3$ & 2.09 & 2.03 & 1.83 & 1.71 & 2.15 & $\infty$\\
\quad  & IFE-EM, $r=3$ & 2.96 & 2.24 & 1.88 & 1.74 & 2.22 & $\infty$\\
\midrule
\multicolumn{8}{@{}l}{\emph{Treated loading at the controls' maximum}}\\
\quad Full conformal & DID & 3.17 & 3.42 & 3.03 & 2.80 & 3.83 & $\infty$\\
\quad  & SC & 3.19 & 2.93 & 1.99 & 1.70 & 2.79 & $\infty$\\
\quad Jackknife+ & DID & 3.17 & 3.42 & 3.32 & 3.20 & 3.83 & $\infty$\\
\quad  & SC & 3.03 & 2.82 & 2.22 & 1.98 & 2.73 & $\infty$\\
\quad  & SDID & 2.31 & 2.21 & 1.94 & 1.79 & 2.29 & $\infty$\\
\quad  & GSC, $r=1$ & 2.57 & 2.66 & 2.43 & 2.29 & 2.83 & $\infty$\\
\quad  & GSC, $r=3$ & 2.10 & 2.04 & 1.83 & 1.72 & 2.15 & $\infty$\\
\quad  & IFE-EM, $r=3$ & 3.00 & 2.27 & 1.88 & 1.74 & 2.22 & $\infty$\\
\midrule
\multicolumn{8}{@{}l}{\emph{Treated loading 0.5 above the maximum}}\\
\quad Full conformal & DID & 3.17 & 3.42 & 3.03 & 2.80 & 3.83 & $\infty$\\
\quad  & SC & 3.19 & 2.99 & 2.03 & 1.71 & 2.86 & $\infty$\\
\quad Jackknife+ & DID & 3.17 & 3.42 & 3.32 & 3.20 & 3.83 & $\infty$\\
\quad  & SC & 3.07 & 2.85 & 2.22 & 1.98 & 2.75 & $\infty$\\
\quad  & SDID & 2.35 & 2.23 & 1.94 & 1.79 & 2.29 & $\infty$\\
\quad  & GSC, $r=1$ & 2.61 & 2.68 & 2.43 & 2.29 & 2.85 & $\infty$\\
\quad  & GSC, $r=3$ & 2.15 & 2.08 & 1.84 & 1.72 & 2.17 & $\infty$\\
\quad  & IFE-EM, $r=3$ & 3.25 & 2.34 & 1.89 & 1.74 & 2.24 & $\infty$\\
\midrule
\multicolumn{8}{@{}l}{\emph{Treated loading 1 above the maximum}}\\
\quad Full conformal & DID & 3.17 & 3.42 & 3.03 & 2.80 & 3.83 & $\infty$\\
\quad  & SC & 3.23 & 3.06 & 2.08 & 1.74 & 2.94 & $\infty$\\
\quad Jackknife+ & DID & 3.17 & 3.42 & 3.32 & 3.20 & 3.83 & $\infty$\\
\quad  & SC & 3.09 & 2.89 & 2.23 & 1.98 & 2.79 & $\infty$\\
\quad  & SDID & 2.41 & 2.26 & 1.95 & 1.80 & 2.31 & $\infty$\\
\quad  & GSC, $r=1$ & 2.68 & 2.71 & 2.44 & 2.29 & 2.86 & $\infty$\\
\quad  & GSC, $r=3$ & 2.23 & 2.13 & 1.85 & 1.72 & 2.18 & $\infty$\\
\quad  & IFE-EM, $r=3$ & 3.51 & 2.45 & 1.90 & 1.74 & 2.26 & $\infty$\\
\midrule
\multicolumn{8}{@{}l}{\emph{Treated loading 2 above the maximum}}\\
\quad Full conformal & DID & 3.17 & 3.42 & 3.03 & 2.80 & 3.83 & $\infty$\\
\quad  & SC & 3.30 & 3.19 & 2.20 & 1.83 & 3.11 & $\infty$\\
\quad Jackknife+ & DID & 3.17 & 3.42 & 3.32 & 3.20 & 3.83 & $\infty$\\
\quad  & SC & 3.14 & 2.92 & 2.23 & 1.98 & 2.83 & $\infty$\\
\quad  & SDID & 2.56 & 2.34 & 1.96 & 1.80 & 2.33 & $\infty$\\
\quad  & GSC, $r=1$ & 2.80 & 2.80 & 2.47 & 2.30 & 2.89 & $\infty$\\
\quad  & GSC, $r=3$ & 2.42 & 2.24 & 1.86 & 1.73 & 2.23 & $\infty$\\
\quad  & IFE-EM, $r=3$ & 4.27 & 2.69 & 1.93 & 1.75 & 2.33 & $\infty$\\
\bottomrule
\end{tabular}
\floatnote{Median width of the full conformal set (exact affine inversion; DID and SC only) and of the Jackknife+ interval, over 1{,}000 replications and the five post-treatment dates. The true effect is 2 and the untreated error has stationary standard deviation 0.37. $\infty$: the whole line, because no candidate can be excluded. Unbounded full conformal sets enter the median with infinite width. Appendix~\ref{app:sims-results} reports their shares.}
\end{table}

%% file: tables/tab_placebo.tex
\begin{table}[!ht]\centering\footnotesize
\caption{Rejection rates of the pre-treatment diagnostics at the diagnostic dates}
\label{tab:sims-placebo}
\setlength{\tabcolsep}{4pt}
\begin{tabular}{@{}llcccccccc@{}}
\toprule
 & & \multicolumn{4}{c}{Exchangeable units} & \multicolumn{4}{c}{Treated loading 2 above the maximum}\\
\cmidrule(lr){3-6}\cmidrule(l){7-10}
Procedure & Estimator & $N=10$ & $N=17$ & $N=39$ & $N=70$ & $N=10$ & $N=17$ & $N=39$ & $N=70$\\
\midrule
Full conformal & \emph{Guaranteed at most} & 0.100 & 0.059 & 0.077 & 0.100 & \multicolumn{4}{c}{(no guarantee)}\\
 & DID & 0.107 & 0.050 & 0.079 & 0.113 & 0.635 & 0.622 & 0.700 & 0.727\\
 & SC & 0.110 & 0.057 & 0.075 & 0.100 & 0.732 & 0.709 & 0.780 & 0.791\\
 & SDID & 0.096 & 0.061 & 0.077 & 0.099 & 0.386 & 0.281 & 0.297 & 0.312\\
\midrule
Jackknife+ & \emph{Guaranteed at most} & 0.100 & 0.059 & 0.077 & 0.100 & \multicolumn{4}{c}{(no guarantee)}\\
 & DID & 0.107 & 0.050 & 0.053 & 0.067 & 0.635 & 0.622 & 0.663 & 0.679\\
 & SC & 0.104 & 0.053 & 0.050 & 0.057 & 0.729 & 0.702 & 0.750 & 0.753\\
 & SDID & 0.086 & 0.062 & 0.053 & 0.059 & 0.307 & 0.236 & 0.237 & 0.241\\
 & GSC, $r=1$ & 0.101 & 0.056 & 0.053 & 0.061 & 0.459 & 0.401 & 0.469 & 0.488\\
 & GSC, $r=3$ & 0.093 & 0.057 & 0.056 & 0.060 & 0.279 & 0.173 & 0.127 & 0.110\\
 & IFE-EM, $r=3$ & 0.076 & 0.052 & 0.055 & 0.061 & 0.145 & 0.126 & 0.126 & 0.111\\
\bottomrule
\end{tabular}
\floatnote{Share of the four diagnostic dates (event times $-3$ to $0$) and 1{,}000 replications at which the diagnostic rejects a zero pseudo-effect at $\gamma=0.10$. With exchangeable units, the rejection probability is at most one minus the guaranteed bound. With the treated loading outside the controls' range, the rejection rate measures the diagnostics' power to detect the shift.}
\end{table}

%% file: tables/tab_position.tex
\begin{table}[!ht]\centering\footnotesize
\caption{Coverage and the treated unit's signal}
\label{tab:sims-position}
\setlength{\tabcolsep}{5pt}
\parbox{\linewidth}{A. Exchangeable design, by whether the treated unit's signal lies within the range of the controls' signals}\\[0.6ex]
\begin{tabular*}{\linewidth}{@{\extracolsep{\fill}}lcccccc@{}}
\toprule
 & \multicolumn{3}{c}{$N=39$} & \multicolumn{3}{c}{$N=70$}\\
\cmidrule(lr){2-4}\cmidrule(l){5-7}
 & All & Inside & Outside & All & Inside & Outside\\
\midrule
\emph{Guaranteed bound} & 0.923 & & & 0.900 & & \\
\emph{Replications} & 1{,}000 & 248 & 752 & 1{,}000 & 448 & 552\\
\multicolumn{7}{@{}l}{\emph{Full conformal inference}}\\
\quad DID & 0.928 & 0.980 & 0.911 & 0.896 & 0.948 & 0.854\\
\quad SC & 0.919 & 0.963 & 0.904 & 0.903 & 0.937 & 0.875\\
\quad SDID & 0.924 & 0.939 & 0.919 & 0.902 & 0.920 & 0.887\\
\multicolumn{7}{@{}l}{\emph{Jackknife+}}\\
\quad GSC, $r=1$ & 0.949 & 0.979 & 0.940 & 0.947 & 0.973 & 0.926\\
\quad GSC, $r=3$ & 0.948 & 0.961 & 0.944 & 0.949 & 0.955 & 0.945\\
\quad IFE-EM, $r=3$ & 0.949 & 0.961 & 0.945 & 0.948 & 0.951 & 0.946\\
\bottomrule
\end{tabular*}

\vspace{2ex}
\parbox{\linewidth}{B. Signals held fixed, synthetic control}\\[0.6ex]
\begin{tabular*}{\linewidth}{@{\extracolsep{\fill}}lcccc@{}}
\toprule
 & \multicolumn{2}{c}{$N=17$} & \multicolumn{2}{c}{$N=39$}\\
\cmidrule(lr){2-3}\cmidrule(l){4-5}
 & Full conformal & Jackknife+ & Full conformal & Jackknife+\\
\midrule
\emph{Guaranteed bound} & 0.941 & 0.941 & 0.923 & 0.923\\
\multicolumn{5}{@{}l}{\emph{Four distinct signals. Controls that share the treated unit's signal}}\\
\quad One & 0.932 & 0.934 & 0.892 & 0.927\\
\quad Two & 0.945 & 0.947 & 0.908 & 0.937\\
\quad Five & 0.958 & 0.959 & 0.920 & 0.950\\
\multicolumn{5}{@{}l}{\emph{Random control signals. The treated unit's signal is}}\\
\quad At the controls' average & 0.992 & 0.993 & 0.975 & 0.988\\
\quad Near the control farthest from that average & 0.973 & 0.971 & 0.956 & 0.975\\
\quad Equal to that control's signal & 0.957 & 0.957 & 0.948 & 0.971\\
\bottomrule
\end{tabular*}
\floatnote{Coverage of the true effect at $\gamma=0.10$, averaged over the five post-treatment dates. The guaranteed bound is that of Table~\ref{tab:levels}; it applies to the exchangeable design only. Panel A splits the replications of the exchangeable design by whether the treated unit's unit effect and loadings lie in the convex hull of the controls'. Panel B uses five draws of the signals and 400 replications of the errors for each. In its first three rows, each unit has one of four signals, which are the first four draws of the exchangeable design's unit effects and loadings. The controls that do not share the treated unit's signal are assigned to the other three signals in turn, so the numbers of controls with each of these three signals differ by at most one. In its last three rows, the treated unit's signal near a control is 0.9 times that control's signal plus 0.1 times the other controls' average.}
\end{table}

%% file: tables/tab_position_studentized.tex
\begin{table}[!ht]\centering\footnotesize
\caption{Coverage with two scores, signals held fixed}
\label{tab:sims-position-stud}
\setlength{\tabcolsep}{5pt}
\parbox{\linewidth}{A. Full conformal inference}\\[0.6ex]
\begin{tabular*}{\linewidth}{@{\extracolsep{\fill}}lcccc@{}}
\toprule
 & \multicolumn{2}{c}{$N=17$} & \multicolumn{2}{c}{$N=39$}\\
\cmidrule(lr){2-3}\cmidrule(l){4-5}
Score & Absolute error & Studentized & Absolute error & Studentized\\
\midrule
\emph{Guaranteed bound} & \multicolumn{2}{c}{0.941} & \multicolumn{2}{c}{0.923}\\
\multicolumn{5}{@{}l}{\emph{Four distinct signals. Controls that share the treated unit's signal}}\\
\quad One & 0.932 & 0.947 & 0.892 & 0.930\\
\quad Two & 0.945 & 0.948 & 0.908 & 0.929\\
\quad Five & 0.958 & 0.943 & 0.920 & 0.923\\
\multicolumn{5}{@{}l}{\emph{Random control signals. The treated unit's signal is}}\\
\quad At the controls' average & 0.992 & 0.932 & 0.975 & 0.909\\
\quad Near the control farthest from that average & 0.973 & 0.956 & 0.956 & 0.955\\
\quad Equal to that control's signal & 0.957 & 0.954 & 0.948 & 0.963\\
\multicolumn{5}{@{}l}{\emph{Four distinct signals. Standard deviation of the treated unit's errors}}\\
\quad Equal to the controls' & 0.949 & 0.945 & 0.927 & 0.920\\
\quad Twice the controls' & 0.748 & 0.956 & 0.675 & 0.943\\
\quad Half the controls' & 0.996 & 0.932 & 0.997 & 0.885\\
\bottomrule
\end{tabular*}

\vspace{2ex}
\parbox{\linewidth}{B. Jackknife+}\\[0.6ex]
\begin{tabular*}{\linewidth}{@{\extracolsep{\fill}}lcccc@{}}
\toprule
 & \multicolumn{2}{c}{$N=17$} & \multicolumn{2}{c}{$N=39$}\\
\cmidrule(lr){2-3}\cmidrule(l){4-5}
Score & Absolute error & Studentized & Absolute error & Studentized\\
\midrule
\emph{Guaranteed bound} & \multicolumn{2}{c}{0.941} & \multicolumn{2}{c}{0.923}\\
\multicolumn{5}{@{}l}{\emph{Four distinct signals. Controls that share the treated unit's signal}}\\
\quad One & 0.934 & 0.948 & 0.927 & 0.955\\
\quad Two & 0.947 & 0.949 & 0.937 & 0.953\\
\quad Five & 0.959 & 0.946 & 0.950 & 0.951\\
\multicolumn{5}{@{}l}{\emph{Random control signals. The treated unit's signal is}}\\
\quad At the controls' average & 0.993 & 0.942 & 0.988 & 0.940\\
\quad Near the control farthest from that average & 0.971 & 0.954 & 0.975 & 0.973\\
\quad Equal to that control's signal & 0.957 & 0.952 & 0.971 & 0.979\\
\multicolumn{5}{@{}l}{\emph{Four distinct signals. Standard deviation of the treated unit's errors}}\\
\quad Equal to the controls' & 0.951 & 0.947 & 0.953 & 0.948\\
\quad Twice the controls' & 0.762 & 0.954 & 0.727 & 0.964\\
\quad Half the controls' & 0.996 & 0.944 & 0.999 & 0.925\\
\bottomrule
\end{tabular*}
\floatnote{Coverage of the true effect at $\gamma=0.10$ with synthetic control, as in Panel B of Table~\ref{tab:sims-position}. The first six rows are the designs of that panel. Absolute error: the score is the absolute prediction error. Studentized: the absolute prediction error is divided by the pre-treatment root mean squared residual of the unit's fit. In the last three rows, each of the four signals is shared by about a quarter of the units, and the treated unit's errors are multiplied by 1, 2 or 0.5. Standard errors are at most 0.005.}
\end{table}

%% file: tables/tab_comparators.tex
\begin{table}[!ht]\centering\footnotesize
\caption{Coverage of unit-level and time-based procedures at $\gamma=0.10$}
\label{tab:sims-comparators}
\setlength{\tabcolsep}{3.5pt}
\begin{tabular}{@{}lcccccc@{}}
\toprule
 & \multicolumn{3}{c}{$N=17$} & \multicolumn{3}{c}{$N=39$}\\
\cmidrule(lr){2-4}\cmidrule(l){5-7}
Procedure & Stationary & Variance shift & Random walk & Stationary & Variance shift & Random walk\\
\midrule
\emph{Guaranteed bound, unit-level} & 0.941 & 0.941 & 0.941 & 0.923 & 0.923 & 0.923\\
\midrule
\multicolumn{7}{@{}l}{\emph{Drifting factors (the main design)}}\\
\quad SC, full conformal & 0.938 & 0.941 & 0.933 & 0.919 & 0.923 & 0.923\\
\quad SC, Jackknife+ & 0.942 & 0.947 & 0.940 & 0.945 & 0.949 & 0.950\\
\quad SC, CWZ & 0.866 & 0.719 & 0.716 & 0.866 & 0.755 & 0.750\\
\quad GSC, Jackknife+ & 0.947 & 0.947 & 0.943 & 0.951 & 0.949 & 0.955\\
\quad GSC, parametric bootstrap & 0.908 & 0.904 & 0.899 & 0.902 & 0.905 & 0.914\\
\midrule
\multicolumn{7}{@{}l}{\emph{Stationary factors}}\\
\quad SC, full conformal & 0.952 & 0.950 & 0.942 & 0.930 & 0.927 & 0.915\\
\quad SC, Jackknife+ & 0.955 & 0.955 & 0.944 & 0.955 & 0.949 & 0.941\\
\quad SC, CWZ & 0.900 & 0.743 & 0.764 & 0.881 & 0.738 & 0.742\\
\quad GSC, Jackknife+ & 0.943 & 0.949 & 0.944 & 0.953 & 0.954 & 0.948\\
\quad GSC, parametric bootstrap & 0.899 & 0.906 & 0.903 & 0.889 & 0.900 & 0.904\\
\bottomrule
\end{tabular}
\floatnote{Coverage of the effect, averaged over the five post-treatment dates and 1{,}000 replications; Monte Carlo standard errors are at most 0.011. The signals are those of the exchangeable design (Appendix~\ref{app:sims}), or, in the lower panel, the same with all three factors stationary AR(1) series. The errors are the main design's stationary AR(1) errors; the same errors doubled for every unit at the post-treatment dates; or random walks with the same innovations. CWZ, the test of \citet{ChernozhukovWuthrichZhu2021}, ranks the treated unit's absolute residual at the target date among its pre-treatment ones, with synthetic control refitted under the null. The bootstrap is \code{fect}'s parametric bootstrap for GSC with one factor (200 draws). Both target $0.90$ and have no finite-sample bound.}
\end{table}

%% file: tables/sc_table_prop99.tex
1985 & $-$3 & $-$3.3 & [$-$16.5, 9.9] & [$-$25.1, 17.9] \\
1986 & $-$2 & $-$2.5 & [$-$22.7, 17.6] & [$-$28.6, 19.0] \\
1987 & $-$1 & $-$6.6 & [$-$19.6, 6.5] & [$-$19.6, 10.7] \\
1988 & 0 & $-$3.2 & [$-$19.7, 13.3] & [$-$24.4, 17.8] \\
1989 & 1 & $-$8.4 & [$-$19.4, 6.3] & [$-$25.6, 9.2] \\
1990 & 2 & $-$9.2 & [$-$26.6, 14.0] & [$-$32.4, 14.4] \\
1991 & 3 & $-$12.6 & [$-$29.0, 3.8] & [$-$30.8, 5.5] \\
1992 & 4 & $-$13.7 & [$-$30.1, 2.7] & [$-$35.7, 8.2] \\
1993 & 5 & $-$17.5 & [$-$35.1, 0.1] & [$-$43.2, 8.1] \\
1994 & 6 & $-$22.0 & [$-$40.1, $-$4.0] & [$-$50.2, 6.1] \\
1995 & 7 & $-$22.9 & [$-$47.9, $-$0.9] & [$-$53.8, 8.0] \\
1996 & 8 & $-$24.0 & [$-$54.0, $-$2.1] & [$-$58.9, 10.9] \\
1997 & 9 & $-$26.3 & [$-$59.4, 2.8] & [$-$59.4, 6.9] \\
1998 & 10 & $-$23.3 & [$-$51.1, 0.4] & [$-$53.9, 7.2] \\
1999 & 11 & $-$27.5 & [$-$53.0, $-$2.1] & [$-$55.9, 0.9] \\
2000 & 12 & $-$26.6 & [$-$51.7, $-$1.5] & [$-$58.8, 5.6] \\
\bottomrule

%% file: tables/sc_table_germany.tex
1986 & $-$3 & $-$0.09 & [$-$2.42, 2.24] & [$-$2.42, 2.24] \\
1987 & $-$2 & $-$0.15 & [$-$2.56, 2.27] & [$-$2.56, 2.27] \\
1988 & $-$1 & $-$0.10 & [$-$2.15, 1.96] & [$-$2.15, 1.96] \\
1989 & 0 & $-$0.08 & [$-$1.99, 1.84] & [$-$1.99, 1.84] \\
1990 & 1 & 0.33 & [$-$1.21, 1.86] & [$-$1.28, 1.84] \\
1991 & 2 & 0.75 & [$-$1.54, 3.03] & [$-$1.54, 3.03] \\
1992 & 3 & 0.56 & [$-$1.84, 2.96] & [$-$1.84, 2.96] \\
1993 & 4 & $-$0.19 & [$-$2.41, 2.02] & [$-$2.57, 1.85] \\
1994 & 5 & $-$0.65 & [$-$3.04, 1.74] & [$-$3.25, 1.53] \\
1995 & 6 & $-$0.79 & [$-$3.57, 1.99] & [$-$3.84, 1.71] \\
1996 & 7 & $-$1.01 & [$-$4.00, 1.97] & [$-$4.31, 1.66] \\
1997 & 8 & $-$1.59 & [$-$4.98, 1.80] & [$-$5.04, 1.84] \\
1998 & 9 & $-$1.77 & [$-$5.05, 1.52] & [$-$5.42, 1.54] \\
1999 & 10 & $-$2.06 & [$-$6.47, 2.36] & [$-$6.97, 2.38] \\
2000 & 11 & $-$2.58 & [$-$11.86, 6.69] & [$-$11.72, 6.84] \\
2001 & 12 & $-$2.76 & [$-$12.14, 6.63] & [$-$11.98, 6.79] \\
2002 & 13 & $-$2.95 & [$-$11.16, 5.26] & [$-$11.01, 5.41] \\
2003 & 14 & $-$3.45 & [$-$11.63, 4.74] & [$-$11.47, 4.89] \\
\bottomrule

%% file: tables/edr_table.tex
ME & 1976 & 10 & 8.4 & 5.5 [0.6, 10.3] & 8.4 [$-$7.8, 24.3] \\
MN & 1976 & 10 & 4.0 & 4.0 [$-$0.9, 8.8] & 5.3 [$-$11.1, 21.3] \\
WI & 1976 & 10 & 9.4 & 6.6 [1.7, 11.4] & 14.3 [$-$2.4, 30.2] \\
ID & 1996 & 5 & $-$5.0 & 1.0 [$-$4.7, 7.0] & $-$7.0 [$-$19.7, 5.2] \\
NH & 1996 & 5 & 7.6 & 2.5 [$-$3.2, 8.5] & 11.3 [$-$1.2, 23.1] \\
WY & 1996 & 5 & 3.9 & 8.6 [2.9, 14.5] & 0.8 [$-$11.7, 12.7] \\
IA & 2008 & 2 & 1.9 & 0.2 [$-$9.2, 9.8] & 3.6 [$-$5.7, 13.0] \\
MT & 2008 & 2 & $-$4.3 & $-$4.1 [$-$13.8, 5.2] & $-$4.5 [$-$14.0, 4.7] \\
CT & 2012 & 1 & $-$0.8 & $-$0.8 [$-$9.1, 7.3] & --- \\
\bottomrule